\documentclass[aps,groupedaddress,amssymb,superscriptaddress,amsmath,amsbsy,amsfonts,twocolumn,pra]{revtex4-1}

\usepackage{amssymb,amsmath,amsthm,color,graphicx,times,graphicx}
\usepackage{hyperref}
\usepackage{subfigure}
\usepackage{times}
\usepackage{bbold}
\usepackage{color}

\usepackage{hyperref}
\usepackage{subfigure}
\usepackage{times}
\usepackage{bbold}
\usepackage{color}
\usepackage{array}
\usepackage{array,multirow,makecell}
\usepackage[table]{xcolor}

\theoremstyle{plain}

\theoremstyle{definition}

\begin{document}

\title{Enhancing the performance of coupled quantum Otto thermal machines without entanglement and quantum correlations}
\author{Abdelkader El Makouri}\affiliation{LPHE-Modeling and Simulation, Faculty of Sciences, Mohammed V University in Rabat, Morocco.}
\author{Abdallah Slaoui}\affiliation{LPHE-Modeling and Simulation, Faculty of Sciences, Mohammed V University in Rabat, Morocco.}\affiliation{Centre of Physics and Mathematics, CPM, Faculty of Sciences, Mohammed V University in Rabat, Rabat, Morocco.}
\author{Mohammed Daoud}\affiliation{Department of Physics, Faculty of Sciences, University Ibn Tofail, Kenitra, Morocco.}

\begin{abstract}
We start with a revision study of two coupled spin-$1/2$ under the influence of Kaplan-Shekhtman-Entin-Wohlman-Aharony (KSEA) interaction and a magnetic field. We first show the role of idle levels, i.e., levels that do not couple to the external magnetic field, when the system is working as a heat engine as well as when it is a refrigerator. Then we extend the results reported in [\href {\doibase https://doi.org/10.1103/PhysRevE.92.022142} {\bibfield  {journal} {\bibinfo  {journal}{PRE.}\ }\textbf {\bibinfo {volume} {92}}, (\bibinfo {year} {2015})\ \bibinfo {pages}{022142}}\BibitemShut {NoStop}] by showing that it is not necessary to change both the magnetic field as well as the coupling parameters to break the extensive property of the work extracted globally from two coupled spin-$1/2$ as has been demonstrated there. Then we study the role of increasing the number of coupled spins on efficiency, extractable work, and coefficient of performance (COP). First, we consider two- and three-coupled spin-$1/2$ Heisenberg $\mathrm{XXX}$-chain. We prove that the latter can outperform the former in terms of efficiency, extractable work, and COP. Then we consider the Ising model, where the number of interacting spins ranges from two to six. We show that only when the number of interacting spins is odd the system can work as a heat engine in the strong coupling regime. The enhancements in efficiency and COP are explored in detail. Finally, this model confirms the idea that entanglement and quantum correlations are not the reasons behind the enhancements observed in efficiency, extracatable work, and COP, but only due to the structure of the energy levels of the Hamiltonian of the working substance. In addition to this, the extensive property of global work as well, is not affected by entanglement and quantum correlations.
\end{abstract}
\date{\today}

\maketitle
\section{Introduction}
Classical thermodynamics \cite{Prigogine} was developed in the $\mathrm{19^{th}}$ century to enhance the efficiency of heat engines \cite{Carnot}. On the other hand, quantum thermodynamics \cite{Kosloff1984, Rezek, Deffner, Gemmer, Binder, Goold, Kosloff, Anders, Alicki, Millen, Quan, Quan1, Allahverdyan, Levy, Skrzypczyk, Mitchison}, is an emergent field that aims to extend thermodynamics to the regime where the working substance is quantum in nature and quantum effects are no longer negligible. In this field, particular attention has been devoted to the study of thermal machines fabricated from tiny quantum systems to improve their performance. This upsurge of interest started with the first work of Scovil and Schulz-DuBois when they showed that a three-level system is equivalent to a quantum heat engine \cite{Scovil}. Since then, a lot of work has been developed to understand thermal machines at the quantum scale. Furthermore, in 2016, the first experimental realization of a quantum heat engine was demonstrated using a trapped ion as a working fluid \cite{Rossnagel1, Rossnagel2, Rossnagel}. In 2017, the same platform was used to fabricate a quantum absorption refrigerator made of only three trapped ions \cite{Maslennikov}. Other platforms to realize quantum thermal machines experimentally could be found in \cite{Bariani, Sothmann, QuanZhang, AltintasHardal, Peterson, Uzdin, Fazio,DVon}, and more recently, all of these platforms have been collected in an article review \cite{MayersDeffner}. When the baths are not thermal \cite{ScullyZubairy, Dillenschneider, HuangWang, AbahLutz, RossnagelAbah, HardalM, Niedenzu, Manzano, Klaers, Agarwalla, ScullyMarlan}, it has been shown that the efficiency of these quantum thermal machines could surpass the thermodynamic Carnot bound. However, there is no violation of the laws of thermodynamics since in this case we have used resources, e.g., coherence, squeezing, and quantum correlations.\par

Many studies have been done since the first work done by Kieu \cite{Kieu, Kieu2} on a quantum heat engine (QHE) composed of one qubit undergoing an Otto cycle composed of two adiabatic and two isochoric stages in which he clarifies some important aspects of the second law of thermodynamics. First, Zhang et al. have extended the same work to two coupled spin-$1/2$ Heisenberg XXX chain \cite{Zhang} where the exchange coupling $\mathrm{J}$ is the parameter changed in the adiabatic stages, and then in Ref.\cite{Johal} where the external magnetic field $\mathrm{h}$ is the parameter changed in the adiabatic stages. The studies in this direction could be stated as follows: in Ref.\cite{Altintas} the authors investigated the case when one of the composite system is a spin-$1/2$ particle and the other is a qudit, in Ref.\cite{Ukraine} when both interacting spins are qudits where the role of negative temperature on the efficiency has been investigated in detail, in Ref.\cite{Mehta} the role of degeneracy, in Ref.\cite{Talkner, Anka, DasGosh} when a thermal bath being replaced by a quantum measurement as a new source of heat, in Ref.\cite{GuoZhang, Zhao, Sodeif} the role of Dzyaloshinski-Moriya interaction, in Ref.\cite{Ramandeep} the authors examine the performance of a two-coupled spins of arbitrary magnitudes (where majorization \cite{Marshall} has been used to find the upper bound of efficiency, which was an open question to be answered), and last but not least, the role of non-adiabaticity (i.e., when the unitary strokes are done in finite time) for one spin and two coupled spin-$1/2$ particles \cite{Johall, Cakmak, CampisiM, Cakmak2, Thomas}.\par

Recently, it has been shown in \cite{Johal} that the efficiency of two coupled spin-$1/2$ particles is greater than the one of the Otto system. And since quantum systems are known for strange phenomena, e.g., quantum coherence, entanglement and quantum correlations, special attention has been devoted to understanding the role of entanglement and quantum correlations in efficiency and extractable work. An investigation of the role of entanglement and quantum correlations has been done in \cite{Chang, Altintas1, Hewgill, Zhang}. However, even with these works, there was no sufficient and clear answer until recently, when Oliveira and Jonathan \cite{Oliveira2020} provided an explanation by demonstrating that the enhancement observed in efficiency is solely due to the presence of idle levels. Moreover, in Ref. \cite{Anka} the same enhancement was proved even for a three-level system, a system that cannot be divided into subsystems. However, in our opinion, a revisited study and more clarifications are needed to gain more insights in this direction.\par

Motivated by these works, we use the Ising model (with and without KSEA interaction) and the Heisenberg model to strengthen and demonstrate the correctness of their reasoning \cite{Anka, Oliveira2020, Johall}. First, we use as a working fluid two coupled spin-$1/2$ particles under the influence of KSEA interaction \cite{Yurischev,Yildirim, Kaplan,Aharony,Shekhtman} and a magnetic field along the $\mathrm{z}$-direction. The KSEA interaction arises from spin-orbit coupling. We will see that when we have only quantum entanglement and working levels (i.e., levels that do couple to the external magnetic field), there is no enhancement in efficiency as it is expected. However, when the $\mathrm{z}$-component of $\mathrm{J}$ is not equal to zero, two idle levels emerge, and we see that there is an enhancement in the efficiency which is due to heat passing through these levels from the cold to the hot bath. Then we show how idle levels affect the COP when the system is working as a refrigerator. We show that they shift the enhancement in COP to large values of $\mathrm{\Gamma_{z}}$. When the system is working as a refrigerator, it has been studied in \cite{Sodeif,Deniz}, but in contrast to them, we explore the role of idle levels as well as the role of increasing the number of interacting spins on the COP, two situations that have not been explored there. Moreover, taking advantage of this model, we generalize the results reported in \cite{Altintas}, by showing that it is not necessary to change both the external magnetic field and the coupling parameters to break the extensive property of the global extractable work.\par

Secondly, we will increase the number of interacting systems to more than two coupled spin-$1/2$ up to six. We first compare the efficiency, extractable work, and COP of two- and three-coupled Heisenberg XXX chain. Our results show that three coupled spins could harvest more work than two coupled spin-$1/2$. Even more, the former could work as a heat engine in the strong coupling regime, which is not the case for the latter. When it comes to the COP, three coupled spins could surpass that of the Otto, which is not the case with two-coupled spin-$1/2$. Then we use the Ising model in which the number of interacting spins ranges from two to six spin-$1/2$ particles. We show that, in contrast to \cite{Altintas}, when we increase the number of coupled spins, we see a remarkable enhancement in the extractable work. The efficiency will be enhanced only for small values of the coupling parameter $\mathrm{J}$. Even more, a new conclusion has been drawn, which is that only when the number of interacting spins is odd, the system can work as a heat engine in the strong coupling regime. When it comes to COP, we found that when the system is ferromagnetic, there is a remarkable enhancement in the COP even though for three- to six-coupled spin-$1/2$ they are nearly coincidable. When the number of interacting spins ranges from two to six, we only consider the $\mathrm{z}$-component of $\mathrm{J}$ to ensure that no entanglement or quantum correlations will build up along the cycle. More precisely, the coupled system will be only in a statistical mixture of factorized states. Therefore, the enhancements in efficiency, extractable work, and COP are only due to the structure of the energy levels of the system. Note that for all models chosen and studied in this work, we have degenerate eigenvalues, where in Ref.\cite{DasGosh} the role of degneracy on efficiency and extractable work has been explored in detail.\par

The rest of the article is organized as follows: In section \ref{A} a brief review of the quantum Otto cycle and the relevant thermodynamic quantities to characterize the Otto heat engine and refrigerator is given. In section \ref{B} we study the role of KSEA interaction on the efficiency, local and global extractable works and COP. The comparison between the efficiency, COP and extractable work from two and three up to six-coupled spin-$1/2$ is done in the section \ref{C}. Then in section \ref{clarify}, we clarify some very important points. And finally, in section \ref{D} we give a summary of our results with future directions.
 
\section{Quantum Otto cycle}\label{A}
Before we start presenting our results, we should provide some necessary definitions and expressions used in this paper. Suppose we have a quantum system $\mathrm{S}$ described by the state $\mathrm{\rho}$ and a Hamiltonian $\mathrm{H}$. The expectation value of the measured energy of $\mathrm{S}$ is
\begin{equation}
\mathrm{U=\langle E\rangle=Tr(\rho H)=\sum_{i}p_{i}E_{i}},
\end{equation}
where $\mathrm{E_{i}}$ are the energy levels and $\mathrm{p_{i}}$ are the corresponding occupation probabilities. The derivative of $\mathrm{U}$ gives
\begin{equation}
\mathrm{dU=\sum_{i}E_{i}dp_{i}+p_{i}dE_{i},}
\end{equation}
which can be divided into heat and work given, respectively, as follows; $\mathrm{\delta Q=\sum_{i}E_{i}dp_{i}}$ and $\mathrm{\delta W=\sum_{i}p_{i}dE_{i}}$ \cite{Quan, Kieu, Alicki}. From these definitions, the average heat and work are: $\mathrm{Q= \sum_{i} \int E_{i}dp_{i}}$ and $\mathrm{W=\sum_{i}\int p_{i}dE_{i}}$. Therefore, the first law of thermodynamics reads
\begin{equation}
\mathrm{dU=\delta Q+\delta W.}
\end{equation}
Mathematically speaking, $\mathrm{dU}$ is an exact differential, however, $\mathrm{\delta Q}$ and $\mathrm{\delta W}$ are not total differentials but are path dependent.\par

The Otto cycle is composed of four stages: two adiabatic stages in which an external controlled parameter is varied, and two isochoric stages in which the system is in contact with a heat bath. These stages are given as follows. $\mathbf{Stage \ 1:}$ The occupation probabilities of each level are $\mathrm{p'_{i}}$. The system is put in contact with a hot bath at tempereture $\mathrm{T_{h}}$ until it reaches thermal equilibrium. Then it is decribed with the new occupation probabilities $\mathrm{p_{i}}$. Since at this stage we have only a change in the occupation probabilities, only heat is exchanged between the working substance and the hot bath. $\mathbf{Stage \ 2:}$ The system is isolated from the hot bath and the magnetic field is changed from $\mathrm{h}$ to $\mathrm{h'}$ (with $\mathrm{h >h'}$). This transformation will be done slowly to ensure the holding of the quantum adiabatic theorem \cite{Kieu}. At this stage, only work is performed since the occupation probabilities stay the same. $\mathbf{Stage \ 3:}$ The system is put in contact with a cold bath at temperature $\mathrm{T_{c}}$($\mathrm{T_{h}>T_{c}}$) until it reaches equilibrium with it. In this case we have a change in the occupation probabilities from $\mathrm{p_{i}}$ to $\mathrm{p'_{i}}$. At this stage, only heat is exchanged. $\mathbf{Stage \ 4:}$ The system is again isolated from the cold bath and the magnetic field is changed back from $\mathrm{h'}$ to $\mathrm{h}$. At this stage, only work is performed. After this stage, the system is again attached to the hot bath at temperature $\mathrm{T_{h}}$ to complete the cycle and return the system to its initial state. Furthermore, note that the Otto cycle is the most commonly used cycle to study quantum thermal machines. The reason is that in this cycle, the system at each stage exchanges either heat or work, not both of them, which leaves no ambiguity in the identification of them correctly. In addition, the definitions of heat and work used in this paper are only valid in the weak coupling regime.\par

When the system is working as a heat engine (for more details see, \cite{Kieu, Zhang, Kieu2}), we must have $\mathrm{Q_{h}}>0$, $\mathrm{Q_{c}}<0$ and $\mathrm{W}>0$. In this case, in $\mathbf{Stage \ 1}$ the system will absorb heat from the hot bath. Its expression is given by
\begin{equation}
\mathrm{Q_{h}=\sum_{i}E_{i}(p_{i}-p_{i}'),}
\end{equation}
where $\mathrm{E_{i}}$ are the eigenvalues of the system at the hot bath side, $\mathrm{p_{i}}$ and $\mathrm{p_{i}'}$ are the populations of the system when it is in contact with the hot bath and the cold bath, respectively. In $\mathbf{Stage \ 3}$ the system will release heat to the cold bath, which is given as follows
\begin{equation}
\mathrm{Q_{c}=\sum_{i}E_{i}'(p_{i}'-p_{i}).}
\end{equation}
From the first law of thermodynamics, we have
\begin{equation}
\mathrm{W=Q_{h}+Q_{c}=\sum_{i}(E_{i}-E_{i}')(p_{i}-p_{i}').}
\end{equation}
Therefore, the expression of efficiency is
\begin{equation}
\mathrm{\eta=\frac{W}{Q_{h}}=1+\frac{Q_{c}}{Q_{h}}.}
\label{eta}
\end{equation}
When the system is working as a refrigerator, we must have $\mathrm{Q_{h}}<0$, $\mathrm{Q_{c}}>0$ and $\mathrm{W}<0$, in other words, the system is running in the reverse of the Otto heat engine cycle. In this case, the system is characterized by its COP, which is given as follows
\begin{equation}
\mathrm{COP=\frac{Q_{c}}{|W|}=\frac{Q_{c}}{|Q_{h}+Q_{c}|}.}
\label{COP}
\end{equation}
We should note that the efficiency of the Otto cycle when the working medium is either a single or (multiple but uncoupled) spin(s) or harmonic oscillator(s) is given by $\mathrm{\eta_{o}=1-h'/h}$. In the same manner, when the system is working as a refrigerator $\mathrm{COP_{o}=\frac{h'}{h-h'}}$, see Refs.\cite{Quan,Quan1}. In this paper, the units are chosen such that $\mathrm{k_{B} =\hbar= 1}$.
\section{Revisited study of a coupled quantum Otto heat engine and refigerator}\label{B}

Our working fluid in this section is a two-coupled spin-$1/2$ $\mathrm{1D}$ Ising model with a $\mathrm{z}$-component KSEA interaction parameter $\mathrm{\Gamma_{z}}$ under the influence of a magnetic field $\mathrm{h}$ in the $\mathrm{z}$-direction. The expression of the Hamiltonian describing this system is \cite{Moriya}
\begin{equation}
\mathrm{H=J_{z}\sigma_{z}^{1}\sigma_{z}^{2}+\Gamma_{z}(\sigma_{x}^{1}\sigma_{y}^{2}+\sigma_{y}^{1}\sigma_{x}^{2})+h(\sigma_{z}^{1}+\sigma_{z}^{2}),}\label{H9}
\end{equation}
where $\mathrm{\sigma_{x,y,z}^{i}}$ are the standard Pauli matrices acting on the site $i\in[1,2]$. The eigenvalues of $\mathrm{H}$ are: $\mathrm{E_{1}=E_{2}=-J_{z}}$, $\mathrm{E_{3}=J_{z}-2\sqrt{h^{2}+\Gamma_{z}^{2}}}$ and $\mathrm{E_{4}=J_{z}+2\sqrt{h^{2}+\Gamma_{z}^{2}}}$. Their corresponding eigenstates in the standard basis $\mathrm{\{|11\rangle,|10\rangle,|01\rangle,|00\rangle\}}$ are: $\mathrm{|\psi_{1}\rangle=|10\rangle}$, $\mathrm{|\psi_{2}\rangle=|01\rangle}$, $|\psi_{3}\rangle=\frac{1}{\sqrt{|\alpha_{1}|^{2}+1}}(\alpha_{1}|11\rangle+|00\rangle)$ and $|\psi_{4}\rangle=\frac{1}{\sqrt{|\alpha_{2}|^{2}+1}}(\alpha_{2}|11\rangle+|00\rangle)$, with $\mathrm{\alpha_{1}=i(-h+\sqrt{h^{2}+\Gamma_{z}^{2}})/\Gamma_{z}}$ and $\mathrm{\alpha_{2}=-i(h+\sqrt{h^{2}+\Gamma_{z}^{2}})/\Gamma_{z}}$. The KSEA interaction has been chosen for two reasons. First, when the $\mathrm{z}$-component of $\mathrm{J}$ is equal to zero, we have entanglement-(between the two coupled spins described by Eq.\ref{H9} when they are in thermal equilibrium with either the hot bath or the cold bath, see Eq.\ref{rhoth})-but no idle levels. More precisely, here we want to test if the presence of only entanglement or quantum correlations could boost the efficiency beyond that of the Otto. Second, for this Hamiltonian, we see that the eigenstates are dependent on the magnetic field $\mathrm{h}$ and $\mathrm{\Gamma_{z}}$ and that the eigenvalues are non-linear in $\mathrm{h}$. This situation is different from the cases considered in \cite{Zhang, Johal, Ukraine, DasGosh}. Below, we see that this influences the extensive property of the work extracted globally. Moreover, as far as we know this interaction has not been considered in the previous works to the authors' knowledge.\par 

When the system in a thermal equilibrium with a heat bath at inverse temperature $\mathrm{\beta}$, it can be described by the density operator $\mathrm{\rho_{th}=\sum_{i}p_{i}|\psi_{i}\rangle\langle \psi_{i}|}$. In the computational basis $\{|11\rangle,|10\rangle,|01\rangle,|00\rangle\}$, this matrix is given as follows
\begin{equation}
\begin{split}
\mathrm{\rho_{th}}=\begin{pmatrix}
\frac{|\alpha_{1}|^{2}}{|\alpha_{1}|^{2}+1} \mathrm{p_{3}}+\frac{|\alpha_{2}|^{2}}{|\alpha_{2}|^{2}+1}\mathrm{p_{4}} & 0 & 0 & \frac{\alpha_{1}^{\star}}{|\alpha_{1}|^{2}+1} \mathrm{p_{3}}+\frac{\alpha_{2}^{\star}}{|\alpha_{2}|^{2}+1}\mathrm{p_{4}} \\
0 & p_{1} & 0 & 0 \\
 0 & 0 & p_{1} & 0 \\
\frac{\alpha_{1}}{|\alpha_{1}|^{2}+1} \mathrm{p_{3}}+\frac{\alpha_{2}}{|\alpha_{2}|^{2}+1}\mathrm{p_{4}} & 0 & 0 & \frac{1}{|\alpha_{1}|^{2}+1} \mathrm{p_{3}}+\frac{1}{|\alpha_{2}|^{2}+1}\mathrm{p_{4}} \\
\end{pmatrix},\label{rhoth}
\end{split}
\end{equation}
with $\mathrm{p_{i}=e^{-\beta E_{i}}/Z}$. The partition function is $\mathrm{Z=2(e^{\beta J_{z}}+e^{-\beta J_{z}}\cosh(2\beta \sqrt{h^{2}+\Gamma_{z}^{2}}))}$, where $\mathrm{\beta=1/k_{B}T}$, $\mathrm{k_{B}}$ is the Boltzmann constant and $T$ is the temperature. Therefore to get the expression of the density matrix of the system at the $\mathbf{Stages}$ $1$ and $3$ we only have to replace $\mathrm{\beta}$ by $\mathrm{\beta_{h}}$ and $\mathrm{\beta_{c}}$, respectively.

To arrive to the expression of this state, we have to follow the next steps. First, we take the expression of $\rm \rho_{th}=p_{1}|\psi_{1}\rangle\langle\psi_{1}|+p_{2}|\psi_{2}\rangle\langle\psi_{2}|+p_{3}|\psi_{3}\rangle\langle\psi_{3}|+p_{4}|\psi_{4}\rangle\langle\psi_{4}|$, then we replace the eigenstates $|\psi_{i}\rangle$ by their expressions, we get
\begin{equation}
\begin{split}
& \hspace{0.2cm} \rm \rho_{th}=\rm p_{1}|10\rangle\langle10|+p_{2}|01\rangle\langle01|
\\ &
+\rm p_{3}\left( \frac{|\alpha_{1}|^{2}|11\rangle\langle11|+\alpha_{1}|11\rangle\langle00|+\alpha_{1}^{\star}|00\rangle\langle11|+|00\rangle\langle00|}{|\alpha_{1}^{2}|+1}\right)
\\ &
\rm +p_{4}\left( \frac{|\alpha_{2}|^{2}|11\rangle\langle11|+\alpha_{2}|11\rangle\langle00|+\alpha_{2}^{\star}|00\rangle\langle11|+|00\rangle\langle00|}{|\alpha_{2}^{2}|+1}\right).
\end{split}
\end{equation}
Then, rearranging this expression, one obtains equation \ref{rhoth}. Furthermore, note that the Hamiltonian given in equation \ref{H9} it only describes the coupling between the spins, i.e., it does not include the properties of the implicitly assumed heat bath, since we are interested in the thermodynamics of the cycle in the long time limit, when the system completely gets thermalized. However, if the bath is included explicitly, the total Hamiltonian will be given as follows: $\mathrm{H_{T}=H_{S}+H_{B}+H_{SB}}$, with $\mathrm{H_{B}}$ is the Hamiltonian of the bath and $\mathrm{H_{BS}}$ is the Hamiltonian describing the interaction between the S and B. If the interaction between the bath and the system is assumed to be weak and markovian then the system would thermalizes to the Gibbs state $\mathrm{\rho_{th}=\frac{e^{-\beta H}}{Tr(e^{-\beta H})}}$. In general, the evolution of the state of the system is given by the next equation,
\begin{equation}
\mathrm{\frac{\partial \rho}{\partial t}=-i[H,\rho]+\mathcal{L}(\rho),}
\label{drho}
\end{equation}
with the first term describing the unitary evolution of the system when there is no interaction with the heat bath and the second describes the effect of the heat bath on the system. The thermal state given in Eq. \ref{rhoth} is the steady state of this equation. Furthermore, since it is not trivial to assume thermalization, one can suppose that each one of the coupled spins is coupled to a heat bath and assuming that both the reservoirs are at the same temperature T and lead to the same decohering rates $\mathrm{\gamma}$, then one will compute the distance between the state of the system as a result of this interaction $\mathrm{\rho}$ (this state will be obtained from Eq. \ref{drho}) and $\mathrm{\rho_{th}}$ (given in Eq. \ref{rhoth}) given by the fidelity $\mathrm{F(\rho_{th},\rho)=[\mathrm{Tr}(\sqrt{\sqrt{\rho_{th}}\rho\sqrt{\rho_{th}}})]^{2}}$, and when $\mathrm{F(\rho_{th},\rho)}\rightarrow$ 1 (depending on the chosen values the parameters $\mathrm{T, J_{z}, \Gamma_{z}, h and \gamma}$), one would be sure that the system is now thermalized (see, Ref. \cite{Altintas}).

Below, when the system is working as a heat engine, the parameter values for the plot are $\mathrm{T_{h}=4}$, $\mathrm{T_{c}=1}$, $\mathrm{h=4}$ and $\mathrm{h'= 3}$, on the other hand, when it is working as a refrigerator, the parameter values for the plot are $\mathrm{T_{h} = 2}$, $\mathrm{T_{c} = 1}$, $\mathrm{h = 5}$ and $\mathrm{h'=2}$. These parameters have been chosen to ensure that $\mathrm{h>h'}$ and $\mathrm{h/T_{h}<h'/T_{c}}$ for the system to function as a heat engine \cite{Kieu, Kieu2}, $\mathrm{h>h'}$ and $\mathrm{h/T_{h}>h'/T_{c}}$ for the system to function as a refrigerator. This is for uncoupled systems, however, for coupled ones, the conditions become more involved because of the coupling.
\subsection{Global description}
Shifting the eigenvalues of the Hamiltonian $\mathrm{H}$ will not alter heats and work \cite{Quan}. They become: $\mathrm{E_{1}=E_{2}=-2J_{z}}$, $\mathrm{E_{3}=-2\sqrt{h^{2}+\Gamma_{z}^{2}}}$ and $\mathrm{E_{4}=2\sqrt{h^{2}+\Gamma_{z}^{2}}}$. Therefore, the expressions of the global thermodynamical quantities, i.e., heats and work, are
\begin{widetext}
\begin{equation}
\mathrm{Q_{h}=\frac{2\sqrt{h^{2}+\Gamma_{z}^{2}} \sinh(2\beta_{c}\sqrt{h'^{2}+\Gamma_{z}^{2}})+2J_{z}e^{2\beta_{c}J_{z}}}{e^{2\beta_{c} J_{z}}+\cosh(2\beta_{c} \sqrt{h'^{2}+\Gamma_{z}^{2}})}-\frac{2\sqrt{h^{2}+\Gamma_{z}^{2}} \sinh(2\beta_{h}\sqrt{h^{2}+\Gamma_{z}^{2}})+2J_{z}e^{2\beta_{h}J_{z}}}{e^{2\beta_{h} J_{z}}+\cosh(2\beta_{h} \sqrt{h^{2}+\Gamma_{z}^{2}})},}
\end{equation}
\begin{equation}
\mathrm{Q_{c}=\frac{-2\sqrt{h'^{2}+\Gamma_{z}^{2}} \sinh(2\beta_{c}\sqrt{h'^{2}+\Gamma_{z}^{2}})-2J_{z}e^{2\beta_{c}J_{z}}}{e^{2\beta_{c} J_{z}}+\cosh(2\beta_{c} \sqrt{h'^{2}+\Gamma_{z}^{2}})}+\frac{2\sqrt{h'^{2}+\Gamma_{z}^{2}} \sinh(2\beta_{h}\sqrt{h^{2}+\Gamma_{z}^{2}})+2J_{z}e^{2\beta_{h}J_{z}}}{e^{2\beta_{h} J_{z}}+\cosh(2\beta_{h} \sqrt{h^{2}+\Gamma_{z}^{2}})},}
\end{equation}
\begin{equation}
\mathrm{W=2(\sqrt{h^{2}+\Gamma_{z}^{2}}-\sqrt{h'^{2}+\Gamma_{z}^{2}})(\frac{\sinh(2\beta_{c}\sqrt{h'^{2}+\Gamma_{z}^{2}})}{e^{2\beta_{c} J_{z}}+\cosh(2\beta_{c} \sqrt{h'^{2}+\Gamma_{z}^{2}})}-\frac{\sinh(2\beta_{h}\sqrt{h^{2}+\Gamma_{z}^{2}})}{e^{2\beta_{h} J_{z}}+\cosh(2\beta_{h} \sqrt{h^{2}+\Gamma_{z}^{2}})}).}
\label{Wo}
\end{equation}
\end{widetext}
From these formulas and Eqs. (\ref{eta}) and (\ref{COP}), it is straightforward to get the formulas of the efficiency as well as the COP. However, since their expressions are too long, we will not write them here. Moreover, since here the thermodynamical quantities are unchanged when substituting $\mathrm{\Gamma_{z}}$ by -$\mathrm{\Gamma_{z}}$ we only plot them for positive values of $\mathrm{\Gamma_{z}}$. For negative values of $\mathrm{J_{z}}$ we found no enhancement, so we plotted them only for positive values of $\mathrm{J_{z}}$.

\begin{figure}[hbtp]
\centering
\includegraphics[scale=0.7]{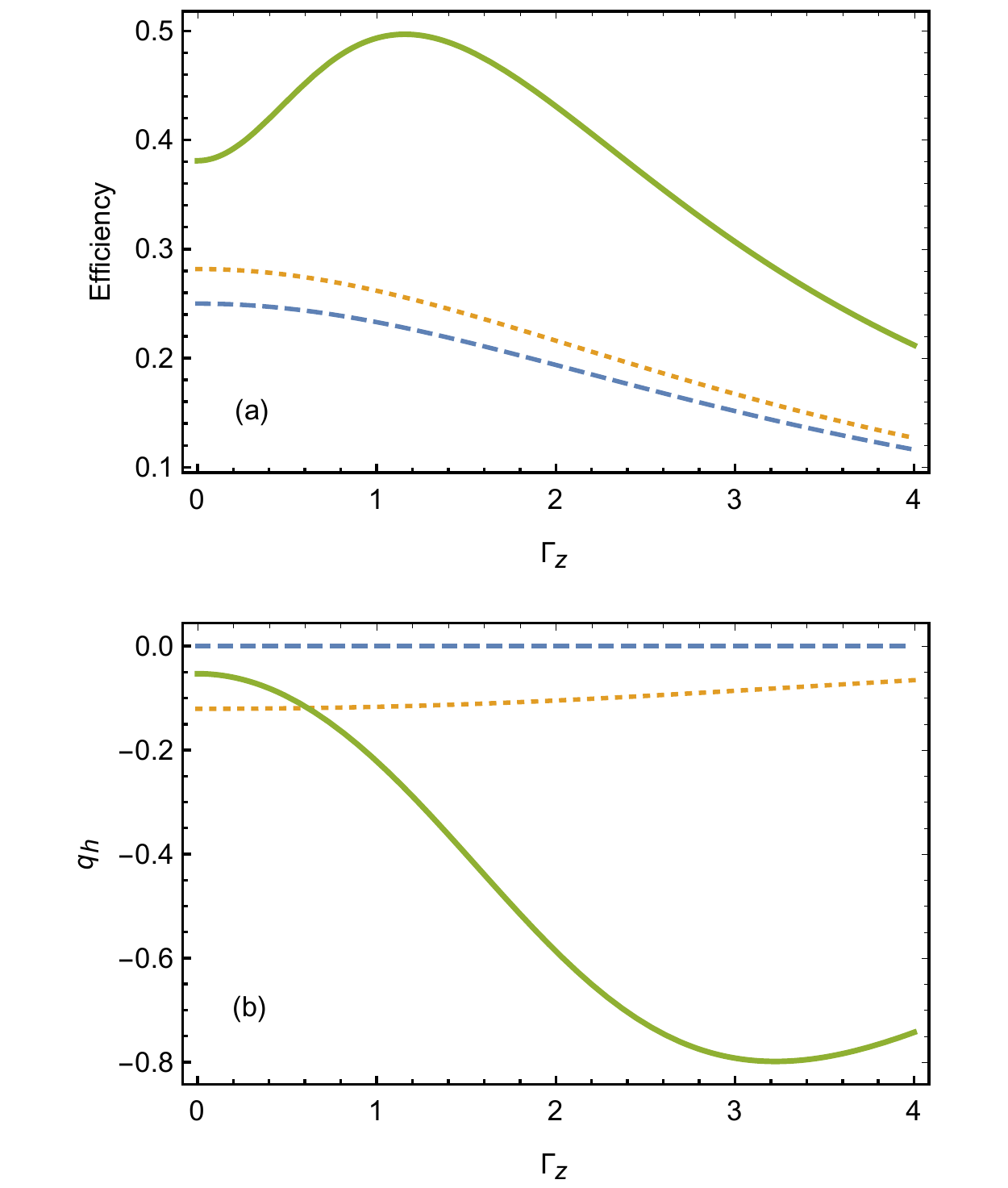}
\caption{(Color online). Plot of (a) efficiency versus $\mathrm{\Gamma_{z}}$ for three different values of $\mathrm{J_{z}=0}$ (blue dashed line), $\mathrm{J_{z}=0.5}$ (orange dotted line) and $\mathrm{J_{z}=2.6}$ (green solid line) (b) the heat absorbed by idle levels $\mathrm{q_{h}}$ from the hot bath as a function of $\mathrm{\Gamma_{z}}$ for $\mathrm{J_{z}=0}$, 0.5 and 2.6. Note that $\mathrm{q_{h}}$ here is the heat absorbed globally from the hot bath by idle levels and should be distinguished from the local heats given by, $\mathrm{q_{h1}}$ and $\mathrm{q_{h2}}$. The parameter values for the plots are $\mathrm{T_{h}=4}$, $\mathrm{T_{c}=1}$, $\mathrm{h=4}$ and $\mathrm{h'=3}$. Note that if one plot $\mathrm{\eta}$ and $\mathrm{q_{h}}$ versus $\mathrm{J_{z}}$ for different values of $\mathrm{\Gamma_{z}}$ will get the same plots.}
\label{1}
\end{figure}
\begin{figure}[hbtp]
\centering
\includegraphics[scale=0.6]{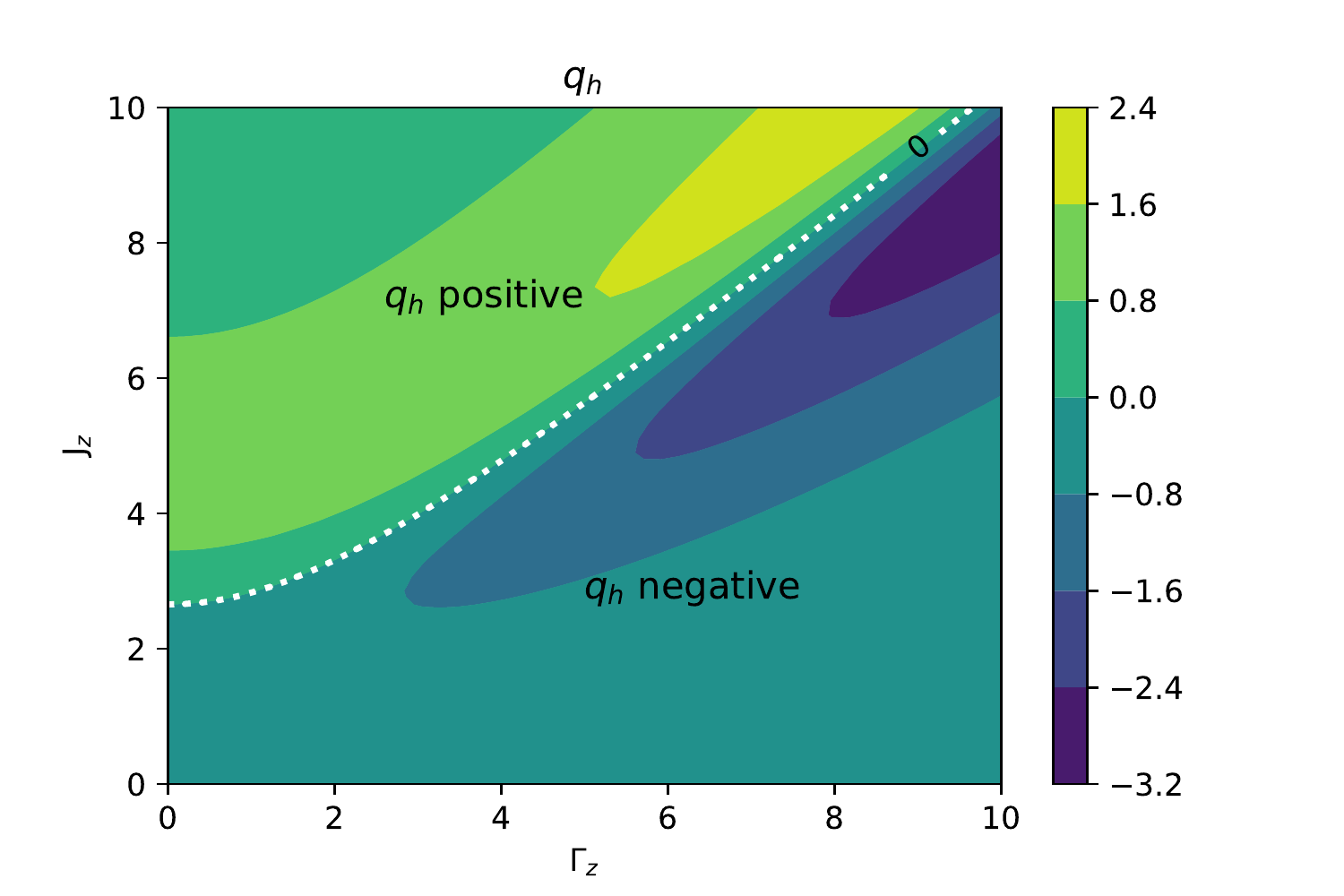}
\caption{(Color online). Contour plot of $\rm{q_{h}}$ as a function of $\rm{J_{z}}$ and $\rm{\Gamma_{z}}$, where the parameter values are the same as in Fig.\ref{1}. We see that for an arbitrary value of $\rm{\Gamma_{z}}$, there is a maximal value of $\rm{J_{z}}$ above which $\rm{q_{h}}$ will become positive. The blue dashed line represents Eq.\ref{star}. For example, at $\rm{\Gamma_{z}}=0$ and when $\rm{J_{z}}$ exceeds 2.65, then $\rm{q_{h}}$ becomes positive. In this case, the efficiency would be less than $\eta_{o}$.}
\label{Idle}
\end{figure}

\begin{figure}[hbtp]
\centering
\includegraphics[scale=0.7]{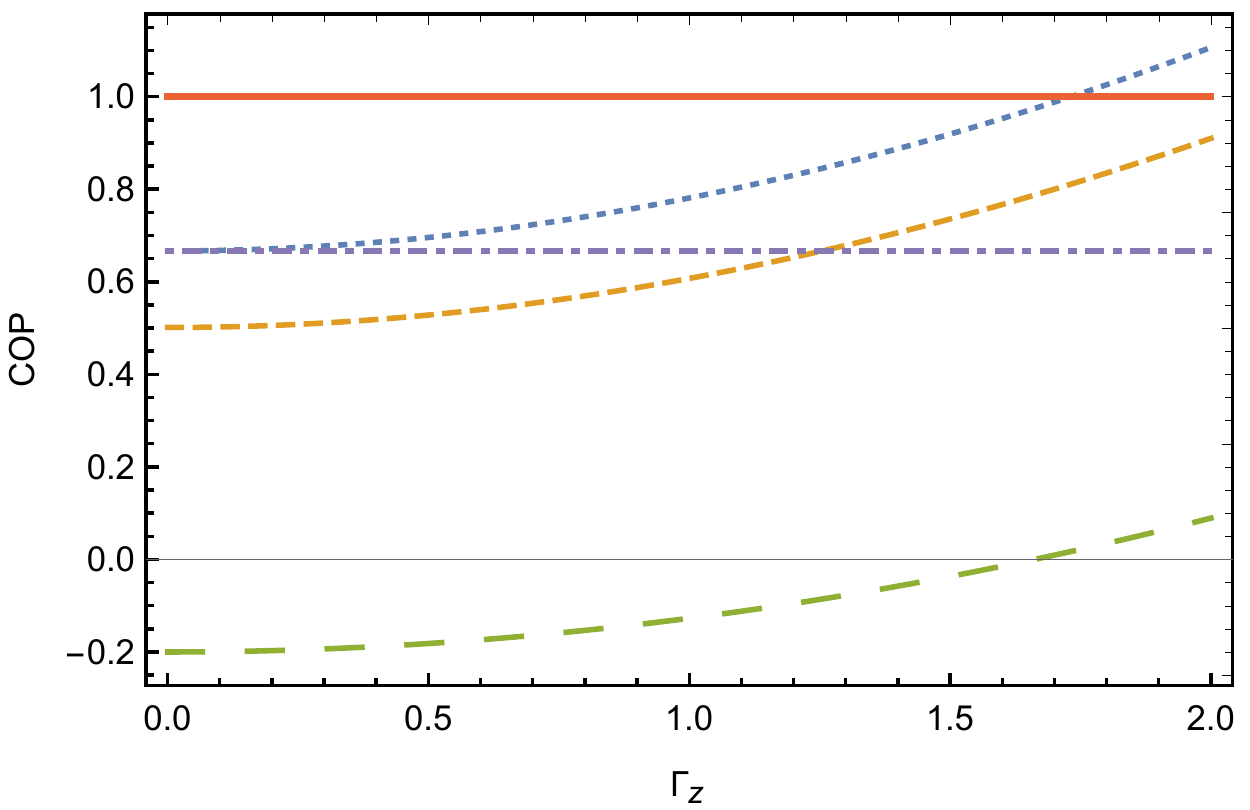}
\caption{(Color online). COP versus $\mathrm{\Gamma_{z}}$ for three different values of $\mathrm{J_{z}=0}$ (blue dotted line), $\mathrm{J_{z}=0.5}$ (orange dashed line), and $\mathrm{J_{z}=2.6}$ (green large-dashing line). The parameter values for the plot are $\mathrm{T_{h}=2}$, $\mathrm{T_{c}=1}$, $\mathrm{h=5}$, and $\mathrm{h'=2}$. The red solid line corresponds to that of Carnot $\mathrm{COP_{c}=T_{c}/(T_{h}-T_{c})}$, and the purple dot-dashed line corresponds to that of Otto $\mathrm{COP_{o}=h'/(h-h')}$. When the COP of the system gets above that of $\mathrm{COP_{c}}$ it is no longer working as a refrigerator. Note that if one plots the COP versus $\mathrm{J_{z}}$ for different values of $\mathrm{\Gamma_{z}}$ will get the same plot.}
\label{2}
\end{figure}

In Fig.\ref{1}(a), the efficiency is plotted as a function of $\mathrm{\Gamma_{z}}$ for three different values of $\mathrm{J_{z}=0}$, $0.5$, and $2.6$. For $\mathrm{J_z =0} $, we see that the efficiency could not surpass that of the Otto $\mathrm{\eta_{o}=1-h'/h}$, since in this case $\rm{q_{h}}=0$. However, for $\mathrm{J_{z}>0}$, e.g., $\mathrm{J_{z}=0.5}$ and $2.6$, we see that the efficiency could surpass $\mathrm{\eta_{o}}$. This could be explained from Fig. \ref{1}(b). When $\rm{J_{z}}\neq0$, we can have $\rm{q_{h}}<0$ depending on $\rm{\Gamma_{z}}$. This is why efficiency can be enhanced. For example, when $\rm{\Gamma_{z}}=0$, $|\rm{q_{h}}|$ is a monotonically increasing function as we increase $\rm{J_{z}}$ and it reaches its maximal value($\approx$0.435) when $\rm{J_{z}}\approx1.788$. However, for $\rm{\Gamma_{z}}=0$ and after $\rm{J_{z}}$ exceeding 1.788, $|\rm{q_{h}}|$ starts decreasing again, and if $\rm{J_{z}}$ is kept increasing, then $\rm{q_{h}}$ can even become positive. This is why we see that $|\rm{q_{h}}|$ for $\mathrm{J_{z}=2.6}$ is less than $|\rm{q_{h}}|$ for $\mathrm{J_{z}=0.5}$ for $\Gamma_{z}\lesssim0.61$. The explicit expression of $\rm{q_{h}}$ in terms of $\rm{J_{z}}$ and $\rm{\Gamma_{z}}$ is given as follows,
\begin{equation}
\begin{split}
\mathrm{q_{h}}=&\mathrm{-2J_{z}(\frac{e^{-J_{z}/4}}{2(e^{J_{z}/4}+e^{-J_{z}/4}\cosh(\sqrt{16+\Gamma_{z}^{2}}/2))}}
\\ &
\mathrm{-\frac{e^{-J_{z}}}{2(e^{J_{z}}+e^{-J_{z}}\cosh(2\sqrt{9+\Gamma_{z}^{2}}))}).}
\end{split}
\end{equation}
Fig.\ref{Idle} displays the contour plot of $\rm{q_{h}}$ as a function of $\rm{J_{z}}$ and $\rm{\Gamma_{z}}$. For an arbitrary value of $\rm{\Gamma_{z}}$ there is a critical value of $\rm{J_{z}}$ above which $\rm{q_{h}}$ would become positive, and it is given by,
\begin{equation}
\mathrm{J_{z}\leq J_{z}^{\star}=\frac{2}{3}\log\left( \frac{\cosh\left(2\sqrt{9+\Gamma_{z}^{2}}\right) }{\cosh\left( \sqrt{16+\Gamma_{z}^{2}}/2\right)}\right).}\label{star}
\end{equation}
For example, when $\rm{\Gamma_{z}=0}$, we have $\rm{J_{z}^{\star}}\approx2.65.$\par

In Fig.\ref{2}, we plot COP as a function of $\mathrm{\Gamma_{z}}$ for different values of $\mathrm{J_{z}}$. We see that, as it is expected, idle levels diminish the COP of the system when it is working as a refrigerator. More precisely, for $\mathrm{J_{z}=0}$ we have no idle levels which makes the COP surpass that of the Otto when $\mathrm{\Gamma_{z}\neq 0}$. However, when $\mathrm{J_{z}\neq 0}$ the COP get enhanced only after certain value of $\mathrm{\Gamma_{z}}$, i.e., the enhancement of COP shift to big values of $\mathrm{\Gamma_{z}}$. This is because for small values of $\mathrm{\Gamma_{z}}$ and when $\mathrm{J_{z}\neq 0}$ we found that idle levels are taking heat from the hot to the cold bath which diminishes the COP. This is due to the fact that they will heat the cold bath we want to cool. For example, when $\mathrm{J_{z}=2.6}$ the COP gets above that of the Otto only when $\mathrm{\Gamma_{z}>3.45}$. This shows that idle levels shift the enhancement in the COP to high values of $\mathrm{\Gamma_{z}}$, more precisely, the enhancement in the performance is present only when idle levels stop taking heat to the cold bath. Therefore, we see that our system without idle levels is a good refrigerator but a bad heat engine. However, this is not the general conclusion, below we give the conditions on the heat absorbed or released by idle levels such that the condition $\mathrm{COP>COP_{o}}$ is met. More precisely, we show that if idle levels channel heat in the same direction as working levels, then the COP can be enhanced. We do not add the plot of entanglement and quantum correlations here since when we plotted entanglement we found no clear relationship between it and the efficiency, the extractable work, or the COP. Moreover, this enhancement will also be seen for coupled but un-correlated spins in the section \ref{D}.

Now we focus on the physics behind the enhancements in the efficiencies. The expressions of heats and work in terms of the eigenvalues and populations are,
\begin{equation}
\mathrm{Q_{h}=-4J_{z}(p_{2}-p_{2}')+2\sqrt{h^{2}+\Gamma_{z}^{2}}\left(p_{4}-p'_{4}+p'_{3}-p_{3}\right),}
\end{equation}
\begin{equation}
\mathrm{Q_{c}=4J_{z}(p_{2}-p_{2}')-2\sqrt{h'^{2}+\Gamma_{z}^{2}}\left(p_{4}-p'_{4}+p'_{3}-p_{3}\right),}
\end{equation}
and,
\begin{equation}
\mathrm{W=2(\sqrt{h^{2}+\Gamma_{z}^{2}}-\sqrt{h'^{2}+\Gamma_{z}^{2}})\left(p_{4}-p'_{4}+p'_{3}-p_{3}\right).}
\end{equation}
These equations could be expressed as well as follows: $\mathrm{Q_{h}=q_{I}+q_{Wh}}$, $\mathrm{Q_{c}=-q_{I}-q_{Wc}}$ and $\mathrm{W=q_{Wh}-q_{Wc}}$. Here $\mathrm{q_{I}}$ and $\mathrm{q_{Wh}}$ are the heats absorbed from the hot bath by idle and working levels respectively. The efficiency is,
\begin{equation}
\mathrm{\eta=\left(1-\frac{\sqrt{h'^{2}+\Gamma_{z}^{2}}}{\sqrt{h^{2}+\Gamma_{z}^{2}}}\right) \left(\frac{1}{1-\frac{4J_{z}}{2\sqrt{h^{2}+\Gamma_{z}^{2}}}\frac{p_{2}-p_{2}'}{p_{4}-p'_{4}+p'_{3}-p_{3}}}\right),}
\end{equation}
which could be written as well as follows, $\mathrm{\eta=\eta_{\Gamma_{z}}\frac{1}{1+\frac{q_{I}}{q_{Wh}}}}$, with $\mathrm{\eta_{\Gamma_{z}}=1-\frac{\sqrt{h'^{2}+\Gamma_{z}^{2}}}{\sqrt{h^{2}+\Gamma_{z}^{2}}}}$. 
\\
One can see that for $\mathrm{\Gamma_{z}=0}$, $\mathrm{\eta}$ will be $\mathrm{\eta_{o}}$ times another term (which has to be bigger than one for $\mathrm{\eta>\eta_{o}}$). Thus, the condition for $\mathrm{\eta>\eta_{o}}$ is that, $\mathrm{0<\frac{4J_{z}}{2h}\frac{p_{2}-p_{2}'}{p_{4}-p'_{4}+p'_{3}-p_{3}}<1}$, or equivalently $\mathrm{-2h(p_{4}-p'_{4}+p'_{3}-p_{3})<q_{I}<0}$. This condition shows that as long as the heat channeled from the hot to the cold bath by idle levels is not positive and not less than $\mathrm{-q_{Wh}}$ then efficiency could be improved. However, if $\mathrm{q_{I}<-q_{Wh}}$ then $\mathrm{Q_{h}}$ will be negative, thus the system will not work as a heat engine. The same thing could be said when $\mathrm{\Gamma_{z}\neq 0}$ even though things get complicated because the eigenvalues are not linear in the magnetic field. Thus, different from $\mathrm{\Gamma_{z}=0}$, in this case we must have $\mathrm{1+\frac{q_{I}}{q_{Wh}}<\frac{\eta_{\Gamma_{z}}}{\eta_{o}}}$. Adding the next condition, $\mathrm{\eta_{o}<\eta<\eta_{c}}$ (even though $\mathrm{\eta}$ may not be able to reach the Carnot bound, but let's take the biggest upper bound allowed by the second law fo thermodynamics), in this case the condition on heat absorbed by idle level from the hot bath is,
\begin{equation}
\mathrm{-q_{Wh}\frac{\eta_{c}-\eta_{\Gamma_{z}}}{\eta_{c}}<q_{I}<-q_{Wh}\frac{\eta_{o}-\eta_{\Gamma_{z}}}{\eta_{o}}<0.}
\end{equation}
We see that $\mathrm{q_{I}}$ not only has to be less than 0 as in the case when $\mathrm{\Gamma_{z}=0}$, but only when it is less than $\mathrm{-q_{Wh}\frac{\eta_{o}-\eta_{\Gamma_{z}}}{\eta_{o}}}$, the efficiency can be enhanced, and this is due to the fact that the eigenvalues are nonlinear in the magnetic field.

The expression of the COP is,
\begin{equation}
\begin{split}
& \rm COP=\frac{\sqrt{h'^{2}+\Gamma_{z}^{2}}}{\sqrt{h^{2}+\Gamma_{z}^{2}}-\sqrt{h'^{2}+\Gamma_{z}^{2}}}
\\  &
\rm -\frac{4J_{z}(p_{2}-p_{2}')}{2(\sqrt{h^{2}+\Gamma_{z}^{2}}-\sqrt{h'^{2}+\Gamma_{z}^{2}})(p_{4}-p'_{4}+p'_{3}-p_{3})}.
\end{split}
\end{equation}
From this equation, for $\mathrm{J_{z}=0}$ we have, $\mathrm{COP=\frac{\sqrt{h'^{2}+\Gamma_{z}^{2}}}{\sqrt{h^{2}+\Gamma_{z}^{2}}-\sqrt{h'^{2}+\Gamma_{z}^{2}}}}$, this term is always bigger than $\mathrm{COP_{o}}$ as long as $\mathrm{\Gamma_{z}\neq0}$. Moreover, if we continue increasing $\mathrm{\Gamma_{z}}$ one can see that this term could reach the Carnot bound. In this case ($\mathrm{J_{z}=0}$) we have no idle levels, but we have two working levels and two degenerate levels with energy 0, that neither contribute to heats nor work. Thus, in this case, our system is equivalent to a two-level system with an energy gap $\mathrm{4\sqrt{h^{2}+\Gamma_{z}^{2}}}$. However, when $\mathrm{J_{z}}$ is not equal to zero things will be different and complicated. For example, when $\mathrm{J_{z}>0}$, the enhancement in this case is dependent on the sign of $\mathrm{p_{2}-p'_{2}}$. More precisely, if $\mathrm{p_{2}-p'_{2}<0}$, in this case we see that the second term will be negative thus it will diminish the COP. However, if $\mathrm{p_{2}-p'_{2}>0}$, in this case the COP will be enhanced. Therefore, in this case, for $\mathrm{COP>COP_{o}}$ the term $\mathrm{q_{c}=4J_{z}(p_{2}-p_{2}')}$ has to be positive, i.e. idle levels has to take heat in the right direction from the cold bath to the hot bath. When $\mathrm{J_{z}=0}$ it does not matter if $\mathrm{p_{2}-p'_{2}<0}$ or $\mathrm{p_{2}-p_{2}'>0}$, since in this case idle levels will channel no heat. However, when 
$\mathrm{J_{z}>0}$ then in this case we must have $\mathrm{p_{2}-p'_{2}\geq0}$, i.e. the probability of occupation of the degenerate idle level $\mathrm{-2J_{z}}$ at the hot bath side has to be equal or bigger than the one at the cold side.

Now we give the theory used to describe the thermodynamics of the global and the local cycles. Let's give the expressions of the global as well as the local work and heats. We start with the global ones.
\begin{equation}
\mathrm{-W_{1}=Tr\left[\rho(H_{g}-H_{g}')\right],}
\end{equation}
this is the work performed during the adiabatic stage 2. $\mathrm{\rho}$ is the state of the system when it is in equilibrium with the hot bath, $\mathrm{H_{g}}$ and $\mathrm{H_{g}}'$ are the global Hamiltonians when the magnetic field is equal to $\mathrm{h}$ and $\mathrm{h}'$ respectively.
\begin{equation}
\mathrm{Q_{h}=Tr\left[H_{g}(\rho-\rho')\right],}
\end{equation}
this is the heat absorbed from the hot bath, $\mathrm{\rho'}$ is the state of the system when it is in equilibrium with the cold bath.
\begin{equation}
\mathrm{-W_{2}=-Tr\left[\rho'(H_{g}-H_{g}')\right],}
\end{equation}
this is the work performed during the adiabatic stage 4. Finally,
\begin{equation}
\mathrm{Q_{c}=-Tr\left[H_{g}'(\rho-\rho')\right],}
\end{equation}
this is the heat released to the cold bath. The first law of thermodynamics reads as $\mathrm{W_{1}+W_{2}+Q_{1}+Q_{2}=0}$, from which we have $\mathrm{W=-(W_{1}+W_{2})=T((\rho-\rho')(H_{g}-H_{g}'))}$. The local ones are given in the same way as follows,
\begin{equation}
\mathrm{-w_{\alpha}=Tr\left[\rho_{\alpha}(H_{l}-H_{l}')\right],}
\end{equation}
\begin{equation}
\mathrm{q_{h\alpha}=Tr\left[H_{l}(\rho_{\alpha}-\rho_{\alpha}')\right],}
\end{equation}
\begin{equation}
\mathrm{-w_{\alpha}'=-Tr\left[\rho_{\alpha}'(H_{l}-H_{l}')\right],}
\end{equation}
\begin{equation}
\mathrm{q_{c\alpha}=-Tr\left[H_{l}'(\rho_{\alpha}-\rho_{\alpha}')\right],}
\end{equation}
with $\mathrm{\alpha=1, \ 2}$, $\mathrm{H_{l}=diag\{h,-h\}}$ is the local Hamiltonian of the spins. The definitions of the global and local works will be compared below for the Ising+KSEA model. Since the reduced state of both spins is the same $\mathrm{\rho_{1}=\rho_{2}}$, the total work extracted locally is,
\begin{equation}
\mathrm{w=2Tr\left[(\rho_{1}-\rho_{1}')(H_{l}-H_{l}')\right],}
\end{equation} 
and if  $\mathrm{\rho_{1}\neq\rho_{2}}$, $\mathrm{w=\sum_{\alpha}Tr((\rho_{\alpha}-\rho_{\alpha}')(H_{l}-H_{l}'))}$. Another definition of local heats and work which one can use as well is the case when the local heats and work are defined with respect to the global Hamiltonian $\mathrm{H_{g}}$ and local state $\mathrm{\rho_{\alpha}}$. In this case we have,
\begin{equation}
\mathrm{-w_{\alpha}=Tr\left[\rho_{\alpha}(H_{g}-H_{g}')\right],}
\end{equation}
\begin{equation}
\mathrm{q_{h\alpha}=Tr\left[H_{g}(\rho_{\alpha}-\rho_{\alpha}')\right],}
\end{equation}
\begin{equation}
\mathrm{-w_{\alpha}'=-Tr\left[\rho_{\alpha}'(H_{g}-H_{g}')\right],}
\end{equation}
\begin{equation}
\mathrm{q_{c\alpha}=-Tr\left[H_{g}'(\rho_{\alpha}-\rho_{\alpha}')\right].}
\end{equation}
If $\mathrm{\rho_{1}=\rho_{2}}$, the total work extracted locally is,
\begin{equation}
\mathrm{w=2Tr\left[(\rho_{1}-\rho_{1}')(H_{g}-H'_{g})\right].}
\end{equation}
In general, we have four cases with respect to which we can compute the average of work and heats: 1) with respect to the global state and the global Hamiltonian, 2) with respect to the local state and the global Hamiltonian, 3) with respect to the global state and the local Hamiltonian, and finally 4) with respect to the local state and the local Hamiltonian. 3 and 4 are equivalent. Furthermore, 1, 2 and 3 may be equal for the total work extracted depending on the model. However, in terms of heats will not be equal. Thus,  we would obtain different efficiencies with these definitions.
\subsection{Local description}
Now let us see how the coupled spin-$1/2$ particles are undergoing the cycle locally. The reduced density matrix of the sub-systems $1$ and $2$ in the standard basis $\{|1\rangle,|0\rangle\}$ when they are in thermal equilibrium with the hot bath is
\begin{equation}
\mathrm{\rho_{1}=\rho_{2} }=\begin{pmatrix}
\frac{1}{2}(1-\frac{\mathrm{h}(\mathrm{p_{3}-p_{4}})}{\mathrm{\sqrt{h^{2}+\Gamma_{z}^{2}}}}) & 0 \\
0 & \frac{1}{2}(1+\frac{\mathrm{h}(\mathrm{p_{3}-p_{4}})}{\mathrm{\sqrt{h^{2}+\Gamma_{z}^{2}}}})  \\
\end{pmatrix}.
\label{rho1}
\end{equation}
The reduced density matrix of the sub-systems $1$ and $2$ in the standard basis $\{|1\rangle,|0\rangle\}$ when they are in thermal equilibrium with the cold bath is
\begin{equation}
\mathrm{\rho'_{1}=\rho_{2}'=}\begin{pmatrix}
\frac{1}{2}(1-\frac{\mathrm{h'}(\mathrm{p_{3}'}-\mathrm{p_{4}'})}{\mathrm{\sqrt{h'^{2}+\Gamma_{z}^{2}}}}) & 0 \\
0 & \frac{1}{2}(1+\frac{\mathrm{h'(p_{3}'-p_{4}')}}{\mathrm{\sqrt{h'^{2}+\Gamma_{z}^{2}}}})  \\
\end{pmatrix}.
\label{rho2}
\end{equation}
Note that Eqs. \ref{rho1} and \ref{rho2}, has been obtained using the normalisation constraint $\mathrm{\sum_{i}p_{i}=\sum_{i}p'_{i}=1}$. Following \cite{Johal}, the local Hamiltonians for the spins are $\mathrm{H={\rm diag}\left(h,-h\right)}$ and $\mathrm{H'={\rm diag}\left(h',-h'\right)}$ at $\mathrm{\mathbf{Stages}}$ $1$ and $3$, respectively. And using the formulas given in the section above, the local heat absorbed from the hot bath $\mathrm{q_{h1,2}}$, the local heat released to the cold bath $\mathrm{q_{c1,2}}$, and the amount of work extracted locally $\mathrm{w_{1,2}}$ for each spin are expressed as follows
\begin{equation}
\mathrm{q_{h1}=q_{h2}=h\left(\frac{h(p_{4}-p_{3})}{\sqrt{h^{2}+\Gamma_{z}^{2}}}+\frac{h'(p_{3}'-p_{4}')}{\sqrt{h'^{2}+\Gamma_{z}^{2}}}\right),}
\end{equation}
\begin{equation}
\mathrm{q_{c1}=q_{c2}=-h'\left(\frac{h(p_{4}-p_{3})}{\sqrt{h^{2}+\Gamma_{z}^{2}}}+\frac{h'(p_{3}'-p_{4}')}{\sqrt{h'^{2}+\Gamma_{z}^{2}}}\right),}
\end{equation}
and
\begin{equation}
\mathrm{w_{1}=w_{2}=(h-h')\left(\frac{h(p_{4}-p_{3})}{\sqrt{h^{2}+\Gamma_{z}^{2}}}+\frac{h'(p_{3}'-p_{4}')}{\sqrt{h'^{2}+\Gamma_{z}^{2}}}\right).}
\end{equation}
The total work extracted locally by both subsystems $1$ and $2$ is given by
\begin{equation}
\mathrm{w=2(h-h')\left(\frac{h(p_{4}-p_{3})}{\sqrt{h^{2}+\Gamma_{z}^{2}}}+\frac{h'(p_{3}'-p_{4}')}{\sqrt{h'^{2}+\Gamma_{z}^{2}}}\right).}
\label{wl}
\end{equation}
\begin{figure}[hbtp]
\centering
\includegraphics[scale=0.7]{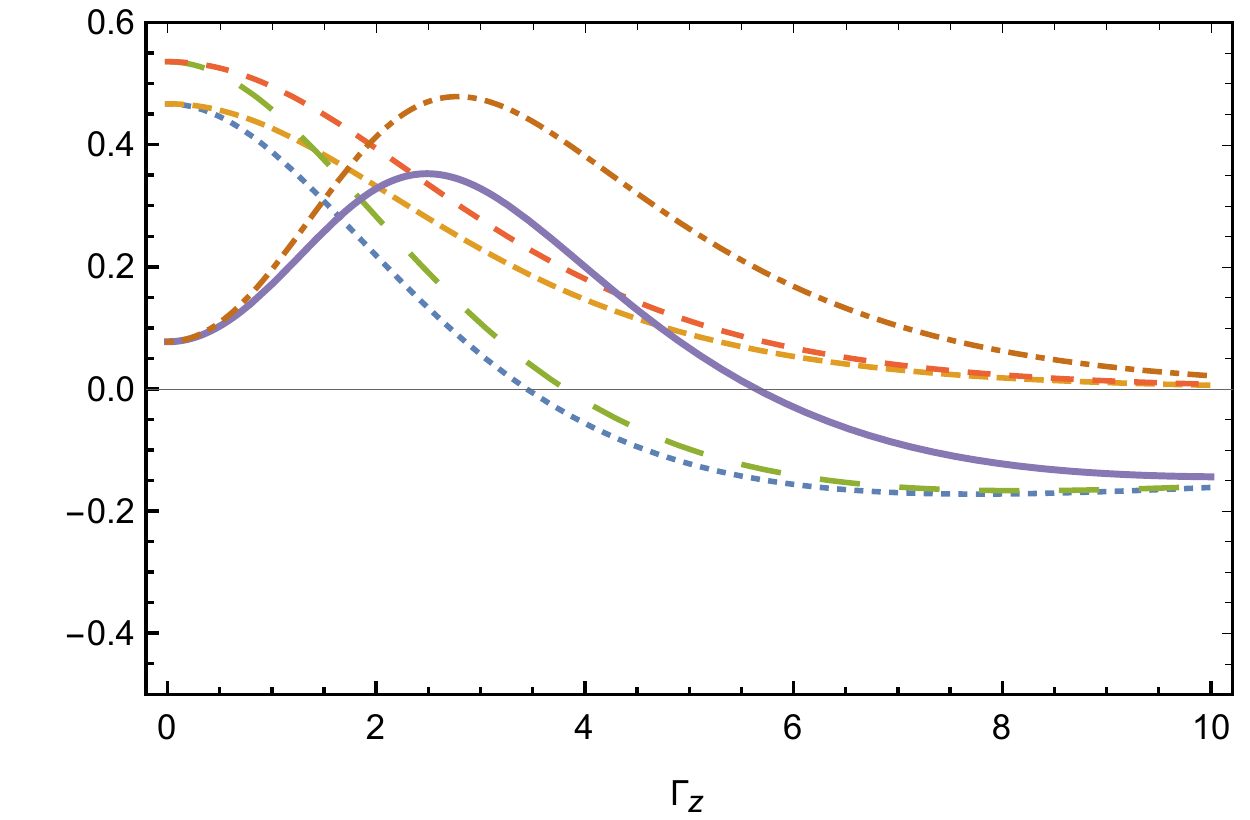}
\caption{(Color online). Eq. \ref{Wo} and Eq. \ref{wl} are plotted versus $\mathrm{\Gamma_{z}}$ for different values of $\mathrm{J_{z}=0}$, 0.5, and 2.6. The parameter values are the same as in Fig. \ref{1}. $\mathrm{w_{J_{z}=0}}$ (blue dotted line), $\mathrm{W_{J_{z}=0}}$ (orange dashed line), $\mathrm{w_{J_{z}=0.5}}$ (green large-dashing line), $\mathrm{W_{J_{z}=0.5}}$ (red medium-dashing line), $\mathrm{w_{J_{z}=2.6}}$ (purple solid line) and $\mathrm{W_{J_{z}=2.6}}$(brown dot-dashed line).}
\label{3}
\end{figure}
This equation and Eq. \ref{Wo} are plotted in Fig. \ref{3} as a function of $\mathrm{\Gamma_{z}}$ for three different values of $\mathrm{J_{z}=0}$, $0.5$ and $2.6$. We see that, in general, the total work extracted locally $\mathrm{w}$ is not equal to the one extracted globally $\mathrm{W}$, and only for very small values of $\mathrm{\Gamma_{z}}$ when they become equal to each other. It is seen that as $\Gamma_{z}$ is increasing, the difference between the local and global works become more pronounced. This is in contrast to \cite{Altintas} in which it was suggested that to violate the extensive property of the work extracted globally, we have to vary the coupling parameters next to the magnetic field $\mathrm{h}$ in the adiabatic stages. This is not the case as our results show clearly. The reason is that the eigenstates of their Hamiltonian were not dependent on the coupling parameters as in the case under consideration and the eigenvalues were linear in the magnetic field. More precisely, it was stated there that when $\mathrm{dH_{ int}=0}$ then we can automatically conclude that the global work will be extensive. This is not true in general, since the total work is also dependent on the state of the total system. That is, if $\mathrm{dW={\rm Tr}(\rho \sum dH_{\rm loc})}$ does not mean the work will be extensive. Put it in another way this does not mean that dW=$\mathrm{\sum dw_{\rm loc}}$ with $\mathrm{dw_{\rm loc}={\rm Tr}_{\rm loc}(\rho_{\rm loc}dH_{\rm loc})}$. The extensive property depends as well on the different parameters entering the eigenstates of the Hamiltonian describing the system. However, when $\mathrm{\Gamma_{z}=0}$ the results are in agreement with each other.

At this point we want to give some clarifications about $\mathrm{W-w\geq0}$. Our study shows that the extensive property is not related to any type of classical or quantum correlations. To make it clear, let's take the above example and divide the expression of the global work into two terms. The expression of the total work extracted globally is given as follows:
\begin{equation}
\mathrm{W=\sum_{i}(E_{i}-E_{i}')(p_{i}-p_{i}').}
\end{equation}
From this expression we have,
\begin{equation}
\mathrm{W=\sum_{i}(E_{i}-E_{i}')p_{i}+\sum_{i}(E_{i}'-E_{i})p_{i}'=-(W_{1}+W_{2}).}
\end{equation}
The first work ($\mathrm{-W_{1}}$) is the work performed when we change the magnetic field from $\mathrm{h}$ to $\mathrm{h}'$ with respect to the state $\mathrm{\rho}$. The second term ($\mathrm{-W_{2}}$) is the work performed when we change back $\mathrm{h}'$ to $\mathrm{h}$. Now let's consider these two works separately and examine the extensive property. We want to compare $\mathrm{-W_{1}}$ and $\mathrm{-W_{2}}$ with their corresponding works invested $\mathrm{-w_{1}}$ and extracted $\mathrm{-w_{2}}$ locally. In the same way we have,
\begin{equation}
\mathrm{w=2\sum_{i}((E_{i1}-E_{i1}')p_{i1}+(E_{i1}'-E_{i1})p_{i1}')=-2(w_{1}+w_{2}).}
\end{equation}
with $\mathrm{E_{i1}=\{h,-h\}}$, and $\mathrm{p_{i\alpha}}$ are the local probabilities. Let's now compare $\mathrm{-W_{1}}$ with $\mathrm{-w_{1}}$. More precisely, we want to compare the works performed globally and locally with each other during stage 2. In this case we have,
\begin{equation}
\mathrm{-W_{1}=2(\sqrt{h^{2}+\Gamma_{z}^{2}}-\sqrt{h'^{2}+\Gamma_{z}^{2}})(p_{4}-p_{3}),}
\end{equation} 
and,
\begin{equation}
\mathrm{-2w_{1}=2(h-h')(\frac{h(p_{4}-p_{3})}{\sqrt{h^{2}+\Gamma_{z}^{2}}}).}
\end{equation}
Since $\mathrm{(p_{4}-p_{3})}<0$, thus $\mathrm{-W_{1}}$ and $\mathrm{-2w_{1}}$ are both negative, therefore they are the works invested globally and locally, respectively. Their difference is given as follows:
\begin{equation}
\begin{split}
\rm -(W_{1}-2w_{1})= & \rm 2((\sqrt{h^{2}+\Gamma_{z}^{2}}-
\sqrt{h'^{2}+\Gamma_{z}^{2}})-\frac{(h-h')h}{\sqrt{h^{2}+\Gamma_{z}^{2}}}) 
\\ 
& \rm (p_{4}-p_{3}).
\end{split}
\end{equation}
One can see clearly that if $\mathrm{\Gamma_{z}=0}$ we get $\mathrm{-W_{1}=-2w_{1}}$, in this case the model reduces to the Ising model. Since $\mathrm{p_{4}-p_{3}}<0$ and that the first term is as well negative when $\mathrm{\Gamma_{z}}\neq0$, thus $\mathrm{-(W_{1}-2w_{1})>0}$. Thus, the total work invested locally during the adiabatic stage 2 is not equal to the one invested globally. In the same way, one can show that $\mathrm{-(W_{2}-2w_{2})>0}$. Note that because $\mathrm{-W_{2}}$ and $\mathrm{-2w_{2}}$ are positive, thus they are the works extracted during stage 4. And because the total works invested (extracted) globally and locally are not equal in stage 2 (stage 4), therefore, the global work will not be extensive. The explanation for this gap is that the eigenvalues of Ising+KSEA model are not linear in the magnetic field and that the eigentates are parameter dependent.

Let's now explain why $\mathrm{W-w\geq0}$ has nothing to do with quantum correlations. If one studies the work extracted by two coupled spins in the Heisenberg, Ising and Ising+KSEA models, one will find that only the first and second models preserve the extensive property, even though both Heisenberg and Ising+KSEA models are quantum mechanically correlated. Thus, the extensive property of the work has nothing to do with quantum correlations, it only depends on the structure of the Hamiltonian. For example, in Ref. \cite{Chang}  the eigenvalues were non-linear in the magnetic field and the eigenstates were parameter dependent. However, there they explain the gap $\mathrm{W-w>0}$ by quantum correlations. In contrast to their conclusion, ours show that quantum correlations are not responsible for this gap. And thus it is difficult to give an interpretation of the extracted work in terms of quantum correlations, since there is no clear relationship between them in our study. However, note that if one includes measurement and feedback in the protocol, in this case, quantum correlations will affect the work output. More precisely, suppose we apply a measurement on one of the correlated spins and then apply some operations on the unmeasured one depending on the result of the measurement(feedback). In this case, the correlations between the spins will be important. However, in our study we were only interested in running the Otto cycle on each spin locally.

Another point that should be clarified is the one concerning the local and global descriptions. More precisely, one may think that the reason behind $\mathrm{W-w>0}$ is that the global cycle is Otto and the local one is not. Let's explain this in detail. This is because from Eqs. \ref{rho1} and \ref{rho2} one may notice that the local probabilities at both sides are magnetic field dependent. Then, one may directly conclude that the cycle is globally Otto and that the quantum adiabatic theorem is valid. However, the adiabatic stages of the local cycle may not be adiabatic and, therefore, the quantum adiabatic theorem is no longer satisfied locally. However, this conclusion is not true. Because according to this conclusion, even the global probabilities will not stay the same in the adiabatic stages. And this is because the global probabilities as well are magnetic field dependent either through $\rm e^{-\beta E_{i}}$ (if $\rm E_{i}$ is magnetic field dependent) or through $\rm Z=2(e^{\beta J_{z}}+e^{-\beta J_{z}}\cos(2\beta\sqrt{h^{2}+\Gamma_{z}^{2}}))$, however, if the the transformation is done slowly there will be no transitions between the states and thus the cycle would satisfy the quantum adiabatic theorem, i.e. there will be no changes in the populations. Therefore, using the same reasoning, the local adiabatic stages would be adiabatic if the transformation is done slowly, thus the system will only exchange work and no heat $\mathrm{q_{ad}=0}$, and the adiabatic theorem is still satisfied. Therefore, the local cycle is also Otto. Furthermore, this model makes it difficult to show that correlations are not necessary to surpass the efficiency of the Otto. To show this, below we use the Ising model, keeping only the z-components which exhibit no quantum correlations at all. The system would be along the cycle in a separable state.
\section{Comparison between multi-coupled spin-$1/2$ particles}\label{C}
Here we first compare the efficiency, the extractable work, and the COP of two- and three-coupled spin-$1/2$ $\mathrm{1D}$ Heisenberg $\mathrm{XXX}$-chain. Then we do the same comparison between two coupled and the systems corresponding up to six spins in the Ising model. The latter has been chosen for two reasons: first to ensure that no entanglement and quantum correlations will be created during the cycle between the interacting spins, as our purpose is to show and confirm that the enhancement is only due to the structure of the energy levels of the system. Secondly, we want to study the role of increasing the number of interacting spins on efficiency, extractable work and COP.
\subsection{Heisenberg model} \label{Heisenberg}
Here, the working fluid is a three-coupled spin-$1/2$ Heisenberg $\mathrm{XXX}$-chain under the influence of a magnetic field $\mathrm{h}$ along the $\mathrm{z}$-axis. We call that the previous works \cite{Zhang, Johal, Ukraine, Altintas, Ramandeep} have considered only two coupled spin-$1/2$. The expression of the Hamiltonian is given by
\begin{equation} 
\mathrm{H=J\sum_{i}^{N}(\sigma_{x}^{i}\sigma_{x}^{i+1}+\sigma_{y}^{i}\sigma_{y}^{i+1}+\sigma_{z}^{i}\sigma_{z}^{i+1})+h\sum_{i}^{N}\sigma_{z}^{i}, }
\end{equation} 
where $\mathrm{\sigma_{x,y,z}^{i}}$ are the standard Pauli matrices acting on the site $\mathrm{i=1,2,3}$. Note that the periodicity is presumed. $\mathrm{J}$ and $\mathrm{h}$ are the exchange coupling and the strength of the external magnetic field, respectively. When $\mathrm{J < 0}$ the model is ferromagnetic, while $\mathrm{J> 0}$ corresponds to an antiferromagnetic system. Here, we have the isotropic situation $\mathrm{J_{x}=J_{y}=J_{z}=J}$. The eigenvalues of $\mathrm{H}$ are given by: $\rm E=\{-(h+3J),-(h+3J),h-3J,h-3J,-3(h-J),3(h+J),-h+3J,h+3J\}$. Their associated eigenstates will not be reported here, since we will not use them. 
\begin{figure}[hbtp]
\centering
\includegraphics[scale=0.7]{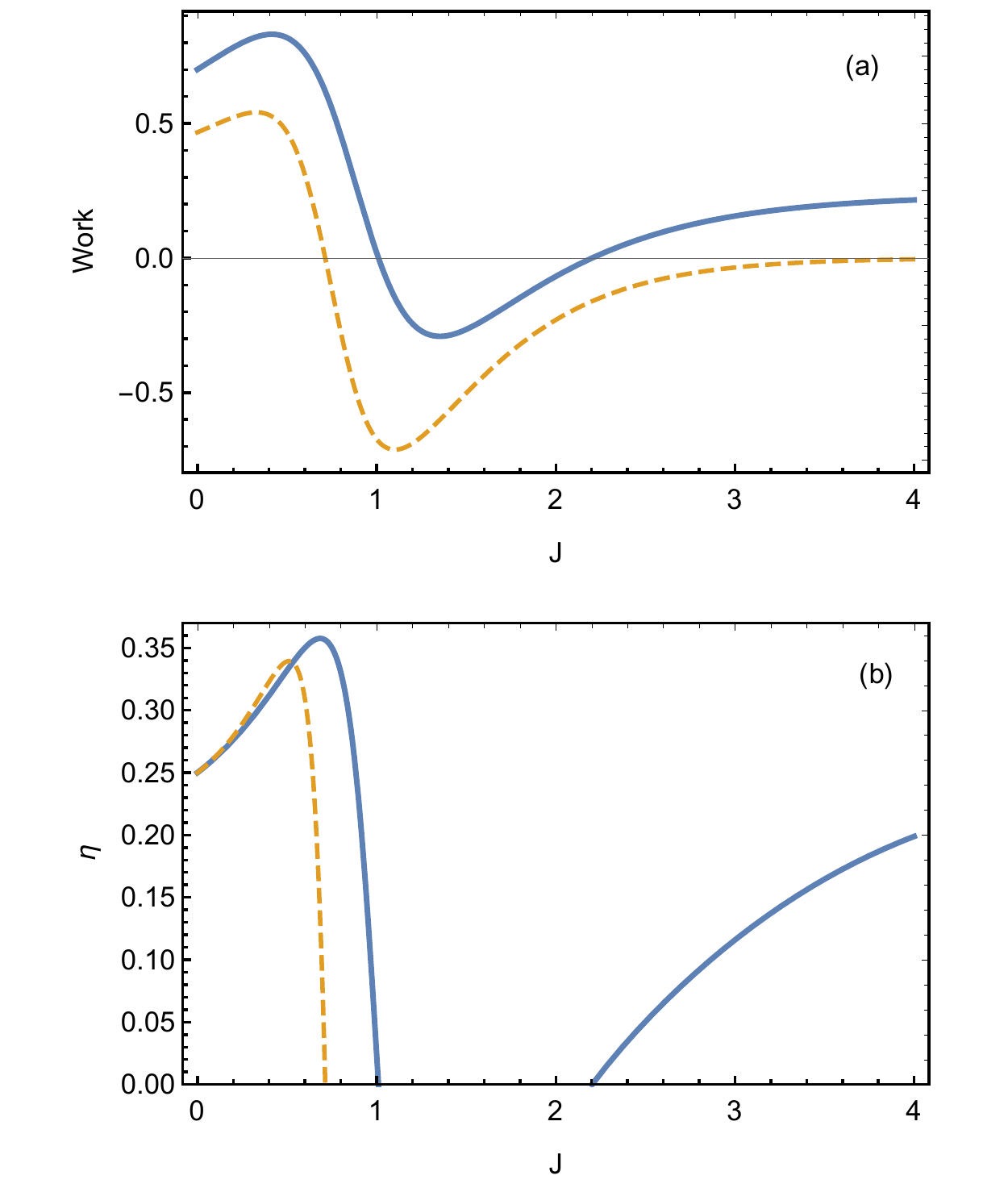}
\caption{(Color online). Plot of (a) the extracted work when $\mathrm{N=2}$ (orange dashed line) and $\mathrm{N=3}$ (blue solid line) as a function of $\mathrm{J}$ (b) the efficiency for both $\mathrm{N=2}$ and $\mathrm{N=3}$ as a function of $\mathrm{J}$. The parameter values are the same as in Fig. \ref{1}.}
\label{4}
\end{figure}
\begin{figure}[hbtp]
\centering
\includegraphics[scale=0.7]{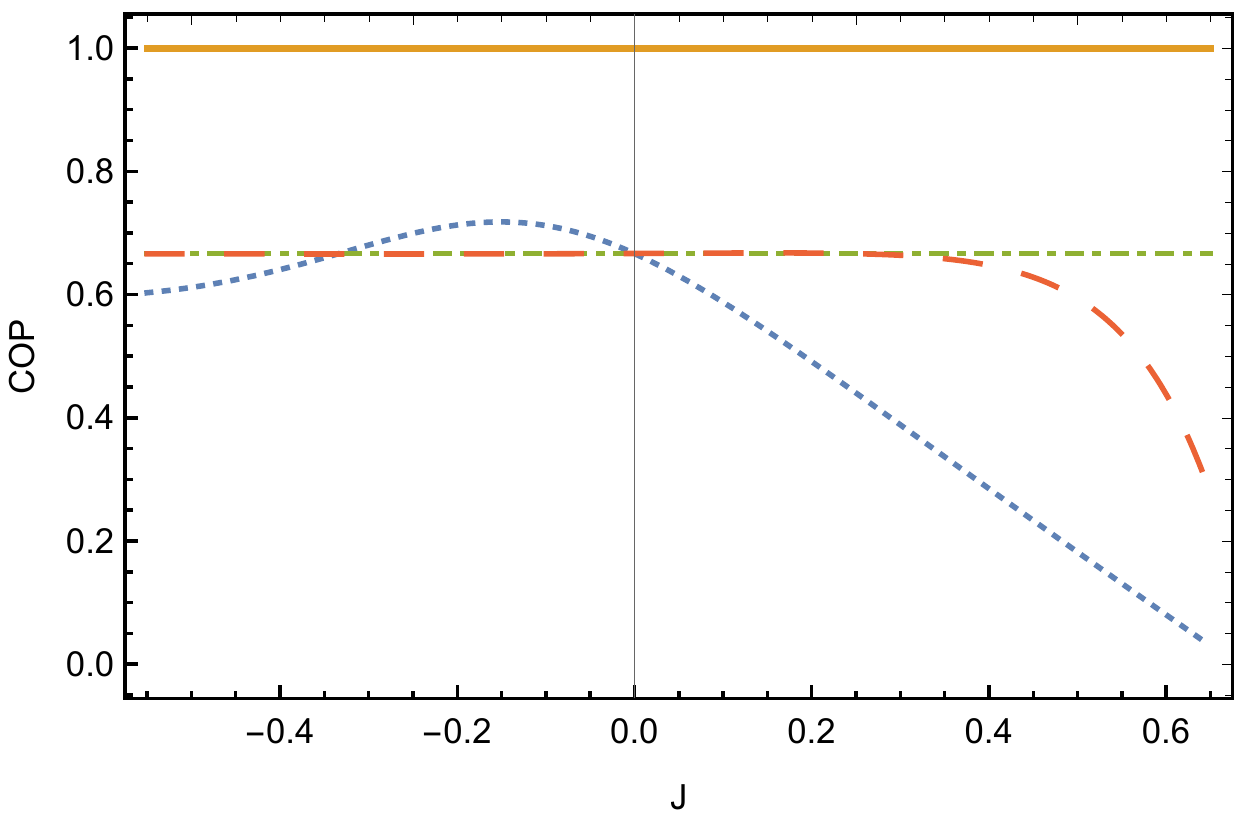}
\caption{(Color online). The plot of COP for $\mathrm{N=2}$ (red large-dashing line) and $\mathrm{N=3}$ (blue dotted line) as a function of $\mathrm{J}$. The parameter values are the same as in Fig. \ref{2}. The green dot-dashed line and the solid orange line correspond to the COP of Otto and Carnot, respectively.}
\label{5}
\end{figure}

In Fig.\ref{4}(a), we plot the work as a function of $\mathrm{J}$ for $\mathrm{N=2}$ and $\mathrm{N=3}$. We see that when $\mathrm{N=3}$, the work extracted globally is bigger than the one from two-coupled spin-$1/2$. Even more, the three coupled spins could still harvest work in the strong coupling regime in contrast to the two-coupled spin-$1/2$. In Fig.\ref{4}(b), the efficiency $\mathrm{\eta}$ is plotted as a function of $\mathrm{J}$. We see that $\mathrm{N=3}$ outperforms $\mathrm{N = 2}$ in terms of the highest value riched of the efficiency. More precisely, for the considered value of the parameters, the maximum value riched when N=3 is $\approx$0.36, and when N=2 it is $\approx$0.34. We should add that for J$\leq$0.52, the efficiency of two coupled spins is higher than that of N=3, as Fig.\ref{4}(b) shows. In Fig.\ref{5} we plot their COPs as well as the ones of Otto $\mathrm{COP_{o}}$ and Carnot $\mathrm{COP_{c}}$. We see that only three-coupled spin-$1/2$ COP could surpass the COP of the Otto $\mathrm{COP_{o}}$, however, the COP of two-coupled spin-$1/2$ is always equal or less than it. We can conclude from those figures that increasing the number of spins enhances the amount of extractable work as well as the maximal values of both the efficiency and the COP. Please, note that here we do not mean that the COP of three coupled spins is always bigger than that of N=2 since, as figure \ref{5} shows, the latter can be higher than that of N=3 for broad values of J, but its maximum value does not exceed the one of N=3. In addition to this, notice that when the system is working as a heat engine for three-coupled spin-$1/2$, its efficiency could surpass that of the Otto $\mathrm{\eta_{o}}$ only when it is anti-ferromagnetic. And when it is a refrigerator, the enhancement is seen only when the system is ferromagnetic. We have also calculated the work extracted locally by the three coupled spins, and we found that it is equal to the one extracted globally. As we said before, this is due to the fact that the eigenstates of the system are parameter independent and that the eigenvalues are linear in the magnetic field \cite{Johal}. However, we didn't include its plot here for brevity.\par

Now let's explain the reasons behind the enhancement in the work extracted. First, we expect that more levels will help extract more work, e.g. a harmonic oscillator could extract more work than a single spin. However, their efficiencies are equal if all energies are shifted by the same amount of energy. More precisely, we see from the figure \ref{4} that when the coupling is turned off, i.e. $\mathrm{J=0}$, then as we increase the number of spins (i.e. when $\mathrm{N=2}$ and $\mathrm{N=3}$ for $\mathrm{J=0}$) more work will be extracted. When turning on the interaction, we see that the work extracted by coupled spins enhances the one from uncoupled ones. Thus we see that both the number of spins N as well as the interaction $\mathrm{J}$ between them enhances the work extracted globally W.

When we compared the efficiency of 3 and 2 coupled Heisenberg spins we found that the latter cannot work as a heat engine in the strong coupling regime, and the reason is that the positive work condition is not satisfied, since in this regime the system consumes work instead of outputing it, thus the system no longer works as a heat engine. More precisely, the work extracted $\mathrm{-W_{2}}$ during the adiabatic stage 4 is less than the one invested $\mathrm{-W_{1}}$ in the adiabatic stage 2, thus $\mathrm{W=-(W_{1}+W_{2})\leq0}$. When it comes to the three coupled spins in the strong coupling regime, the eigenvalues of the system become equidistant, thus with efficiency equal to the one of the Otto $\mathrm{\eta_{o}=1-\frac{h'}{h}}$.

Concerning the local and global thermodynamics of the three spins Heisenberg model, we have the preservation of the extensive property of the work, i.e.,
\begin{equation}
\mathrm{W=w_{1}+w_{2}+w_{3}},
\end{equation}
with the local work extracted by each spin is $\mathrm{w_{\alpha}=Tr((H_{l}-H'_{l})(\rho_{\alpha}-\rho_{\alpha}'))}$, with $\mathrm{\alpha=1}$, 2, and 3. The local states are, $\mathrm{\rho_{1}=(a,1-a)}$, $\mathrm{\rho_{2}=(b,1-b)}$ and $\mathrm{\rho_{3}=(c,1-c)}$. The parameters a, b and c are functions of the populations and they are different, not like the case of two coupled spins, because in this case the state is no longer symmetric with respect to the partial trace. Therefore, the local extractable work is,
\begin{equation}
\mathrm{w_{1}+w_{2}+w_{3}=2(h-h')(a-a'+b-b'+c-c')}.
\end{equation}
If we replace a, a', b, b', c, and c' with their expressions in terms of populations, we get that the global work W is equal to the total local one. It would be very important to compute the upper bound of the efficiency when we are no longer interested in two interacting systems but N s-spins, and to investigate if the number of coupled spins could affect the upper bound of the efficiency in addition to the coupling and the spin of the particles (see, \cite{Ramandeep}), i.e. if $\mathrm{\eta_{up}}$ is a function of N.
\subsection{Ising model} \label{Ising}
The Hamiltonian of the $\mathrm{N}$ spin-$\frac{1}{2}$ Ising model under the influence of a magnetic field $\mathrm{h}$ is 
\begin{equation} 
\mathrm{H=J\sum_{i=1}^{N}\sigma_{z}^{i}\sigma_{z}^{i+1}+h\sum_{i=1}^{N}\sigma_{z}^{i}, }
\label{H} 
\end{equation} 
where $\mathrm{\sigma_{z}^{i}}$ is the $\mathrm{z}$-component of the Pauli spin matrices acting on the site $\mathrm{i\in [1,N]}$. The periodicity is here as well as assumed, i.e., $\mathrm{\sigma_{z}^{N+1} = \sigma_{z}^{1} }$. $\mathrm{N}$ is the total number of sites, and here it will range from $2$ to $6$. The eigenvalues of this Hamiltonian are given in the appendix. Note that this Hamiltonian has been used in Ref.\cite{Oliveira2020}, to show that the enhancement is only a matter of the structure of the system and it is not quantum correlations. Here, we also study the case when the system is working as a refrigerator  which was not considered in \cite{Oliveira2020}. 

\begin{figure}[hbtp]
\includegraphics[scale=0.7]{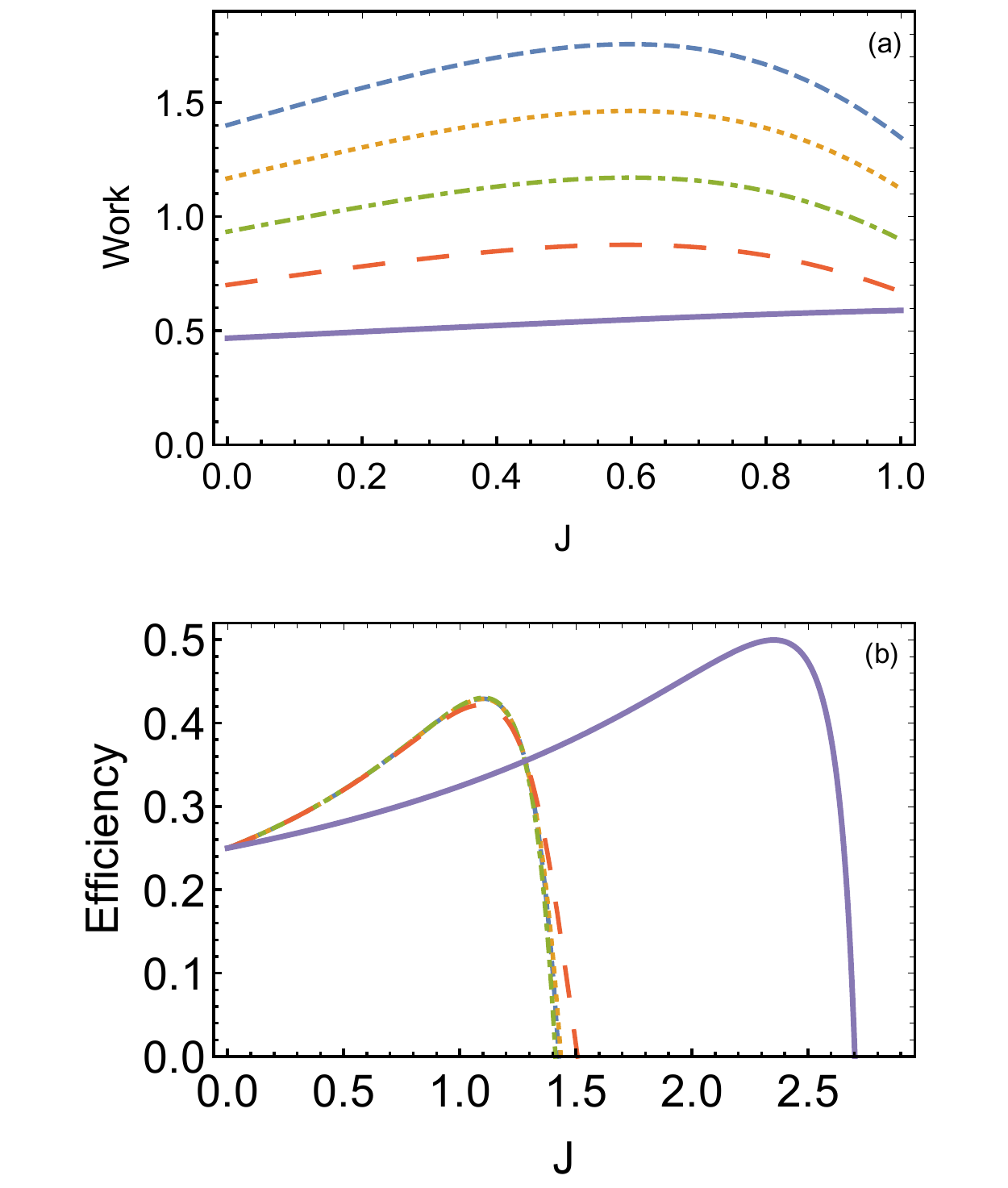}
\caption{(Color online). Plot of (a) the extracted work from $\mathrm{N}$ coupled spins as a function of $\mathrm{J}$, where $\mathrm{N}$ ranges from $2$ to $6$ (b) the efficiency of coupled spin-$1/2$ as a function of $\mathrm{J}$. The parameter values are the same as in Fig.\ref{1}. It is worth noting that the efficiencies of three to six coupled spins are nearly coincidable. $\mathrm{N=6}$ (blue dashed line), $\mathrm{N=5}$ (orange dotted line), $\mathrm{N=4}$ (green dot-dashed line), $\mathrm{N=3}$ (red large-dashing line), $\mathrm{N=2}$ (purple solid line).}
\label{6}
\end{figure}
\begin{figure}[hbtp]
\includegraphics[scale=0.7]{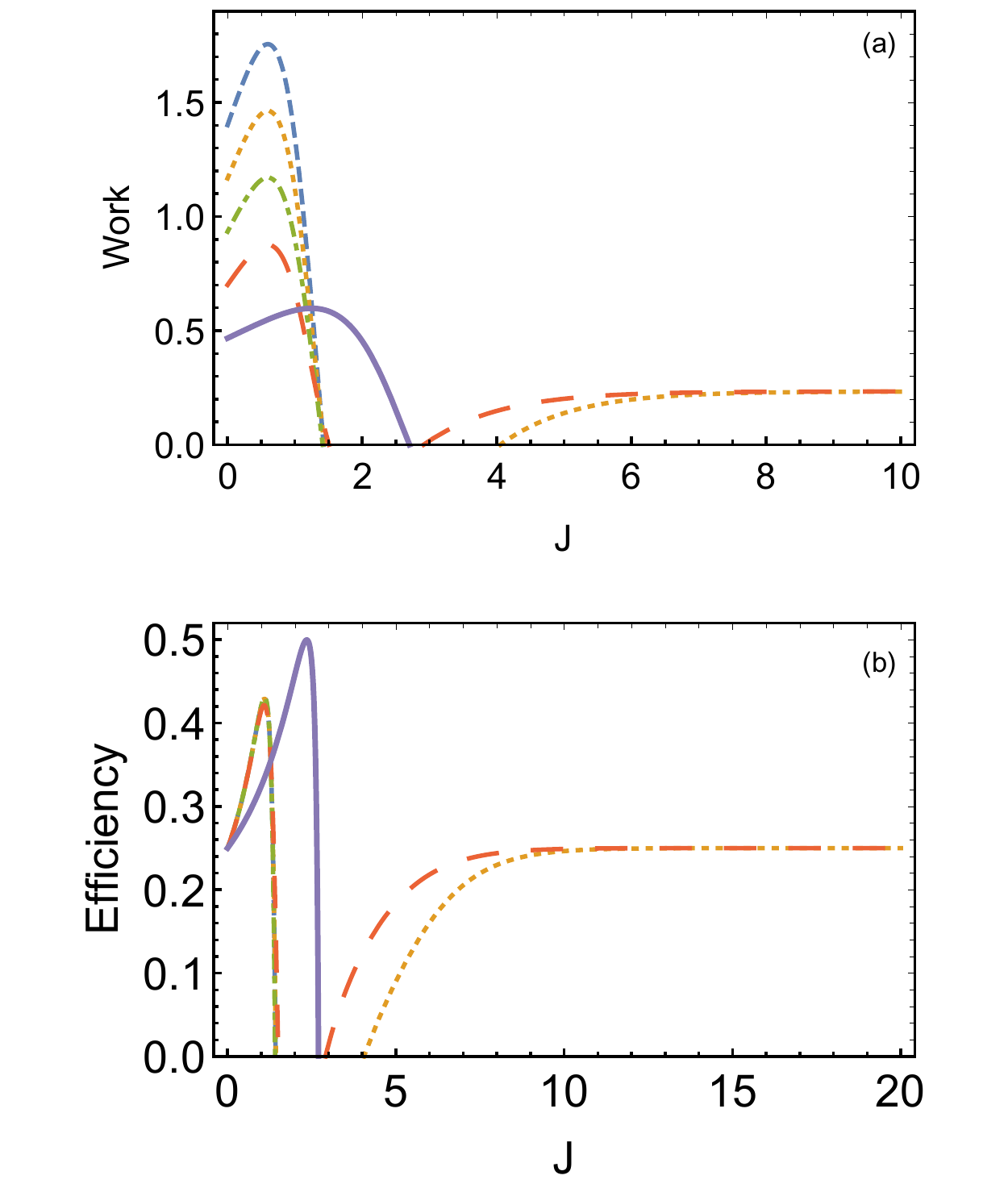}
\caption{(Color online). Plot of (a) the extracted work from $\mathrm{N}$ coupled spin-$1/2$ as a function of $\mathrm{J}$, where $\mathrm{N}$ ranges from $2$ to $6$ (b) the efficiency of coupled spin-$1/2$ as a function of $\mathrm{J}$. Here, in contrast to Fig.\ref{6} the work and efficiency are plotted in the weak as well as in the strong coupling regimes. The parameter values are the same as in Fig. \ref{1}. $\mathrm{N=6}$ (blue dashed line), $\mathrm{N=5}$ (orange dotted line), $\mathrm{N=4}$ (green dot-dashed line), $\mathrm{N=3}$ (red large-dashing line), $\mathrm{N=2}$ (purple solid line).}
\label{7}
\end{figure}
\begin{figure}[hbtp]
\centering
\includegraphics[scale=0.55]{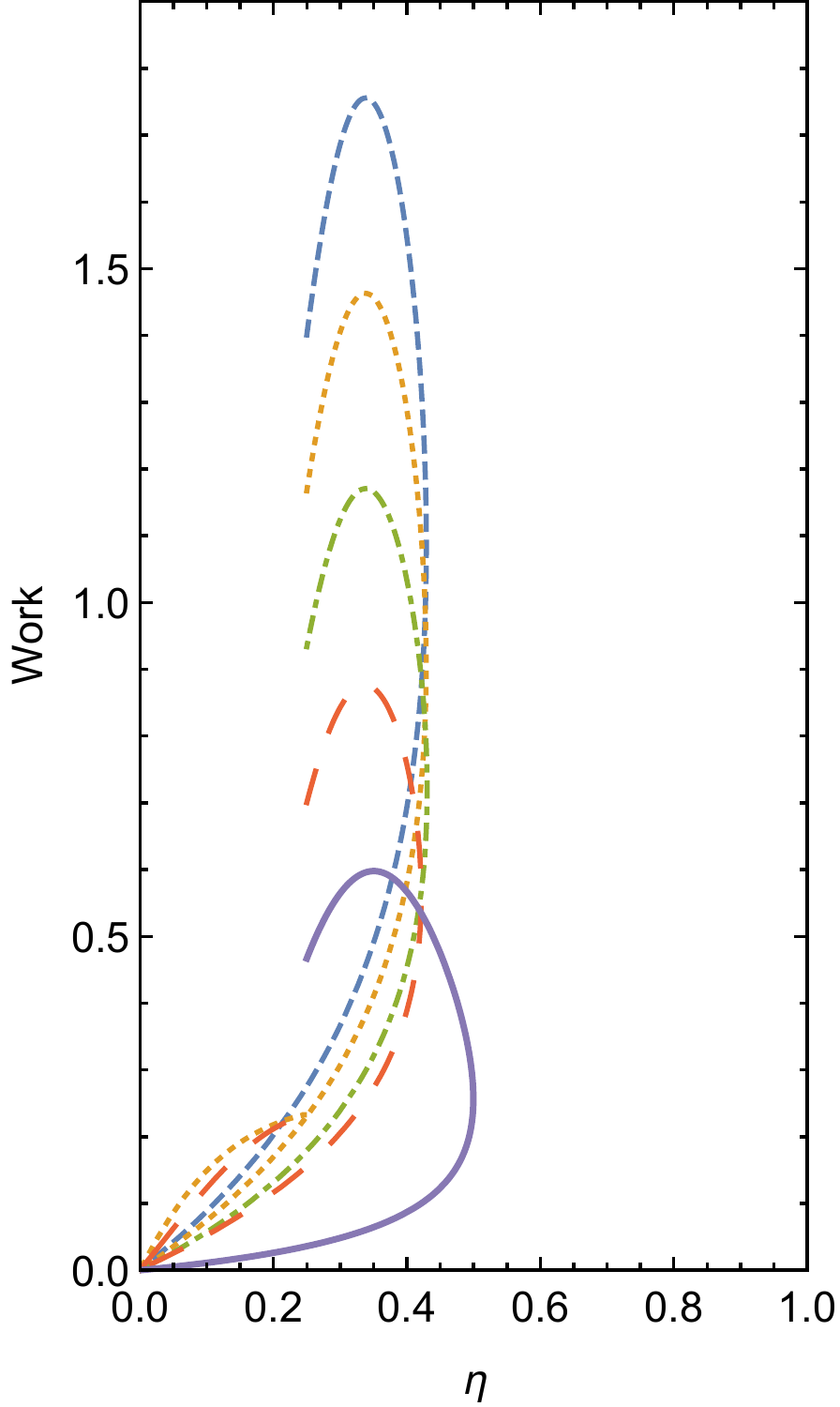}
\caption{(Color online). The work extracted versus the efficiency, where $\mathrm{J}$ is in the interval $[0,10]$, and $\mathrm{N}$ ranges from $2$ to $6$. The parameter values are the same as in Fig.\ref{1}. $\mathrm{N=6}$ (blue dashed line), $\mathrm{N=5}$ (orange dotted line), $\mathrm{N=4}$ (green dot-dashed line), $\mathrm{N=3}$ (red large-dashing line), $\mathrm{N=2}$ (purple solid line).}
\label{8}
\end{figure}
\begin{figure}[hbtp]
\centering
\includegraphics[scale=0.7]{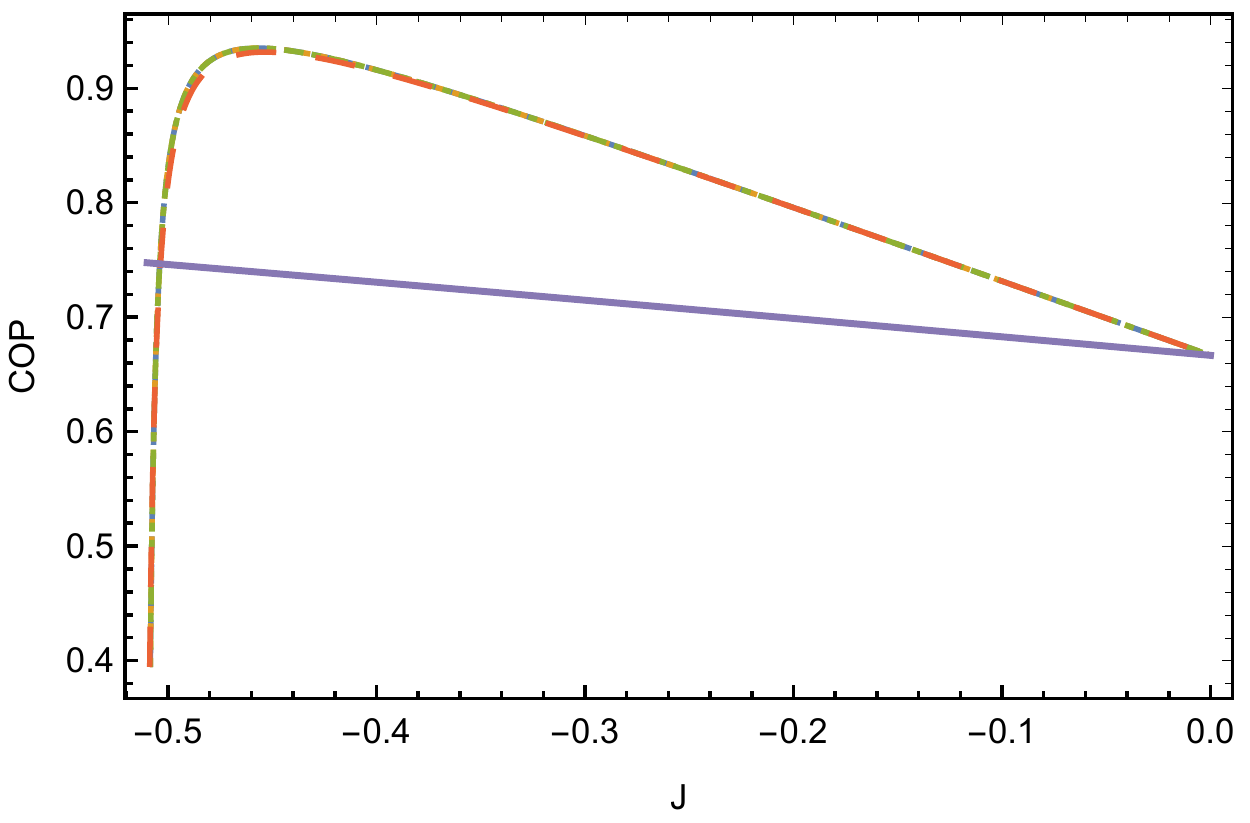}
\caption{(Color online). The COP of $\mathrm{N}$ coupled spin-$1/2$ as a function of $\mathrm{J}$. It is only in the interval $[-0.5,0]$ that we have an enhancement in the COP for three to six-coupled spin-$1/2$. The parameter values are the same as in Fig.\ref{2}. $\mathrm{N=6}$ (blue dashed line), $\mathrm{N=5}$ (orange dotted line), $\mathrm{N=4}$ (green dot-dashed line), $\mathrm{N=3}$ (red large-dashing line), $\mathrm{N=2}$ (purple solid line).}
\label{9}
\end{figure}

In Fig. \ref{6}(a), we plot the work extracted from two coupled spins up to six. We see that as we increase their number, more work can be extracted. As explained before, it is the number of spins N and the interaction between them J which enhances the work extracted. In Fig.\ref{6}(b), we plot the efficiency as a function of $\mathrm{J}$ for $\mathrm{N = 2}$ up to $\mathrm{N = 6}$. We see that increasing the number of coupled spins enhances the efficiency, but only for small values of $\mathrm{J}$. However, the efficiency of three up to six coupled spins is nearly coincidable. In \cite{Altintas} the authors studied a coupled spin-$1/2$ with another $s$ spin where $s$ ranges from $1$ to $3$, and they found that all of these interacting spins can work as heat engines in the strong coupling regime. However, in our case, we see from Fig.\ref{7}, that only three and five interacting spins could work as heat engines in the strong regime. Furthermore, in our case, in contrast to them, we see a remarkable enhancement in the extractable work, as in Ref.\cite{DasGosh}, since in our case as well, we have degeneracy. In Fig.\ref{8} we plot work versus efficiency for $\mathrm{J}$ in the interval $[0,10]$. We can see that the efficiency at maximum work is not altered by the number of interacting spins. However, in contrast to \cite{Altintas}, the work at maximum efficiency is affected by the number of interacting spins.\par 

Now let us study the case when our coupled spins are working as a refrigerator. For this, the COP is plotted in Fig.\ref{9}. It is seen that when the system is ferromagnetic, the COP of $3$ to $6$ coupled spins is above that of two-coupled spin-$1/2$. However, all of them surpass the COP of the Otto engine. In addition to this, notice that when the system is working as a heat engine (Figs.\ref{6} and \ref{7}), its efficiency can surpass Otto one only when the system is anti-ferromagnetic. And when it is a refrigerator, the enhancement is seen when the system is ferromagnetic (Fig.\ref{9}). We now briefly compare our results with the ones reported in \cite{GuoZhang, Zhang, Altintas}. It was shown in \cite{GuoZhang, Zhang} that entanglement between the two-coupled spins at the end of the hot isochoric stage is not necessary for harvesting work. They show that we only need to have entanglement at the end of the cold isochotic stage. In contrast, in \cite{Altintas} , it was shown that we can harvest work even if we have no entanglement on the cold isochoric side and the hot isochoric side, but quantum discord was present. In our case, we have neither entanglement nor quantum correlations, and even more, the system could harvest work. To resume, from Figs. \ref{6}, \ref{7}, \ref{8} and \ref{9} we see that the extractable work, the efficiency, as well as the COP of these coupled but not correlated systems could surpass the corresponding ones of the Otto without exploiting entanglement and quantum correlations.\par 

Here, when the system is working as a heat engine, we plot work and efficiency only for positive values of $\mathrm{J}$. This is because only when the system is anti-ferromagnetic do we have an enhancement in work as well as efficiency. When the system is working as a refrigerator, the enhancement is observed in the COP only when the system is ferromagnetic. One can explain this just from the eigenvalues and the populations of the interacting spins. For $\mathrm{N=2}$, the enhancement in the COP is approximately in the interval $[-2,0]$. For $\mathrm{N = 3}$ to $6$, the enhancement in COP is approximately in the interval $[-0.5,0]$.

As for the Ising model, the eigenvalues are linear in the magnetic field and the eigenstates are parameter independent, thus one can easily demonstrate that we have the preservation of the extensive property of the work, i.e. $\mathrm{W_{g}=Nw_{local}}$ which means that the work extracted from N coupled spins globally will be equal to N times the one extracted locally from one spin $\mathrm{w}$. This is due to the fact that, for the Ising model considered here, all the local spins have the same reduced state independently on N, thus $\mathrm{W=Nw}$. We conclude that as long as the eigenvalues are linear in the magnetic field and the eigenstates are parameter independent, the global work would be extensive.

Let's now focus on the local and the global thermodynamics of the two coupled spin-1/2 Ising model. We recall that the eigenvalues of this system are $\rm \{E_{1}=2h,E_{2}=E_{3}=-2J,E_{4}=-2h\}$ and their corresponding probabilities are, $\rm \rho=\{p_{1},p_{2},p_{2},p_{4}\}$. The local reduced states of the subsystems are equal, $\rm \rho_{1}=\rho_{2}=diag\{p_{1}+p_{2},p_{4}+p_{2}\}$. The expression of the global work in terms of the energies and the populations is,
\begin{equation}
\mathrm{W=(E_{1}-E'_{1})(p_{1}-p'_{1}+p'_{4}-p_{4}).}
\end{equation}
The local work extracted from each subsystem is,
\begin{equation}
\mathrm{w_{1}=(E_{1}-E'_{1})(p_{1}-p'_{1}+p'_{4}-p_{4})/2.}
\end{equation}
Thus, the global work is extensive. In addition, one should note that even though the state is only classically correlated, the local effective temperatures are not equal to the ones of the baths, and this is because of the coupling J. The effective temperatures \cite{Johal} are given by,
\begin{equation}
\mathrm{T_{eff,1}=T_{eff,2}=2h(\log(\frac{p_{2}+p_{4}}{p_{1}+p_{2}}))^{-1},}
\end{equation}
at the hot bath side and,
\begin{equation}
\mathrm{T_{eff,1}'=T_{eff,2}'=2h'(\log(\frac{p_{2}'+p_{4}'}{p_{1}'+p_{2}'}))^{-1},}
\end{equation}
at the cold bath side. This tells us that even though the global extracted work is the same as the local one and that the spins are only classically correlated, the heats absorbed from and released to the heat baths are not equal. Therefore, the operation modes will not be the same. This is because the local temperatures are not equal to the ones of the heat baths. The global heats are,
\begin{equation}
\mathrm{Q_{h}=E_{1}(p_{1}-p'_{1}+p'_{4}-p_{4})+2E_{2}(p_{2}-p'_{2}),}
\end{equation}
\begin{equation}
\mathrm{Q_{c}=-E'_{1}(p_{1}-p'_{1}+p'_{4}-p_{4})-2E_{2}(p_{2}-p'_{2}),}
\end{equation}
and the total local ones are,
\begin{equation}
\mathrm{q_{h,12}=E_{1}(p_{1}-p'_{1}+p'_{4}-p_{4}),}
\end{equation}
\begin{equation}
\mathrm{q_{c,12}=-E'_{1}(p_{1}-p'_{1}+p'_{4}-p_{4}).}
\end{equation}
We see that they are not equal, and are different by the quantity $\mathrm{2E_{2}(p_{2}-p'_{2})}$, which is the responsible for th enhancement of the efficiency. More precisely, this term, even though makes no difference between the work extracted by a global agent and a local one, the global $\mathrm{\eta_{g}}$ and $\mathrm{COP_{g}}$ can exceed the ones of the Otto $\mathrm{\eta_{o}}$ and $\mathrm{COP_{o}}$, which is not the case for the local efficiency ($\mathrm{\eta_{l}=1-h'/h}$) and local COP ($\mathrm{COP_{l}=h'/h-h'}$) which are equal to the ones of the Otto. This means that a global agent can extract the same work as a local one, but by absorbing less amount of heat from the hot bath. And for refrigeration, for the same amount of consumed work, the global agent can absorb a bigger amount of heat than the local one. Thus, the global agent will cool more efficiently than a local one, and this is independently if the system is classically or quantum mechanically correlated.

Finally, let's explain the enhancement observed in the COP of the Ising model. The COP is given as follows:
\begin{equation}
\mathrm{COP=\frac{h'}{h-h'}+\frac{-4J_{z}(p_{2}-p_{2}')}{2(h-h')(p_{4}-p'_{4}+p'_{3}-p_{3})}}.
\end{equation}
$\mathrm{q_{c}=4J_{z}(p_{2}-p_{2}')}$ this is the heat absorbed from the cold bath by the degenerate idle level $\mathrm{-2J_{z}}$. In this case, one sees that if $\mathrm{J_{z}=0}$ the COP is equal to the one of the Otto. And to exceed it, the second term has to be positive, which is equivalent to $\mathrm{-4J_{z}(p_{2}-p_{2}')<0}$, since the denominator is negative. This means that idle levels have to absorb heat from the cold bath in addition to working ones, otherwise the COP will not be degraded. Thus, if $\mathrm{J_{z}>0}$ then $\mathrm{p_{2}-p'_{2}}$ has to be positive. However, for $\mathrm{J_{z}>0}$, we have $\mathrm{p_{2}-p'_{2}<0}$, thus the second term will degrade the COP. This expalins why the enhancement in the COP is observed only when the system is ferromagnetic. Note that in contrast to Ising+KSEA model in which idle levels are not needed to enhance the COP, here we need them, but they have to channel heat in the same direction as working levels. Finally, following the same arguments, we can explain the enhancement in the efficiency.

\section{Clarifying some important points}\label{clarify}
At this point, before we conclude this paper, we should clarify some important points:

First, in the previous works \cite{Johal,Ukraine,Altintas} we have seen that the eigenvalues were linear in $\mathrm{h}$ for the local as well as for the global Hamiltonian. However, we see here (Sec. \ref{B}) that the eigenvalues of the global Hamiltonian are not linear in the $\mathrm{h}$ in contrast to the local one. Therefore, the extensive property of the work is dependent on the eigenvalues as well as on the eigenstates of the Hamiltonian. Moreover, we can safely say that the extensive property of the work is not related to any type of quantum or classical correlations. It is only related to the coupling parameters. More precisely, if the chosen interaction between the spins does still preserve the linearity of the eigenvalues in the magnetic field, then in this case the extensive property of the work would still be preserved. However, if it violates the linearity, then we can no longer ensure that the global work is equal to the total local one. This is because the local Hamiltonian is linear in h.
\\
Note that one may think that the preservation of the extensive property has to do with the assumption that the system is in equilibrium with a heat bath apositive temperature $\mathrm{\beta>0}$, thus the state is passive, i.e. higher levels of energy are occupied with less probabilities (for $\mathrm{E_{i}\geq E_{j}}$ we have $\mathrm{p_{i}\leq p_{j}}$, see ref.\cite{Ergotropy}), which means that state ergotropy is zero. However, this is not the case. More precisely, we can let the system thermalizes with a heat bath at negative temperature $\mathrm{\beta}<0$, thus the state in this case is not passive because higher levels will be occupied with high populations, i.e. for $\mathrm{E_{i}\geq E_{j}}$ we have $\mathrm{p_{i}\geq p_{j}}$, thus the state would have non-zero ergotropy. However, even in this case, the extensive property is still preserved for the Heisneberg model but not for the Ising+KSEA model. And this is because the temperature of the bath will only affect the value of populations, but not the expression of the global as well as the local works in terms of energies and populations. Therefore, $\mathrm{W-w\geq0}$ is neither related to quantum correlations nor to ergotropy, i.e. is the state being passive or not, it is only dependent on the Hamiltonian $\mathrm{H}$.

Second, actually one may ask that even though if entanglement is not present, however there will be quantum discord which could affect the thermodynamical quantities, i.e. heats, work and efficiency. However, the Ising model shows us that this enhancement in efficiencies, i.e. $\mathrm{\eta>\eta_{o}}$ and $\mathrm{COP>COP_{o}}$ could be observed as well in the absence of discord.

Third, note that when one sees the eigenvalues of the interacting spins, either for the Heisenberg or the Ising model, one may still not be convinced if it is the presence of idle levels or something else that is the reason behind this enhancement. In this case, one can simply shift the eigenvalues and will get two sets of levels, i.e. working and idle levels. Let's take two examples: the first is the case of three coupled spins XXX chain. In this case, shifting the eigenvalues $\rm E=\{-(h+3J),-(h+3J),h-3J,h-3J,-3(h-J),3(h+J),-h+3J,h+3J\}$ by +h one gets the new eigenvalues: $\rm E=\{-3J,-3J,2h-3J,2h-3J,-2h+3J,4h+3J,-3J,2h+3J\}$. Thus, we obtain three idle levels. These levels would not contribute to work, but only to heat. The other example is the Ising model of two interacting spins which are, $\rm \{-2h,-2J,-2J,2h\}$. And if one plots the heat absorbed by the idle levels, one sees that they are channeling heat in the wrong direction of the one by the working levels. However, it is not necessary that all idle levels take heat in the opposite direction, but only the average of heat should be in the opposite direction of the working levels. Furthermore, one should note that even though there is a part of working levels that does not depend on the magnetic field, i.e. J, this part only contributes to $\mathrm{Q_{h}}$ and $\mathrm{Q_{c}}$, but work is only dependent on the working part, i.e. the part that depends on the magnetic field. Thus, it is not necessary to divide the levels into two sets: idle and working levels to explain the enhancements. 
\\
Let's take the case when the eigenvalues are linear in the h and J, i.e. $\mathrm{E_{i}=a_{i}h+b_{i}J}$ and $\mathrm{E'_{i}=a_{i}h'+b_{i}J}$. $\mathrm{a_{i}}$ and $\mathrm{b_{i}}$ are just numbers, they can be positive or negative. Then the heats and work expressions are,
\begin{equation}
\mathrm{Q_{h}=\sum_{i}(a_{i}h+b_{i}J)(\frac{e^{-\beta_{h}(a_{i}h+b_{i}J)}}{Z_{h}}-\frac{e^{-\beta_{c}(a_{i}h'+b_{i}J)}}{Z_{c}})},
\end{equation}
\begin{equation}
\mathrm{\mathrm{Q_{c}=-\sum_{i}(a_{i}h'+b_{i}J)(\frac{e^{-\beta_{h}(a_{i}h+b_{i}J)}}{Z_{h}}-\frac{e^{-\beta_{c}(a_{i}h'+b_{i}J)}}{Z_{c}}})},
\end{equation}
and work is,
\begin{equation}
\mathrm{W=\sum_{i}a_{i}(h-h')(\frac{e^{-\beta_{h}(a_{i}h+b_{i}J)}}{Z_{h}}-\frac{e^{-\beta_{c}(a_{i}h'+b_{i}J)}}{Z_{c}})}.
\end{equation}
With $\mathrm{Z=\sum_{i}e^{-\beta E_{i}}}$. We see that the heats are affected by both, the part of eigenvalues that depend oh h and the one of J. However, work is affected by J only through the populations, i.e. the idle part of $\mathrm{E_{i}}$ disappears because of the difference $\mathrm{E_{i}-E'_{i}}$. Thus, if J is tuned in the right way, more work will be extracted than in the case $\mathrm{J=0}$. The same thing can be said when the eigenvalues are linear in h but not in J. However, when $\mathrm{E_{i}}$ are nonlinear in h things become more complicated.\par
The above expressions of heats and work can be rewritten as,
\begin{equation}
\mathrm{Q_{h}=ah+bJ}, \ \mathrm{Q_{c}=-ah'-bJ, \ and \ W=a(h-h')}.
\end{equation}
With $\mathrm{a=\sum_{i}a_{i}(e^{-\beta_{h}(a_{i}h+b_{i}J)}/Z_{h}-e^{-\beta_{c}(a_{i}h'+b_{i}J)}/Z_{c})}$ and $\mathrm{b=\sum_{i}b_{i}(e^{-\beta_{h}(a_{i}h+b_{i}J)}/Z_{h}-e^{-\beta_{c}(a_{i}h'+b_{i}J)}/Z_{c})}$. In this case the efficiency will be,
\begin{equation}
\mathrm{\eta=\eta_{o}\frac{1}{1+\frac{bJ}{ah}}}.
\end{equation}
For $\mathrm{\eta>\eta_{o}}$ the second term has to be bigger than one. From $\mathrm{W>0}$ we have $\mathrm{a>0}$
(we have assumed that $\mathrm{h>h'}$), thus depending on the sign of $\mathrm{bJ}$, $\mathrm{\eta}$ either can be enhanced or degraded. More precisely, if $\mathrm{bJ>0}$ then $\mathrm{\eta<\eta_{o}}$ and if $\mathrm{bJ<0}$ then $\mathrm{\eta>\eta_{o}}$. Thus, when the term of global heat $\mathrm{Q_{h}}$ absorbed from the hot bath multiplied by h takes heat from the hot bath to the cold bath, the heat multiplied by J has to take it in the other direction. Now, let's explain the enhancement in the COP. The expression of the COP is,
\begin{equation}
\mathrm{COP=COP_{o}+\frac{bJ}{a(h-h')}.}
\end{equation}
Since $\mathrm{a(h-h')<0}$, in this case bJ has to be negative as well. And because -bJ is the amount of heat absorbed from the cold bath, thus it will enhance the COP. Therefore, it is necessary in this case that idle levels take heat from the bath we want to cool otherwise the COP will be degraded.\par
Third, in Ref.\cite{Chang} the authors found a nice relation between the work extracted and entropies. This relationship is given as follows:
\begin{equation}
\mathrm{W=(T_{h}-T_{c})(S(\rho)-S(\rho'))-T_{h}H[p'_{i}|p_{i}]-T_{c}H[p_{i}|p'_{i}]}.\label{WW}
\end{equation}
H is the relative entropy and it is positive. $\rho$ and $\rho'$ are the equilibrium states of the system at the hot and cold baths, and p and p' are the corresponding populations, respectively. This relationship can be easily proven starting from,
\begin{equation}
\mathrm{W=\sum_{i}(E_{i}-E'_{i})(p_{i}-p'_{i})}.
\end{equation}
Then using $\mathrm{E_{i}=-T_{h}\log(p_{i}Z_{h})}$ and $\mathrm{E'_{i}=-T_{c}\log(p'_{i}Z_{c})}$, we get this relation. Therefore, for $\mathrm{W>0}$ we have to ensure not only that $\mathrm{T_{h}>T_{c}}$ but as well $\mathrm{S(\rho)>S(\rho')}$, and an addition to this $\mathrm{(T_{h}-T_{c})(S(\rho)-S(\rho'))>T_{h}H[p'_{i}|p_{i}]+T_{c}H[p_{i}|p'_{i}}$. More precisely, this relationship tells us that for the positive work condition to be satisfied the entropy of the system at the hot bath side has to be greater than the one at the cold bath side. And according to the second law, this is true, as heat will flow spontaneously from a high entropy system to a less entropy one. Moreover, this relation is useful either for a coupled or only one system, and it's difficult to relate it to quantum correlations, since it tells us for $\mathrm{W>0}$ we have only to compare the entropies of the system at the hot and cold bath sides independtly of the system being correlated or just a one system. In the same manner, if one is interested in the local thermodynamics of correlated systems, one can have an expression of the local work in terms of entropies and relative entropies. In this case, the expression of the local work in terms of entropies is,
\begin{equation}
\begin{split}
\rm w_{\alpha}= & \mathrm{(T_{h}-T_{c})(S(\rho_{\alpha})-S(\rho'_{\alpha}))-T_{h}H[p'_{i,\alpha}|p_{i,\alpha}]}
\\ &
\mathrm{-T_{c}H[p_{i,\alpha}|p'_{i,\alpha}].} \label{ww}
\end{split}
\end{equation}
with $\mathrm{\alpha=1}$ and 2. $\mathrm{\rho_{\alpha}}$ is the local reduced state of the system, $\mathrm{p_{i,\alpha}}$ and $\mathrm{p'_{i,\alpha}}$ are the local populations of the system when it is in thermal equilibrium with a hot bath and cold bath, respectively. Therefore, according to our study, the equations \ref{WW} and \ref{ww} tell us that the condition for a positive work either globally or locally can be explained only from the entropies of the system at the hot and cold bath sides without mentioning quantum correlations and entanglement. Furthermore, one should note that Eqs. \ref{WW} and \ref{ww} are not used to explain efficiency enhancement, but they are only related to the positive work condition.\par
Fourth, our results do not claim that quantum correlations should be excluded at all to enhance the performance of thermal machines. But it shows that surpassing the Otto efficiencies could be possible even without using them. Therefore, one may wonder if the Carnot efficiency could be surpassed using idle levels, and our answer would be no. And the reason is that even though some levels take heat in the opposite direction to the average of heat flow, the effective temperature associated with each of the two levels of the system is equal to the heat baths, i.e. $\mathrm{T_{h}}$ and $\mathrm{T_{c}}$, thus $\mathrm{\eta\leq \eta_{C}=1-\frac{T_{c}}{T_{h}}}$. In addition to this, one should note that the enhancement in efficiency comes at a cost which is due to creating such non-uniform eigenvalues.\par

Finally, our studies show that neither the enhancement in work, efficiency, COP nor the extensive property of work are related to the presence of quantum correlations. We have seen as well that idle levels are needed to take heat in the wrong direction to enhance efficiency. However, when the system is working as a refrigerator, we saw that depending on the model, they may not be necessary to contribute to the heats to enhance the COP or if their contribution is necessary they have to take heat in the same direction as the working levels, otherwise they will degrade the heat absorbed from the cold bath, thus $\mathrm{COP\leq COP_{o}}$.
\section{Conclusions}\label{D}
By studying two coupled spin-$1/2$ Ising model under the influence of KSEA interaction in a magnetic field along the $\mathrm{z}$-axis, we show that the enhancement observed in the efficiency is only due to the structure of the energy levels of the system and not entanglement or quantum correlations. Furthermore, we reexamined the results reported in \cite{Altintas} according to which to break the extensive property of the work extracted globally, we have to change the coupling parameters next to the magnetic field. In our case, we show that this is not necessary and we can break it even when only the magnetic field is changed in the adiabatic stages. Then we study the effect of increasing the number of interacting spin-$1/2$ on efficiency, extractable work, and COP. In contrast to Ref.\cite{Altintas}, when they extend the dimension of one spin, they have a little enhancement in efficiency and the amount of extractable work, in our case we see a remarkable enhancement in the extractable work, though the maximum of the efficiencies is still bounded by the one of two-coupled spin-$1/2$. In addition, we see that only when the number of the interacting spins is odd, the system could work as a heat engine in the strong coupling regime. The enhancement in COP is seen as well when the system is working as a refrigerator. Moreover, note here that we did not consider the $\mathrm{XY}$ components of $\mathrm{J}$, only the $\mathrm{z}$-component of $\mathrm{J}$ to ensure that no entanglement or quantum correlations will build up between the interacting spins along the cycle. Our results support the expectation, put forward in Refs.\cite{ Anka, Oliveira2020, Johall}, that the structure of the Hamiltonian is a resource and that this is the reason behind the enhancement observed in the efficiency and the extractable work, as well as the COP.\par

It is worth noting that the enhancement observed in efficiency and COP was also observed in refs.\cite{Mendes, Sourav}. In Ref.\cite{Sourav} the role of anharmonicity on COP has been investigated. The authors there showed that the COP of a quantum harmonic oscillator after introducing anharmonicity could surpass that of the Otto. Moreover, this was shown even for a two-level system, as in our case, when we have two coupled spin-$1/2$ under the influence of KSEA interaction in the absence of idle levels, i.e., when $\mathrm{J_{z}=0}$. In Ref. \cite{Mendes}, the role of Kerr-nonlinearity on efficiency and COP has been studied in detail, showing that Kerr-nonlinearity could boost them above the ones of Otto. To resume, our work supports the idea that the enhancement in the performance of coupled spins is only a matter of the structure of the Hamiltonian. It would be useful to look at the role of thermal fluctuations \cite{Jarzynski1997, Campisi, Crooks}, since here and in the previous works \cite{Zhang, Johal, Altintas, Sodeif, Deniz, Ukraine, Mehta, Anka, Zhao, Ramandeep} the interest was only in the average of the thermodynamical quantities, i.e., heat, work, efficiency, and COP. Furthermore, our work and the previous ones will pave the way for efficient quantum thermal machines in which the structure of the system is used as a resource. Inspired by other quantum resource theories, \cite{Brandao, Brandao2, Winter, Baumgratz, Gour, Spekkens}, a resource theory of the structure of the energy levels of quantum systems, will also be interesting to study \cite{Mauro}.\par
Our results go in the same way as in \cite{Kurizki23,GWatanabe,Latune,Kamimura,Souza}, which show the effect of the coupling on the performance of quantum heat engines. We hope to investigate further our study concerning the global and the local thermodynamics more thoroughly, or in general, when both the unitary and the thermalization strokes are in finite time.\par
{\bf Acknowledgment:} We gratefully thank anonymous reviewers for their constructive remarks and for their insightful comments and suggestions, which have improved the quality of the paper.
\appendix
\section{EIGENVALUES OF THE WORKING MEDIUM}
Here we give the eigenvalues of the Hamiltonian \ref{H}. Their corresponding eigenstates are the elements of the computational basis. The eigenvalues of the Hamiltonian when $\mathrm{N=2}$ are $\rm \{2h,-2J,-2J,-2h\}$ \cite{Johal, Anka}.

The eigenvalues of the Hamiltonian when $\mathrm{N=3}$ are $\rm \{3h+3J,h-J,h-J,-h-J,h-J,-h-J,-h-J,-3h+3J\}$. The partition function of the system when it is in equilibrium with a heat bath at inverse temperature $\mathrm{\beta}$ is: $\rm Z=2e^{-3\beta J}\cosh(3\beta h)+6e^{\beta J}\cosh(\beta h)$.

The eigenvalues of the Hamiltonian when $\mathrm{N=4}$ are $\rm \{4h+4J,2h,2h,0,2h,-4J,0,-2h,2h,0,-4J,-2h,0,-2h,-2h, \\ -4h+4J\} $. The partition function of the system when it is in equilibrium with a heat bath at inverse temperature $\mathrm{\beta}$ is: $\rm Z=2(e^{4\beta J}+e^{-4\beta J}\cosh(4\beta h))+4(1+2\cosh(2\beta h))$.

The eigenvalues of the Hamiltonian when $\mathrm{N=5}$ are $\rm \{5h+5J,3h+J,3h+J,h+J,3h+J,h-3J,h+J,-h+J,3h+J,h-3J,h-3J,-h-3J,h+J,-h-3J,-h+J,-3h+J,3h+J,h+J,-h-3J,-3h+J,h+J,-h+J,-h-3J,-3h+J,h+J,-h+J,-h-3J,-3h+J,-h+J,-3h+J,-3h+J,-5h+5J\}$. The partition function of the system when it is in equilibrium with a heat bath at inverse temperature $\mathrm{\beta}$ is: $\rm Z=2e^{-5\beta J}\cosh(5\beta h)+10e^{-\beta J}(\cosh(\beta h)+\cosh(3\beta h))+10e^{3\beta J} \cosh(\beta h)$.

The eigenvalues of the Hamiltonian when $\mathrm{N=6}$ are $\rm \{6h+6J,4h+2J,4h+2J,4h+2J,4h+2J,4h+2J,4h+2J,2h+2J, 2h+2J,2h+2J,2h+2J,2h+2J,2h+2J,2h-2J,2h-2J,2h-2J,2h-2J, 2h-2J,2h-2J,2h-2J,2h-2J,2h-2J,2J,2J,2J,2J,2J,2J,-2J,-2J,-2J,-2J,-2J,-2J,\\-2J,-2J,-2J,-2J,-2J,-2J,-2h+2J,-2h+2J,-2h+2J,-2h+2J,-2h+2J,-2h+2J,-2h-2J,-2h-2J,-2h-2J,-2h-2J,-2h-2J,-2h-2J,-2h-2J,-2h-2J, -2h-2J,-4h+2J,-4h+2J,-4h+2J,-4h+2J,-4h+2J,-4h+2J,-6J,-6J,-6h+6J\}$. The partition function when the system is in thermal equilibrium with a heat bath at inverse temperature $\beta$ is: $\rm Z=2(e^{-6\beta J}\cosh(6\beta h)+e^{6\beta J})+12e^{-2\beta J}(\cosh(4\beta h)+\cosh(2\beta h))+e^{2\beta J}(18 \cosh(2\beta h)+6)+12 \cosh(2\beta J)$.

\begin{thebibliography}{88}%
\makeatletter
\providecommand \@ifxundefined [1]{%
 \@ifx{#1\undefined}
}%
\providecommand \@ifnum [1]{%
 \ifnum #1\expandafter \@firstoftwo
 \else \expandafter \@secondoftwo
 \fi
}%
\providecommand \@ifx [1]{%
 \ifx #1\expandafter \@firstoftwo
 \else \expandafter \@secondoftwo
 \fi
}%
\providecommand \natexlab [1]{#1}%
\providecommand \enquote  [1]{``#1''}%
\providecommand \bibnamefont  [1]{#1}%
\providecommand \bibfnamefont [1]{#1}%
\providecommand \citenamefont [1]{#1}%
\providecommand \href@noop [0]{\@secondoftwo}%
\providecommand \href [0]{\begingroup \@sanitize@url \@href}%
\providecommand \@href[1]{\@@startlink{#1}\@@href}%
\providecommand \@@href[1]{\endgroup#1\@@endlink}%
\providecommand \@sanitize@url [0]{\catcode `\\12\catcode `\$12\catcode
  `\&12\catcode `\#12\catcode `\^12\catcode `\_12\catcode `\%12\relax}%
\providecommand \@@startlink[1]{}%
\providecommand \@@endlink[0]{}%
\providecommand \url  [0]{\begingroup\@sanitize@url \@url }%
\providecommand \@url [1]{\endgroup\@href {#1}{\urlprefix }}%
\providecommand \urlprefix  [0]{URL }%
\providecommand \Eprint [0]{\href }%
\providecommand \doibase [0]{http://dx.doi.org/}%
\providecommand \selectlanguage [0]{\@gobble}%
\providecommand \bibinfo  [0]{\@secondoftwo}%
\providecommand \bibfield  [0]{\@secondoftwo}%
\providecommand \translation [1]{[#1]}%
\providecommand \BibitemOpen [0]{}%
\providecommand \bibitemStop [0]{}%
\providecommand \bibitemNoStop [0]{.\EOS\space}%
\providecommand \EOS [0]{\spacefactor3000\relax}%
\providecommand \BibitemShut  [1]{\csname bibitem#1\endcsname}%
\let\auto@bib@innerbib\@empty
\bibitem [{\citenamefont {Prigogine}\ \emph {}(2015)\citenamefont {Prigogine}}]{Prigogine}
\BibitemOpen
\bibfield  {author} {\bibinfo {author} {\bibfnamefont {D.}\ \bibnamefont {Kondepudi}} and \bibinfo {author} {\bibfnamefont {I.}\ \bibnamefont {Prigogine},}\
}\bibfield{title} {{\bibinfo {title}{Modern thermodynamics: from heat engines to dissipative structures, 2nd ed. (John Wiley \& Sons Ltd, Chichester, 2015)}}\ }\BibitemShut {NoStop}
\bibitem [{\citenamefont {Carnot}\ \emph {}(1824)\citenamefont {Carnot}}]{Carnot}
\BibitemOpen
\bibfield  {author} {\bibinfo {author} {\bibfnamefont {S.}\ \bibnamefont {Carnot},}\
}\bibfield  {title} { {\bibinfo {title} {Réflexions sur la puissance motrice du feu et sur les machines propres a développer cette puissance, (Bachelier, Paris, 1824)}}\ }\BibitemShut {NoStop}
\bibitem [{\citenamefont {Alicki}\ \emph {et~al.}(1979)\citenamefont {Alicki} }]{Alicki}
\BibitemOpen
\bibfield  {author} {\bibinfo {author} {\bibfnamefont {R.}\ \bibnamefont {Alicki},}\
}\bibfield  {title} { {\bibinfo {title} {The quantum open system as a model of the heat engine,}}\ }\href {\doibase
	https://iopscience.iop.org/article/10.1088/0305-4470/12/5/007} {\bibfield  {journal} {\bibinfo  {journal}
		{J. Phys. A: Math. Gen.}\ }\textbf {\bibinfo {volume} {12}},\ \bibinfo {pages}
	{L103} (\bibinfo {year} {1979})}\BibitemShut {NoStop}
	\bibitem [{\citenamefont {Kosloff1984}\ \emph {et~al.}(1984)\citenamefont {Kosloff1984} }]{Kosloff1984}
\BibitemOpen
\bibfield  {author} {\bibinfo {author} {\bibfnamefont {R.}\ \bibnamefont {Kosloff},}\
}\bibfield  {title} { {\bibinfo {title} {A quantum mechanical open system as a model of a heat engine,}}\ }\href {\doibase
	https://doi.org/10.1063/1.446862} {\bibfield  {journal} {\bibinfo  {journal}
		{J. Chem. Phys.}\ }\textbf {\bibinfo {volume} {80}},\ \bibinfo {pages}
	{1625} (\bibinfo {year} {1984})}\BibitemShut {NoStop}
	\bibitem [{\citenamefont {Allahverdyan}\ \emph {et~al.}(2015) \citenamefont {Allahverdyan}}]{Allahverdyan}
\BibitemOpen
\bibfield  {author} {\bibinfo {author} {\bibfnamefont {A. E.}\ \bibnamefont
		{Allahverdyan}}, \bibinfo {author} {\bibfnamefont {R.}~\bibnamefont
		{Balian}}, and \bibinfo {author} {\bibfnamefont {T. M.}~\bibnamefont
		{Nieuwenhuizen,}} \
}\bibfield  {title} { {\bibinfo {title} {Quantum thermodynamics: thermodynamics at the nanoscale,}}\ }\href {\doibase
https://doi.org/10.1080/09500340408231829
} {\bibfield  {journal} {\bibinfo  {journal}
		{J. Mod. Opt.}\ }\textbf {\bibinfo {volume} {51}},\ \bibinfo {pages}
	{2703} (\bibinfo {year} {2004})}\BibitemShut {NoStop}
	\bibitem [{\citenamefont {Quan}\ \emph {et~al.}(2007)\citenamefont {Quan}}]{Quan}
\BibitemOpen
\bibfield  {author} {\bibinfo {author} {\bibfnamefont {H. T.}\ \bibnamefont
		{Quan}}, \bibinfo {author} {\bibfnamefont {Yi-xi}~\bibnamefont
		{Liu}}, \bibinfo {author} {\bibfnamefont {C. P.}~\bibnamefont
		{Sun}}, and \bibinfo {author} {\bibfnamefont {F.}~\bibnamefont
		{Nori}},\
}\bibfield  {title} { {\bibinfo {title} {Quantum thermodynamic cycles and quantum heat engines,}}\ }\href {\doibase
https://doi.org/10.1103/PhysRevE.76.031105
} {\bibfield  {journal} {\bibinfo  {journal}
		{Phys. Rev. E.}\ }\textbf {\bibinfo {volume} {76}},\ \bibinfo {pages}
	{031105} (\bibinfo {year} {2007})}\BibitemShut {NoStop}
	\bibitem [{\citenamefont {Quan1}\ \emph {et~al.}(2009) \citenamefont {Quan1}}]{Quan1}
\BibitemOpen
\bibfield  {author} {\bibinfo {author} {\bibfnamefont {H. T.}\ \bibnamefont
		{Quan},}\
}\bibfield  {title} { {\bibinfo {title} {Quantum thermodynamic cycles and quantum heat engines. II,}}\ }\href {\doibase
https://doi.org/10.1103/PhysRevE.79.041129} {\bibfield  {journal} {\bibinfo  {journal}
		{Phys. Rev. E}\ }\textbf {\bibinfo {volume} {79}},\ \bibinfo {pages}
	{041129} (\bibinfo {year} {2009})}\BibitemShut {NoStop}
\bibitem [{\citenamefont {Gemmer}\ \emph {et~al.}(2009)\citenamefont {Gemmer} }]{Gemmer}
\BibitemOpen
\bibfield  {author} {\bibinfo {author} {\bibfnamefont {J.}\ \bibnamefont {Gemmer}}, \bibinfo {author} {\bibfnamefont {M.}\ \bibnamefont {Michel}}, and \bibinfo {author} {\bibfnamefont {G.}\ \bibnamefont {Mahler},}\
}\bibfield  {title} { {\bibinfo {title} {Quantum thermodynamics, 2nd ed. (Springer-Verlag, Berlin Heidelberg, 2009)}}\ }\href {} {}\BibitemShut {NoStop} 
\bibitem [{\citenamefont {Skrzypczyk}\ \emph {et~al.}(2012) \citenamefont {Skrzypczyk}}]{Skrzypczyk}
\BibitemOpen
\bibfield  {author} {\bibinfo {author} {\bibfnamefont {N.}\ \bibnamefont
		{Brunner}}, \bibinfo {author} {\bibfnamefont {N.}~\bibnamefont
		{Linden}}, \bibinfo {author} {\bibfnamefont {S.}~\bibnamefont
		{Popescu}}, and \bibinfo {author} {\bibfnamefont {P.}~\bibnamefont
		{Skrzypczyk,}}\
}\bibfield  {title} { {\bibinfo {title} {Virtual qubits, virtual temperatures, and the foundations of thermodynamics,}}\ }\href {\doibase
https://doi.org/10.1103/PhysRevE.85.051117
} {\bibfield  {journal} {\bibinfo  {journal}
		{Phys. Rev. E}\ }\textbf {\bibinfo {volume} {85}},\ \bibinfo {pages}
	{05111} (\bibinfo {year} {2012})}\BibitemShut {NoStop}
\bibitem [{\citenamefont {Kosloff}\ \emph {et~al.}(2013)  \citenamefont {Kosloff}}]{Kosloff}
\BibitemOpen
\bibfield  {author} {\bibinfo {author} {\bibfnamefont {R.}\ \bibnamefont
		{Kosloff},}\
}\bibfield  {title} { {\bibinfo {title} {Quantum thermodynamics: A dynamical viewpoint,}}\ }\href {\doibase
	https://doi.org/10.3390/e15062100} {\bibfield  {journal} {\bibinfo  {journal}
		{Entropy}\ }\textbf {\bibinfo {volume} {15}},\ \bibinfo {pages}
	{2100} (\bibinfo {year} {2013})}\BibitemShut {NoStop}
	\bibitem [{\citenamefont {Levy}\ \emph {et~al.}(2013)  \citenamefont {Levy}}]{Levy}
\BibitemOpen
\bibfield  {author} {\bibinfo {author} {\bibfnamefont {R.}\ \bibnamefont
		{Kosloff},} and \bibinfo {author} {\bibfnamefont {A.}\ \bibnamefont
		{Levy},}\
}\bibfield  {title} { {\bibinfo {title} {Quantum heat engines and refrigerators: Continuous devices,}}\ }\href {\doibase
https://doi.org/10.1146/annurev-physchem-040513-103724} {\bibfield  {journal} {\bibinfo  {journal}
		{Annu. Rev. Phys. Chem.}\ }\textbf {\bibinfo {volume} {65}},\ \bibinfo {pages}
	{365} (\bibinfo {year} {2013})}\BibitemShut {NoStop}
	\bibitem [{\citenamefont {Goold}\ \emph {et~al.}(2016) \citenamefont {Goold}}]{Goold}
\BibitemOpen
\bibfield  {author} {\bibinfo {author} {\bibfnamefont {J.}\ \bibnamefont
		{Goold}}, \bibinfo {author} {\bibfnamefont {M.}\ \bibnamefont {Huber}}, \bibinfo {author} {\bibfnamefont {A.}\ \bibnamefont {Riera}}, \bibinfo {author} {\bibfnamefont {L.}\ \bibnamefont {del Rio}}, and \bibinfo {author} {\bibfnamefont {P.}\ \bibnamefont {Skrzypczyk}},\
}\bibfield  {title} { {\bibinfo {title} {The role of quantum information in thermodynamics—a topical review,}}\ }\href {\doibase  https://doi.org/10.1088/1751-8113/49/14/143001
} {\bibfield  {journal} {\bibinfo  {journal}
		{J. Phys. A: Math. Theor.}\ }\textbf {\bibinfo {volume} {49}},\ \bibinfo {pages}
	{143001} (\bibinfo {year} {2016})}\BibitemShut {NoStop}		
\bibitem [{\citenamefont {Anders}\ \emph {et~al.}(2016)  \citenamefont {Anders}}]{Anders}
\BibitemOpen
\bibfield  {author} {\bibinfo {author} {\bibfnamefont {S.}\ \bibnamefont
		{Vinjanampathy}}, and \bibinfo {author} {\bibfnamefont {J.}~\bibnamefont
		{Anders}},\
}\bibfield  {title} { {\bibinfo {title} {Quantum thermodynamics,}}\ }\href {\doibase 
https://doi.org/10.1080/00107514.2016.1201896} {\bibfield  {journal} {\bibinfo  {journal}
		{Contemp. Phys.}\ }\textbf {\bibinfo {volume} {57}},\ \bibinfo {pages}
	{545} (\bibinfo {year} {2016})}\BibitemShut {NoStop}
	\bibitem [{\citenamefont {Millen}\ \emph {et~al.}(2016) \citenamefont {Millen}}]{Millen}
\BibitemOpen
\bibfield  {author} {\bibinfo {author} {\bibfnamefont {J.}\ \bibnamefont
		{Millen}}, and \bibinfo {author} {\bibfnamefont {A.}~\bibnamefont
		{Xuereb}},\
}\bibfield  {title} { {\bibinfo {title} {Perspective on quantum thermodynamics,}}\ }\href {\doibase 
https://doi.org/10.1088/1367-2630/18/1/011002} {\bibfield  {journal} {\bibinfo  {journal}
		{New J. Phys.}\ }\textbf {\bibinfo {volume} {18}},\ \bibinfo {pages}
	{011002} (\bibinfo {year} {2016})}\BibitemShut {NoStop}
	\bibitem [{\citenamefont {Rezek}\ \emph {et~al.}(2017) \citenamefont {Rezek}}]{Rezek}
\BibitemOpen
\bibfield  {author} {\bibinfo {author} {\bibfnamefont {R.}\ \bibnamefont
		{Kosloff}}, and \bibinfo {author} {\bibfnamefont {Y.}~\bibnamefont
		{Rezek}},\
}\bibfield  {title} { {\bibinfo {title} {The quantum harmonic Otto cycle}},\ }\href {\doibase 
https://doi.org/10.3390/e19040136} {\bibfield  {journal} {\bibinfo  {journal}
		{Entropy}\ }\textbf {\bibinfo {volume} {19}},\ \bibinfo {pages}
	{136} (\bibinfo {year} {2017})}\BibitemShut {NoStop}	
\bibitem [{\citenamefont {Binder}\ \emph {}(2018)\citenamefont {Binder}}]{Binder}
\BibitemOpen
\bibfield  {author} {\bibinfo {author} {\bibfnamefont {F.}\ \bibnamefont {Binder}}, \bibinfo {author} {\bibfnamefont {L. A.}~\bibnamefont {Correa}}, \bibinfo {author} {\bibfnamefont {C.}~\bibnamefont {Gogolin}}, \bibinfo {author} {\bibfnamefont {J.}~\bibnamefont {Anders}}, \ and \ \bibinfo {author} {\bibfnamefont {G.}~\bibnamefont {Adesso},}\
}\bibfield  {title} { {\bibinfo {title} {Thermodynamics in the quantum regime,}}\ } \href {\doibase
https://doi.org/10.1007/978-3-319-99046-0} {\bibfield  {journal} {\bibinfo  {journal}{Fundamental Theories of Physics,}\ }\textbf {\bibinfo {volume} {195}},\ \bibinfo {pages}
{1-2} (\bibinfo {year} {2018})}\BibitemShut {NoStop}
\bibitem [{\citenamefont {Deffner}\ \emph {}(2019)\citenamefont {Deffner}}]{Deffner}
\BibitemOpen
\bibfield  {author} {\bibinfo {author} {\bibfnamefont {S.}\ \bibnamefont {Deffner}}, \ and \ \bibinfo {author} {\bibfnamefont {S.}~\bibnamefont {Campbell},}\
}\bibfield  {title} { {\bibinfo {title} {Quantum Thermodynamics: An introduction to the thermodynamics of quantum information, (Morgan \& Claypool Publishers, 2019)}}\ }\BibitemShut {NoStop} 
\bibitem [{\citenamefont {Mitchison}\ \emph {}(2019)\citenamefont {Mitchison}}]{Mitchison}
\BibitemOpen
\bibfield  {author} {\bibinfo {author} {\bibfnamefont {M.T.}\ \bibnamefont
{Mitchison},}\
}\bibfield {title}{{\bibinfo {title} {Quantum thermal absorption machines: refrigerators, engines and clocks,}}\ }\href {\doibase
https://doi.org/10.1080/00107514.2019.1631555
} {\bibfield  {journal} {\bibinfo  {journal}{Contemp. Phys.}\ }\textbf {\bibinfo {volume} {60}},\ \bibinfo {pages}{164} (\bibinfo {year} {2019})}\BibitemShut {NoStop} 
\bibitem [{\citenamefont {Scovil}\ \emph {}(1959)\citenamefont {Scovil}}]{Scovil}
\BibitemOpen
\bibfield  {author} {\bibinfo {author} {\bibfnamefont {H. E. D.}\ \bibnamefont
		{Scovil}}, and \bibinfo {author} {\bibfnamefont {E. O.}~\bibnamefont
		{Schulz-Dubois}},\
}\bibfield  {title} { {\bibinfo {title} {Three-level masers as heat engines,}}\ } \href {\doibase
https://doi.org/10.1103/PhysRevLett.2.262
} {\bibfield  {journal} {\bibinfo  {journal}
		{Phys. Rev. Lett.}\ }\textbf {\bibinfo {volume} {2}},\ \bibinfo {pages}
	{262} (\bibinfo {year} {1959})}\BibitemShut {NoStop} 
\bibitem [{\citenamefont {Rossnagel}\ \emph {}(2012)\citenamefont {Rossnagel}}]{Rossnagel1}
\BibitemOpen
\bibfield  {author} {\bibinfo {author} {\bibfnamefont {O.}\ \bibnamefont
		{Abah}}, \bibinfo {author} {\bibfnamefont {J.}~\bibnamefont
		{Ro\ss nagel}}, \bibinfo {author} {\bibfnamefont { G.}~\bibnamefont
		{Jacob}}, \bibinfo {author} {\bibfnamefont {S.}~\bibnamefont
		{Deffner}}, \bibinfo {author} {\bibfnamefont {F.}~\bibnamefont
		{Schmidt-Kaler}}, \bibinfo {author} {\bibfnamefont {K.}~\bibnamefont
		{Singer}}, and \bibinfo {author} {\bibfnamefont {E.}~\bibnamefont
		{Lutz}}, \
}\bibfield  {title} { {\bibinfo {title} {Single-ion heat engine at maximum power,}}\ }\href {\doibase
https://doi.org/10.1103/PhysRevLett.109.203006
} {\bibfield  {journal} {\bibinfo  {journal}
		{Phys. Rev. Lett.}\ }\textbf {\bibinfo {volume} {109}},\ \bibinfo {pages}
	{203006} (\bibinfo {year} {2012})}\BibitemShut {NoStop} 
	\bibitem [{\citenamefont {Rossnagel2}\ \emph {}(2014)\citenamefont {Rossnagel2}}]{Rossnagel2}
\BibitemOpen
\bibfield  {author} {\bibinfo {author} {\bibfnamefont {J.}\ \bibnamefont
		{Ro\ss nagel}}, \bibinfo {author} {\bibfnamefont {O.}~\bibnamefont
		{Abah}}, \bibinfo {author} {\bibfnamefont { F.}~\bibnamefont
		{Schmidt-Kaler}}, \bibinfo {author} {\bibfnamefont {K.}~\bibnamefont
		{Singer}}, and \bibinfo {author} {\bibfnamefont {E.}~\bibnamefont
		{Lutz}}, \
}\bibfield  {title} { {\bibinfo {title} {Nanoscale heat engine beyond the Carnot limit,}}\ }\href {\doibase
https://doi.org/10.1103/PhysRevLett.112.030602
} {\bibfield  {journal} {\bibinfo  {journal}
		{Phys. Rev. Lett.}\ }\textbf {\bibinfo {volume} {112}},\ \bibinfo {pages}
	{030602} (\bibinfo {year} {2014})}\BibitemShut {NoStop} 
\bibitem [{\citenamefont {Rossnagel}\ \emph {}(2016)\citenamefont {Rossnagel}}]{Rossnagel}
\BibitemOpen
\bibfield  {author} {\bibinfo {author} {\bibfnamefont {J.}\ \bibnamefont
		{Ro\ss nagel}}, \bibinfo {author} {\bibfnamefont {S. T.}~\bibnamefont
		{Dawkins}}, \bibinfo {author} {\bibfnamefont {K. N.}~\bibnamefont
		{Tolazzi}}, \bibinfo {author} {\bibfnamefont {O.}~\bibnamefont
		{Abah}}, \bibinfo {author} {\bibfnamefont {E.}~\bibnamefont
		{Lutz}}, \bibinfo {author} {\bibfnamefont {F.}~\bibnamefont
		{Schmidt-Kaler}}, and \bibinfo {author} {\bibfnamefont {K.}~\bibnamefont
		{Singer}}, \
}\bibfield  {title} { {\bibinfo {title} {A single-atom heat engine,}}\ }\href {\doibase
https://doi.org/10.1126/science.aad6320
} {\bibfield  {journal} {\bibinfo  {journal}
		{Science}\ }\textbf {\bibinfo {volume} {352}},\ \bibinfo {pages}
	{325} (\bibinfo {year} {2016})}\BibitemShut {NoStop} 
\bibitem [{\citenamefont {Maslennikov}\ \emph {}(2017)\citenamefont {Maslennikov}}]{Maslennikov}
\BibitemOpen
\bibfield  {author} {\bibinfo {author} {\bibfnamefont {G.}\ \bibnamefont
		{Maslennikov}}, \bibinfo {author} {\bibfnamefont {S.}~\bibnamefont
		{Ding}}, \bibinfo {author} {\bibfnamefont {R}~\bibnamefont
		{Hablutzel}}, \bibinfo {author} {\bibfnamefont {J.}~\bibnamefont
		{Gan}}, \bibinfo {author} {\bibfnamefont {A.}~\bibnamefont
		{Roulet}}, \bibinfo {author} {\bibfnamefont {S.}~\bibnamefont
		{Nimmrichter}}, \bibinfo {author} {\bibfnamefont {J.}~\bibnamefont
		{Dai}}, \bibinfo {author} {\bibfnamefont {V.}~\bibnamefont
		{Scarani}}, and \bibinfo {author} {\bibfnamefont {D.}~\bibnamefont
		{Matsukevich},}\
}\bibfield  {title} { {\bibinfo {title} {Quantum absorption refrigerator with trapped ions,}}\ }\href {\doibase
https://doi.org/10.1038/s41467-018-08090-0
} {\bibfield  {journal} {\bibinfo  {journal}
		{Nat. Commun.}\ }\textbf {\bibinfo {volume} {10}},\ \bibinfo {pages}
	{202} (\bibinfo {year} {2017})}\BibitemShut {NoStop} 
	\bibitem [{\citenamefont {QuanZhang}\ \emph {}(2006)\citenamefont {QuanZhang}}]{QuanZhang}
\BibitemOpen
\bibfield  {author} {\bibinfo {author} {\bibfnamefont {H. T.}\ \bibnamefont
		{Quan}}, \bibinfo {author} {\bibfnamefont {P.}~\bibnamefont
		{Zhang}}, and \bibinfo {author} {\bibfnamefont {C. P.}~\bibnamefont
		{Sun},} \
}\bibfield  {title} { {\bibinfo {title} {Quantum-classical transition of photon-Carnot engine induced by quantum decoherence,}}\ }\href {\doibase
https://doi.org/10.1103/PhysRevE.73.036122
} {\bibfield  {journal} {\bibinfo  {journal}
		{Phys. Rev. E}\ }\textbf {\bibinfo {volume} {73}},\ \bibinfo {pages}
	{036122} (\bibinfo {year} {2006})}\BibitemShut {NoStop}	
	\bibitem [{\citenamefont {Sothmann}\ \emph {}(2012)\citenamefont {Sothmann}}]{Sothmann}
\BibitemOpen
\bibfield  {author} {\bibinfo {author} {\bibfnamefont {B.}\ \bibnamefont
		{Sothmann}}, and \bibinfo {author} {\bibfnamefont {M.}~\bibnamefont
		{Büttiker}}, \
}\bibfield{title}{{\bibinfo{title} {Magnon-driven quantum-dot heat engine,}}\ } \href {\doibase
https://doi.org/10.1209/0295-5075/99/27001
} {\bibfield  {journal} {\bibinfo  {journal}
		{EPL}\ }\textbf {\bibinfo {volume} {99}},\ \bibinfo {pages}
	{27001} (\bibinfo {year} {2012})}\BibitemShut {NoStop}
	\bibitem [{\citenamefont {Fazio}\ \emph {}(2013)\citenamefont {Fazio}}]{Fazio}
\BibitemOpen
\bibfield  {author} {\bibinfo {author} {\bibfnamefont {D.}\ \bibnamefont
		{Venturelli}}, \bibinfo {author} {\bibfnamefont {R.}~\bibnamefont
		{Fazio}}, and \bibinfo {author} {\bibfnamefont {V.}~\bibnamefont
		{Giovannetti},}\
}\bibfield  {title} { {\bibinfo {title} {Minimal self-contained quantum refrigeration machine based on four quantum dots,}}\ }\href {\doibase
https://doi.org/10.1103/PhysRevLett.110.256801
} {\bibfield  {journal} {\bibinfo  {journal}
		{Phys. Rev. Lett.}\ }\textbf {\bibinfo {volume} {110}},\ \bibinfo {pages}
	{256801} (\bibinfo {year} {2013})}\BibitemShut {NoStop}
\bibitem [{\citenamefont {Bariani}\ \emph {}(2014)\citenamefont {Bariani}}]{Bariani}
\BibitemOpen
\bibfield  {author} {\bibinfo {author} {\bibfnamefont {K.}\ \bibnamefont
		{Zhang}}, \bibinfo {author} {\bibfnamefont {F.}~\bibnamefont
		{Bariani}}, and \bibinfo {author} {\bibfnamefont {P.}~\bibnamefont
		{Meystre}}\
}\bibfield  {title} { {\bibinfo {title} {Quantum optomechanical heat engine,}}\ }\href {\doibase
https://doi.org/10.1103/PhysRevLett.112.150602
} {\bibfield  {journal} {\bibinfo  {journal}
		{Phys. Rev. Lett.}\ }\textbf {\bibinfo {volume} {112}},\ \bibinfo {pages}
	{150602} (\bibinfo {year} {2014})}\BibitemShut {NoStop}
\bibitem [{\citenamefont {AltintasHardal}\ \emph {}(2015)\citenamefont {AltintasHardal}}]{AltintasHardal}
\BibitemOpen
\bibfield  {author} {\bibinfo {author} {\bibfnamefont {F.}\ \bibnamefont
		{Altintas}}, \bibinfo {author} {\bibfnamefont {A. Ü. C.}~\bibnamefont
		{Hardal}}, and \bibinfo {author} {\bibfnamefont {Ö E.}~\bibnamefont
		{Müstecaplıoğlu},}\
}\bibfield  {title} { {\bibinfo {title} {Rabi model as a quantum coherent heat engine: From quantum biology to superconducting circuits,}}\ }\href {\doibase
https://doi.org/10.1103/PhysRevA.91.023816
} {\bibfield  {journal} {\bibinfo  {journal}
		{Phys. Rev. A.}\ }\textbf {\bibinfo {volume} {91}},\ \bibinfo {pages}
	{023816} (\bibinfo {year} {2015})}\BibitemShut {NoStop}
\bibitem [{\citenamefont {Peterson}\ \emph {}(2019)\citenamefont {Peterson}}]{Peterson}
\BibitemOpen
\bibfield  {author} {\bibinfo {author} {\bibfnamefont {J. P. S.}\ \bibnamefont
		{Peterson}}, \bibinfo {author} {\bibfnamefont {T. B.}~\bibnamefont
		{Batalhão}},  \bibinfo {author} {\bibfnamefont {M.}~\bibnamefont
		{Herrera}},  \bibinfo {author} {\bibfnamefont {A. M.}~\bibnamefont
		{Souza}},  \bibinfo {author} {\bibfnamefont {R. S.}~\bibnamefont
		{Sarthour}},  \bibinfo {author} {\bibfnamefont {I. S.}~\bibnamefont
		{Oliveira}},  and  \bibinfo {author} {\bibfnamefont {R. M.}~\bibnamefont
		{Serra},}\
}\bibfield {title}{{\bibinfo {title} {Experimental characterization of a spin quantum heat engine,}}\ }\href {\doibase
https://doi.org/10.1103/PhysRevLett.123.240601
} {\bibfield  {journal} {\bibinfo  {journal}
		{Phys. Rev. Lett.}\ }\textbf {\bibinfo {volume} {123}},\ \bibinfo {pages}
	{240601} (\bibinfo {year} {2019})}\BibitemShut {NoStop}
\bibitem [{\citenamefont {Uzdin}\ \emph {}(2019)\citenamefont {Uzdin}}]{Uzdin}
\BibitemOpen
\bibfield  {author} {\bibinfo {author} {\bibfnamefont {J.}\ \bibnamefont
		{Klatzow}}, \bibinfo {author} {\bibfnamefont {J. N.}~\bibnamefont
		{Becker}},  \bibinfo {author} {\bibfnamefont {P. M.}~\bibnamefont
		{Ledingham}},  \bibinfo {author} {\bibfnamefont {C.}~\bibnamefont
		{Weinzetl}},  \bibinfo {author} {\bibfnamefont {K. T.}~\bibnamefont
		{Kaczmarek}},  \bibinfo {author} {\bibfnamefont {D. J.}~\bibnamefont
		{Saunders}},  \bibinfo {author} {\bibfnamefont {J.}~\bibnamefont
		{Nunn}},  \bibinfo {author} {\bibfnamefont {I. A.}~\bibnamefont
		{Walmsley}},  \bibinfo {author} {\bibfnamefont {R.}~\bibnamefont
		{Uzdin}}, and  \bibinfo {author} {\bibfnamefont {E.}~\bibnamefont
		{Poem},}\
}\bibfield {title}{{\bibinfo {title} {Experimental demonstration of quantum effects in the operation of microscopic heat engines,}}\ }\href {\doibase
https://doi.org/10.1103/PhysRevLett.122.110601
} {\bibfield  {journal} {\bibinfo  {journal}
		{Phys. Rev. Lett.}\ }\textbf {\bibinfo {volume} {112}},\ \bibinfo {pages}
	{110601} (\bibinfo {year} {2019})}\BibitemShut {NoStop}
	\bibitem [{\citenamefont {DVon}\ \emph {}(2019)\citenamefont {DVon}}]{DVon}
\BibitemOpen
\bibfield  {author} {\bibinfo {author} {\bibfnamefont {D.}\ \bibnamefont
		{D. von Lindenfels et al}}, \bibinfo {author} {\bibfnamefont {J. N.}~\bibnamefont
		{Becker}},  \bibinfo {author} {\bibfnamefont {P. M.}~\bibnamefont
		{Ledingham}},  \bibinfo {author} {\bibfnamefont {C.}~\bibnamefont
		{Weinzetl}},  \bibinfo {author} {\bibfnamefont {K. T.}~\bibnamefont
		{Kaczmarek}},  \bibinfo {author} {\bibfnamefont {D. J.}~\bibnamefont
		{Saunders}},  \bibinfo {author} {\bibfnamefont {J.}~\bibnamefont
		{Nunn}},  \bibinfo {author} {\bibfnamefont {I. A.}~\bibnamefont
		{Walmsley}},  \bibinfo {author} {\bibfnamefont {R.}~\bibnamefont
		{Uzdin}}, and  \bibinfo {author} {\bibfnamefont {E.}~\bibnamefont
		{Poem},}\
}\bibfield {title}{{\bibinfo {title} {Experimental demonstration of quantum effects in the operation of microscopic heat engines,}}\ }\href {\doibase
https://doi.org/10.1103/PhysRevLett.122.110601
} {\bibfield  {journal} {\bibinfo  {journal}
		{Phys. Rev. Lett.}\ }\textbf {\bibinfo {volume} {123}},\ \bibinfo {pages}
	{080602} (\bibinfo {year} {2019})}\BibitemShut {NoStop}
\bibitem [{\citenamefont {MayersDeffner}\ \emph {}(2022)\citenamefont {MayersDeffner}}]{MayersDeffner}
\BibitemOpen
\bibfield  {author} {\bibinfo {author} {\bibfnamefont {N. M.}\ \bibnamefont
		{Myers}}, \bibinfo {author} {\bibfnamefont {O.}~\bibnamefont
		{Abah}} and \bibinfo {author} {\bibfnamefont {S.}~\bibnamefont
		{Deffner},}\
}\bibfield{title}{{\bibinfo {title} {Quantum thermodynamic devices: from theoretical proposals to experimental reality,}}\ }\href {\doibase
https://doi.org/10.48550/arXiv.2201.01740
} {\bibfield  {journal} {\bibinfo  {journal}
		{AVS quantum science}\ }\textbf {\bibinfo {volume} {4}},\ \bibinfo {pages}
	{027101} (\bibinfo {year} {2022})}\BibitemShut {NoStop}	 
 \bibitem [{\citenamefont {ScullyMarlan}\ \emph {}(2001)\citenamefont {ScullyMarlan}}]{ScullyMarlan}
\BibitemOpen
\bibfield  {author} {\bibinfo {author} {\bibfnamefont {M. O.}\ \bibnamefont
		{Scully},}\
}\bibfield  {title} { {\bibinfo {title} {Extracting work from a single thermal bath via quantum negentropy,}}\ }\href {\doibase
10.1103/PhysRevLett.87.220601
} {\bibfield  {journal} {\bibinfo  {journal}
		{Phys. Rev. Lett.}\ }\textbf {\bibinfo {volume} {87}},\ \bibinfo {pages}
	{220601} (\bibinfo {year} {2001})}\BibitemShut {NoStop}
\bibitem [{\citenamefont {ScullyZubairy}\ \emph {}(2003)\citenamefont {ScullyZubairy}}]{ScullyZubairy}
\BibitemOpen
\bibfield  {author} {\bibinfo {author} {\bibfnamefont {M. O.}\ \bibnamefont
		{Scully}}, \bibinfo {author} {\bibfnamefont {M. S.}~\bibnamefont
		{Zubairy}}, \bibinfo {author} {\bibfnamefont {G. S.}~\bibnamefont
		{Agarwal}}, and \bibinfo {author} {\bibfnamefont {H. }~\bibnamefont
		{Walther},} \
}\bibfield  {title} { {\bibinfo {title} {Extracting work from a single heat bath via vanishing quantum coherence
,}}\ } \href {\doibase
10.1126/science.1078955
} {\bibfield  {journal} {\bibinfo  {journal}
		{Science}\ }\textbf {\bibinfo {volume} {299}},\ \bibinfo {pages}
	{826} (\bibinfo {year} {2003})}\BibitemShut {NoStop}
\bibitem [{\citenamefont {Dillenschneider}\ \emph {}(2009)\citenamefont {Dillenschneider}}]{Dillenschneider}
\BibitemOpen
\bibfield  {author} {\bibinfo {author} {\bibfnamefont {R.}\ \bibnamefont
		{Dillenschneider}}, and \bibinfo {author} {\bibfnamefont {E.}~\bibnamefont
		{Lutz},}\
}\bibfield {title}{{\bibinfo {title} {Energetics of quantum correlations,}}\ }\href {\doibase
https://doi.org/10.1209/0295-5075/88/50003
} {\bibfield  {journal} {\bibinfo  {journal}
		{EPL}\ }\textbf {\bibinfo {volume} {88}},\ \bibinfo {pages}
	{50003} (\bibinfo {year} {2009})}\BibitemShut {NoStop}
\bibitem [{\citenamefont {HuangWang}\ \emph {}(2012)\citenamefont {HuangWang}}]{HuangWang}
\BibitemOpen
\bibfield  {author} {\bibinfo {author} {\bibfnamefont {X. L.}\ \bibnamefont
		{Huang}}, \bibinfo {author} {\bibfnamefont {T.}~\bibnamefont
		{Wang}}, and \bibinfo {author} {\bibfnamefont {X. X.}~\bibnamefont
		{Yi}},\
}\bibfield{title} {{\bibinfo {title} {Effects of reservoir squeezing on quantum systems and work extraction,}}\ }\href {\doibase
https://doi.org/10.1103/PhysRevE.86.051105
} {\bibfield  {journal} {\bibinfo  {journal}
		{Phys. Rev. E}\ }\textbf {\bibinfo {volume} {86}},\ \bibinfo {pages}
	{051105} (\bibinfo {year} {2012})}\BibitemShut {NoStop}
\bibitem [{\citenamefont {AbahLutz}\ \emph {}(2014)\citenamefont {AbahLutz}}]{AbahLutz}
\BibitemOpen
\bibfield  {author} {\bibinfo {author} {\bibfnamefont {O.}\ \bibnamefont
		{Abah}}, and \bibinfo {author} {\bibfnamefont {E.}~\bibnamefont
		{Lutz},}\
}\bibfield  {title} { {\bibinfo {title} {Efficiency of heat engines coupled to nonequilibrium reservoirs,}}\ }\href {\doibase
https://doi.org/10.1209/0295-5075/106/20001
} {\bibfield  {journal} {\bibinfo  {journal}
		{EPL}\ }\textbf {\bibinfo {volume} {106}},\ \bibinfo {pages}
	{20001} (\bibinfo {year} {2014})}\BibitemShut {NoStop}
\bibitem [{\citenamefont {RossnagelAbah}\ \emph {}(2014)\citenamefont {RossnagelAbah}}]{RossnagelAbah}
\BibitemOpen
\bibfield  {author} {\bibinfo {author} {\bibfnamefont {J.}\ \bibnamefont
		{Rossnagel}}, \bibinfo {author} {\bibfnamefont {O.}~\bibnamefont
		{Abah}}, \bibinfo {author} {\bibfnamefont {F.}~\bibnamefont
		{Schmidt-Kaler}}, \bibinfo {author} {\bibfnamefont {K.}~\bibnamefont
		{Singer}}, and \bibinfo {author} {\bibfnamefont {E.}~\bibnamefont
		{Lutz},}\
}\bibfield  {title} { {\bibinfo {title} {Efficiency of heat engines coupled to nonequilibrium reservoirs,}}\ }\href {\doibase
https://doi.org/10.1103/PhysRevLett.112.030602
} {\bibfield  {journal} {\bibinfo  {journal}
		{Phys. Rev. Lett.}\ }\textbf {\bibinfo {volume} {112}},\ \bibinfo {pages}
	{030602} (\bibinfo {year} {2014})}\BibitemShut {NoStop}	
\bibitem [{\citenamefont {HardalM}\ \emph {}(2015)\citenamefont {HardalM}}]{HardalM}
\BibitemOpen
\bibfield  {author} {\bibinfo {author} {\bibfnamefont {A. Ü. C.}\ \bibnamefont
		{Hardal}}, and \bibinfo {author} {\bibfnamefont {Ö E.}~\bibnamefont
		{Müstecaplıoğlu},}\
}\bibfield{title} { {\bibinfo {title} {Superradiant quantum heat engine,}}\ }\href {\doibase
https://doi.org/10.1038/srep12953
} {\bibfield  {journal} {\bibinfo  {journal}
		{Sci. Rep.}\ }\textbf {\bibinfo {volume} {5}},\ \bibinfo {pages}
	{12953} (\bibinfo {year} {2015})}\BibitemShut {NoStop}	
\bibitem [{\citenamefont {Niedenzu}\ \emph {}(2016)\citenamefont {Niedenzu}}]{Niedenzu}
\BibitemOpen
\bibfield  {author} {\bibinfo {author} {\bibfnamefont {W.}\ \bibnamefont
		{Niedenzu}}, \bibinfo {author} {\bibfnamefont {D. }~\bibnamefont
		{Gelbwaser-Klimovsky}}, \bibinfo {author} {\bibfnamefont {A. G. }~\bibnamefont
		{Kofman}}, and \bibinfo {author} {\bibfnamefont {G. }~\bibnamefont
		{Kurizki},}\
}\bibfield  {title} { {\bibinfo {title} {On the operation of machines powered by quantum non-thermal baths,}}\ }\href {\doibase
https://doi.org/10.1088/1367-2630/18/8/083012
} {\bibfield  {journal} {\bibinfo  {journal}
		{New J. Phys.}\ }\textbf {\bibinfo {volume} {18}},\ \bibinfo {pages}
	{083012} (\bibinfo {year} {2016})}\BibitemShut {NoStop}	
\bibitem [{\citenamefont {Manzano}\ \emph {}(2016)\citenamefont {Manzano}}]{Manzano}
\BibitemOpen
\bibfield  {author} {\bibinfo {author} {\bibfnamefont {G.}\ \bibnamefont
		{Manzano}}, \bibinfo {author} {\bibfnamefont {F.}~\bibnamefont
		{Galve}}, \bibinfo {author} {\bibfnamefont {R.}~\bibnamefont
		{Zambrini}} and \bibinfo {author} {\bibfnamefont {J. M. R.}~\bibnamefont
		{Parrondo},}\
}\bibfield {title}{{\bibinfo {title} {Entropy production and thermodynamic power of the squeezed thermal reservoir,}}\ }\href {\doibase
https://doi.org/10.1103/PhysRevE.93.052120
} {\bibfield  {journal} {\bibinfo  {journal}
		{Phys. Rev. E}\ }\textbf {\bibinfo {volume} {93}},\ \bibinfo {pages}
	{052120} (\bibinfo {year} {2016})}\BibitemShut {NoStop}	
\bibitem [{\citenamefont {Klaers}\ \emph {}(2017)\citenamefont {Klaers}}]{Klaers}
\BibitemOpen
\bibfield  {author} {\bibinfo {author} {\bibfnamefont {J.}\ \bibnamefont
		{Klaers}}, \bibinfo {author} {\bibfnamefont {S.}~\bibnamefont
		{Faelt}}, \bibinfo {author} {\bibfnamefont {A.}~\bibnamefont
		{Imamoglu}}, and \bibinfo {author} {\bibfnamefont {E.}~\bibnamefont
		{Togan}}\
}\bibfield  {title} { {\bibinfo {title} {Squeezed thermal reservoirs as a resource for a nanomechanical engine beyond the carnot limit,}}\ }\href {\doibase
https://doi.org/10.1103/PhysRevX.7.031044
} {\bibfield  {journal} {\bibinfo  {journal}
		{Phys. Rev. X}\ }\textbf {\bibinfo {volume} {7}},\ \bibinfo {pages}
	{031044} (\bibinfo {year} {2017})}\BibitemShut {NoStop}	
\bibitem [{\citenamefont {Agarwalla}\ \emph {}(2017)\citenamefont {Agarwalla}}]{Agarwalla}
\BibitemOpen
\bibfield  {author} {\bibinfo {author} {\bibfnamefont {B. K.}\ \bibnamefont
		{Agarwalla}}, \bibinfo {author} {\bibfnamefont {J.-H.}~\bibnamefont
		{Jiang}}, and \bibinfo {author} {\bibfnamefont {D.}~\bibnamefont
		{Segal},}\
}\bibfield{title}{{\bibinfo {title} {Quantum efficiency bound for continuous heat engines coupled to noncanonical reservoirs,}}\ }\href {\doibase
https://doi.org/10.1103/PhysRevB.96.104304
} {\bibfield  {journal} {\bibinfo  {journal}
		{Phys. Rev. B}\ }\textbf {\bibinfo {volume} {96}},\ \bibinfo {pages}
	{104304} (\bibinfo {year} {2017})}\BibitemShut {NoStop}
\bibitem [{\citenamefont {Kieu}\ \emph {et~al.}(2004) \citenamefont {Kieu}}]{Kieu}
\BibitemOpen
\bibfield  {author} {\bibinfo {author} {\bibfnamefont {T. D.}\ \bibnamefont
		{Kieu},}\
}\bibfield {title}{{\bibinfo {title} {The second law, Maxwell's demon, and work derivable from quantum heat engines,}}\ }\href {\doibase
https://doi.org/10.1103/PhysRevLett.93.140403
} {\bibfield  {journal} {\bibinfo  {journal}
		{Phys. Rev. Lett.}\ }\textbf {\bibinfo {volume} {93}},\ \bibinfo {pages}
	{140403} (\bibinfo {year} {2004})}\BibitemShut {NoStop}
\bibitem [{\citenamefont {Kieu2}\ \emph {et~al.}(2006) \citenamefont {Kieu2}}]{Kieu2}
\BibitemOpen
\bibfield  {author} {\bibinfo {author} {\bibfnamefont {T. D.}\ \bibnamefont
		{Kieu},}\
}\bibfield  {title} { {\bibinfo {title} {Quantum heat engines, the second law and Maxwell's daemon,}}\ }\href {\doibase
10.1140/epjd/e2006-00075-5
} {\bibfield  {journal} {\bibinfo  {journal}
		{Eur. Phys. J. D}\ }\textbf {\bibinfo {volume} {39}},\ \bibinfo {pages}
	{115} (\bibinfo {year} {2006})}\BibitemShut {NoStop}	
\bibitem [{\citenamefont {Zhang}\ \emph {et~al.}(2007) \citenamefont {Zhang}}]{Zhang}
\BibitemOpen
\bibfield  {author} {\bibinfo {author} {\bibfnamefont {T.}\ \bibnamefont
		{Zhang}}, \bibinfo {author} {\bibfnamefont {W.-T.}~\bibnamefont
		{Liu}} \bibinfo {author} {\bibfnamefont {P.-X.}~\bibnamefont
		{Chen}} and \bibinfo {author} {\bibfnamefont {C.-Z.}~\bibnamefont
		{Li},}\
}\bibfield{title}{{\bibinfo {title} {Four-level entangled quantum heat engines,}}\ }\href {\doibase
https://doi.org/10.1103/PhysRevA.75.062102
} {\bibfield  {journal} {\bibinfo  {journal}
		{Phys. Rev. A}\ }\textbf {\bibinfo {volume} {75}},\ \bibinfo {pages}
	{062102} (\bibinfo {year} {2007})}\BibitemShut {NoStop}
	\bibitem [{\citenamefont {Johal}\ \emph {}(2011)\citenamefont {Johal}}]{Johal}
\BibitemOpen
\bibfield  {author} {\bibinfo {author} {\bibfnamefont {G.}\ \bibnamefont
		{Thomas}}, and \bibinfo {author} {\bibfnamefont {R. S.}~\bibnamefont
		{Johal}},\
}\bibfield{title}{{\bibinfo {title} {A coupled quantum Otto cycle,}}\ }\href {\doibase
https://doi.org/10.1103/PhysRevE.83.031135
} {\bibfield  {journal} {\bibinfo  {journal}
		{Phys. Rev. E}\ }\textbf {\bibinfo {volume} {83}},\ \bibinfo {pages}
	{031135} (\bibinfo {year} {2011})}\BibitemShut {NoStop}
	\bibitem [{\citenamefont {Altintas}\ \emph {et~al.}(2015) \citenamefont {Altintas}}]{Altintas}
\BibitemOpen
\bibfield  {author} {\bibinfo {author} {\bibfnamefont {F.}\ \bibnamefont
		{Altintas}}, and \bibinfo {author} {\bibfnamefont {Ö. E.}~\bibnamefont
		{ Müstecaplıoğlu},}\
}\bibfield{title}{{\bibinfo{title}{General formalism of local thermodynamics with an example: Quantum Otto engine with a spin-$1/2$ coupled to an arbitrary spin}}\ }\href {\doibase
	https://doi.org/10.1103/PhysRevE.92.022142} {\bibfield  {journal} {\bibinfo  {journal}
		{Phys. Rev. E}\ }\textbf {\bibinfo {volume} {92}},\ \bibinfo {pages}
	{022142} (\bibinfo {year} {2015})}\BibitemShut {NoStop}
	\bibitem [{\citenamefont {Ukraine}\ \emph {et~al.}(2015) \citenamefont {Ukraine}}]{Ukraine}
\BibitemOpen
\bibfield  {author} {\bibinfo {author} {\bibfnamefont {E. A.}\ \bibnamefont
		{Ivanchenko1},}\
}\bibfield{title}{{\bibinfo{title}{Quantum Otto cycle efficiency on coupled qudits,}}\ }\href {\doibase
https://doi.org/10.1103/PhysRevE.92.032124
} {\bibfield  {journal} {\bibinfo  {journal}
		{Phys. Rev. E}\ }\textbf {\bibinfo {volume} {92}},\ \bibinfo {pages}
	{032124} (\bibinfo {year} {2015})}\BibitemShut {NoStop}
	\bibitem [{\citenamefont {Mehta}\ \emph {et~al.}(2017)}]{Mehta}
\BibitemOpen
\bibfield  {author} {\bibinfo {author} {\bibfnamefont {V.}\ \bibnamefont
		{Mehta}}, and \bibinfo {author} {\bibfnamefont {R. S.}\ \bibnamefont
		{Johal}},\
}\bibfield{title}{{\bibinfo{title}{Quantum Otto engine with exchange coupling in the presence of level degeneracy,}}\ }\href {\doibase
https://doi.org/10.1103/PhysRevE.96.032110} {\bibfield  {journal} {\bibinfo  {journal}
		{Phys. Rev. E}\ }\textbf {\bibinfo {volume} {96}},\ \bibinfo {pages}
	{032110} (\bibinfo {year} {2017})}\BibitemShut {NoStop}
	\bibitem [{\citenamefont {Talkner}\ \emph {et~al.}(2017) \citenamefont {Talkner}}]{Talkner}
\BibitemOpen
\bibfield  {author} {\bibinfo {author} {\bibfnamefont {J.}\ \bibnamefont
		{Yi}}, \bibinfo {author} {\bibfnamefont {P.}\ \bibnamefont
		{Talkner}}, and \bibinfo {author} {\bibfnamefont {K. W.}\ \bibnamefont {Kim},}\
}\bibfield{title}{{\bibinfo{title}{Single-temperature quantum engine without feedback control,}}\ }\href {\doibase
https://doi.org/10.1103/PhysRevE.96.022108} {\bibfield  {journal} {\bibinfo  {journal}{Phys. Rev. E}\ }\textbf {\bibinfo {volume} {96}},\ \bibinfo {pages}
	{022108} (\bibinfo {year} {2017})}\BibitemShut {NoStop}
	\bibitem [{\citenamefont {DasGosh}\ \emph {et~al.}(2019) \citenamefont {DasGosh}}]{DasGosh}
\BibitemOpen
\bibfield  {author} {\bibinfo {author} {\bibfnamefont {A.}\ \bibnamefont
		{Das}}, and \bibinfo {author} {\bibfnamefont {S.}\ \bibnamefont {Ghosh},}\
}\bibfield{title}{{\bibinfo {title}{Measurement based quantum heat engine with coupled working medium,}}\ }\href {\doibase
https://doi.org/10.3390/e21111131} {\bibfield  {journal} {\bibinfo  {journal}
		{Entropy}\ }\textbf {\bibinfo {volume} {21}},\ \bibinfo {pages}
	{1131} (\bibinfo {year} {2019})}\BibitemShut {NoStop}
	\bibitem [{\citenamefont {Anka}\ \emph {et~al.}(2021) \citenamefont {Anka}}]{Anka}
\BibitemOpen
\bibfield  {author} {\bibinfo {author} {\bibfnamefont {M. F.}\ \bibnamefont
		{Anka}}, \bibinfo {author} {\bibfnamefont {T. R.}\ \bibnamefont {de Oliveira},} and \bibinfo {author} {\bibfnamefont {D.}\ \bibnamefont {Jonathan},}\
}\bibfield{title}{{\bibinfo{title} {Measurement-based quantum heat engine in a multilevel system,}}\ }\href {\doibase
https://doi.org/10.1103/PhysRevE.104.054128} {\bibfield  {journal} {\bibinfo  {journal}
		{Phys. Rev. E}\ }\textbf {\bibinfo {volume} {104}},\ \bibinfo {pages}
	{5} (\bibinfo {year} {2021})}\BibitemShut {NoStop}
	\bibitem [{\citenamefont {GuoZhang}\ \emph {et~al.}(2008) \citenamefont {GuoZhang}}]{GuoZhang}
\BibitemOpen
\bibfield  {author} {\bibinfo {author} {\bibfnamefont {G.-F.}\ \bibnamefont
		{Zhang},}\
}\bibfield{title}{{\bibinfo {title} {Entangled quantum heat engines based on two two-spin systems with Dzyaloshinski-Moriya anisotropic antisymmetric interaction,}}\ }\href {\doibase
https://doi.org/10.1140/epjd/e2008-00133-0} {\bibfield  {journal} {\bibinfo  {journal}
		{Eur. Phys. J. D}\ }\textbf {\bibinfo {volume} {49}},\ \bibinfo {pages}
	{123} (\bibinfo {year} {2008})}\BibitemShut {NoStop}
	\bibitem [{\citenamefont {Zhao}\ \emph {et~al.}(2017) \citenamefont {Zhao}}]{Zhao}
\BibitemOpen
\bibfield  {author} {\bibinfo {author} {\bibfnamefont {L.-M. }\ \bibnamefont
		{Zhao}}, and \bibinfo {author} {\bibfnamefont {G.-F.}~\bibnamefont
		{Zhang},}\
}\bibfield{title}{{\bibinfo{title}{Entangled quantum Otto heat engines based on two-spin systems with the Dzyaloshinski–Moriya interaction,}}\ }\href {\doibase
https://doi.org/10.1007/s11128-017-1665-0} {\bibfield  {journal} {\bibinfo  {journal}
		{Quant. Inf. Process.}\ }\textbf {\bibinfo {volume} {16}},\ \bibinfo {pages}
	{216} (\bibinfo {year} {2017})}\BibitemShut {NoStop}
\bibitem [{\citenamefont {Sodeif}\ \emph {et~al.}(2021) \citenamefont {Sodeif}}]{Sodeif}
\BibitemOpen
\bibfield  {author} {\bibinfo {author} {\bibfnamefont {S.}\ \bibnamefont
		{Ahadpour}}, and \bibinfo {author} {\bibfnamefont {F.}~\bibnamefont
		{Mirmasoudi},}\
}\bibfield{title}{{\bibinfo {title}{Coupled two-qubit engine and refrigerator in Heisenberg model,}}\ }\href {\doibase
	10.1007/s11128-021-03019-x} {\bibfield  {journal} {\bibinfo  {journal}
		{Quant. Inf. Process.}\ }\textbf {\bibinfo {volume} {20}},\ \bibinfo {pages}
	{63} (\bibinfo {year} {2021})}\BibitemShut {NoStop}	
	\bibitem [{\citenamefont {Ramandeep}\ \emph {et~al.}(2021) \citenamefont {Ramandeep}}]{Ramandeep}
\BibitemOpen
\bibfield  {author} \bibinfo {author} {\bibfnamefont {R. S.}\ \bibnamefont {Johal}} and {\bibinfo {author} {\bibfnamefont {V.}\ \bibnamefont{Mehta}},\
}\bibfield{title}{{\bibinfo{title}{Quantum heat engines with complex working media, complete Otto cycles and heuristics,}}\ }\href {\doibase
https://doi.org/10.3390/e23091149} {\bibfield  {journal} {\bibinfo  {journal}
		{Entropy}\ }\textbf {\bibinfo {volume} {23}},\ \bibinfo {pages}
	{1149} (\bibinfo {year} {2021})}\BibitemShut {NoStop}
	\bibitem [{\citenamefont {Marshall}\ \emph {et~al.}(2011) \citenamefont {Marshall}}]{Marshall}
\BibitemOpen
\bibfield  {author} {\bibinfo {author} {\bibfnamefont {A. W.}\ \bibnamefont
		{Marshall}}, \bibinfo {author} {\bibfnamefont {I.}\ \bibnamefont {Olkin}}, and \bibinfo {author} {\bibfnamefont {B. C.}\ \bibnamefont {Arnold}},\
}\bibfield  {title} { {\bibinfo {title} {Inequalities: Theory of Majorization and Its Applications, Springer Series in Statistics (Springer, 2011)}}\ } \BibitemShut {NoStop}
\bibitem [{\citenamefont {Thomas}\ \emph {et~al.}(2013) \citenamefont {Thomas}}]{Thomas}
\BibitemOpen
\bibfield  {author} {\bibinfo {author} {\bibfnamefont {G.}\ \bibnamefont
		{Thomas},} and \bibinfo {author} {\bibfnamefont {R. S.}\ \bibnamefont
		{Johal},}\
}\bibfield{title}{{\bibinfo{title}{Friction due to inhomogeneous driving of coupled spins in a quantum heat engine,}}\ }\href {\doibase
https://doi.org/10.1140/epjb/e2014-50231-1
} {\bibfield  {journal} {\bibinfo  {journal}
		{Eur. Phys. J. B}\ }\textbf {\bibinfo {volume} {87}},\ \bibinfo {pages}
	{166} (\bibinfo {year} {2013})}\BibitemShut {NoStop}
\bibitem [{\citenamefont {Cakmak}\ \emph {et~al.}(2019)\citenamefont {Cakmak}}]{Cakmak}
\BibitemOpen
\bibfield  {author} {\bibinfo {author} {\bibfnamefont {B.}\ \bibnamefont
		{Çakmak},} and \bibinfo {author} {\bibfnamefont {Ö. E.}\ \bibnamefont
		{Müstecaplıoğlu},}\
}\bibfield{title}{{\bibinfo{title}{Spin quantum heat engines with shortcuts to adiabaticity,}}\ }\href {\doibase
https://doi.org/10.1103/PhysRevE.99.032108
} {\bibfield  {journal} {\bibinfo  {journal}
		{Phys. Rev. E }\ }\textbf {\bibinfo {volume} {99}},\ \bibinfo {pages}
	{032108} (\bibinfo {year} {2019})}\BibitemShut {NoStop}
\bibitem [{\citenamefont {CampisiM}\ \emph {et~al.}(2020) \citenamefont {CampisiM}}]{CampisiM}
\BibitemOpen
\bibfield{author} {\bibinfo {author} {\bibfnamefont {A.}\ \bibnamefont
		{Solfanelli},} \bibinfo {author} {\bibfnamefont {M.}\ \bibnamefont
		{Falsetti},} and \bibinfo {author} {\bibfnamefont {M.}\ \bibnamefont
		{Campisi},}\
}\bibfield{title}{{\bibinfo{title}{Nonadiabatic single-qubit quantum Otto engine,}}\ }\href {\doibase
https://doi.org/10.1103/PhysRevB.101.054513
}{\bibfield{journal}{\bibinfo  {journal}
		{Phys. Rev. B}\ }\textbf {\bibinfo {volume} {101}},\ \bibinfo {pages}
	{054513} (\bibinfo {year} {2020})}\BibitemShut {NoStop}
\bibitem [{\citenamefont {Cakmak2}\ \emph {et~al.}(2021)\citenamefont {Cakmak2}}]{Cakmak2}
\BibitemOpen
\bibfield{author}{\bibinfo{author}{\bibfnamefont {B.}\ \bibnamefont
		{Çakmak},}\
}\bibfield{title}{{\bibinfo {title}{Finite-time two-spin quantum Otto engines: Shortcuts to adiabaticity vs. irreversibility,}}\ }\href {\doibase
https://doi.org/10.3906/fiz-2101-10
} {\bibfield{journal}{\bibinfo{journal}{Turk. J. Phys,}\ }\textbf {\bibinfo {volume} {45}},\ \bibinfo {pages}{59} (\bibinfo {year}{2021})}\BibitemShut {NoStop}
\bibitem [{\citenamefont {Johall}\ \emph {et~al.}(2022) \citenamefont {Johall}}]{Johall}
\BibitemOpen
\bibfield{author} {\bibinfo {author} {\bibfnamefont {C.}\ \bibnamefont
		{Cherubim}}, \bibinfo {author} {\bibfnamefont {Th. R. de.}\ \bibnamefont {Oliveira}}, and \bibinfo {author} {\bibfnamefont {D.}\ \bibnamefont {Jonathan}},\
}\bibfield{title}{{\bibinfo {title}{Nonadiabatic coupled-qubit Otto cycle with bidirectional operation and efficiency gains,}}\ }\href {\doibase
https://doi.org/10.48550/arXiv.2201.01664} {\bibfield  {journal} {\bibinfo{journal}{Phy. Rev. E}\ }\textbf {\bibinfo {volume} {105}},\ \bibinfo {pages}{044120} (\bibinfo {year}{2022})}\BibitemShut {NoStop}
	\bibitem [{\citenamefont {Chang}\ \emph {et~al.}()\citenamefont {Chang} }]{Chang}
\BibitemOpen
\bibfield  {author} {\bibinfo {author} {\bibfnamefont {Y.}\ \bibnamefont
		{Yeo}}, and \bibinfo {author} {\bibfnamefont {C. C.}~\bibnamefont
		{Kwong}},\ }\bibfield{title}{{\bibinfo {title}{Quantum heat engines and information,}}\ }\href {\doibase
https://doi.org/10.48550/arXiv.0708.2480
} {\bibfield  {journal} {\bibinfo  {journal}
		{arXiv:0708.2480v1}\ } }\BibitemShut {NoStop}
	\bibitem [{\citenamefont {Altintas1}\ \emph {et~al.}(2014) \citenamefont {Altintas1}}]{Altintas1}
\BibitemOpen
\bibfield  {author} {\bibinfo {author} {\bibfnamefont {F.}\ \bibnamefont
		{Altintas}}, \bibinfo {author} {\bibfnamefont {A. Ü. }~\bibnamefont
		{Hardal}}, and \bibinfo {author} {\bibfnamefont {Ö. E.}~\bibnamefont
		{Müstecaplıoğlu}},\
}\bibfield{title}{{\bibinfo {title}{Quantum correlated heat engine with spin squeezing,}}\ }\href {\doibase
	https://doi.org/10.1103/PhysRevE.90.032102} {\bibfield  {journal} {\bibinfo  {journal}
		{Phys. Rev. E}\ }\textbf {\bibinfo {volume} {90}},\ \bibinfo {pages}
	{032102} (\bibinfo {year} {2014})}\BibitemShut {NoStop}
	\bibitem [{\citenamefont {Hewgill}\ \emph {et~al.}(2018)\citenamefont {Hewgill} }]{Hewgill}
\BibitemOpen
\bibfield  {author} {\bibinfo {author} {\bibfnamefont {A. }\ \bibnamefont {Hewgill}}, \bibinfo {author} {\bibfnamefont {A.}~\bibnamefont
		{Ferraro}} and \bibinfo {author} {\bibfnamefont {G.}~\bibnamefont
		{De Chiara}},\
}\bibfield{title}{{\bibinfo{title}{Quantum correlations and thermodynamic performances of two-qubit engines with local and common baths,}}\ }\href {\doibase
https://doi.org/10.1103/PhysRevA.98.042102} {\bibfield{journal} {\bibinfo  {journal}{Phys. Rev. A,}\ }\textbf {\bibinfo {volume} {98}},\ \bibinfo {pages}
	{042102} (\bibinfo {year} {2018})}\BibitemShut {NoStop}
	\bibitem [{\citenamefont {Oliveira2020}\ \emph {et~al.}(2021)\citenamefont {Oliveira2020}}]{Oliveira2020}
\BibitemOpen
\bibfield  {author} {\bibinfo {author} {\bibfnamefont {T. R.}\ \bibnamefont {de Oliveira}}, and \bibinfo {author} {\bibfnamefont {D.}\ \bibnamefont {Jonathan},} \ }\bibfield{title}{{\bibinfo {title} {Efficiency gain and bidirectional operation of quantum engines with decoupled internal levels,}}\ }\href {\doibase
https://doi.org/10.1103/PhysRevE.104.044133} {\bibfield  {journal} {\bibinfo  {journal}
		{Phys. Rev. E}\ }\textbf {\bibinfo {volume} {104}},\ \bibinfo {pages}
	{044133} (\bibinfo {year} {2020})}\BibitemShut {NoStop}	
	\bibitem [{\citenamefont {Kaplan}\ \emph {et~al.}(1983)\citenamefont {Kaplan} }]{Kaplan}
\BibitemOpen
\bibfield  {author} {\bibinfo {author} {\bibfnamefont {T. A.}\ \bibnamefont {Kaplan}},\ }\bibfield{title}{{\bibinfo {title} {Single-band Hubbard model with spin-orbit coupling,}}\ }\href {\doibase
	10.1007/BF01301591} {\bibfield  {journal} {\bibinfo  {journal}
		{Zeitschrift Für Physik B Condens. Matter}\ }\textbf {\bibinfo {volume} {49}},\ \bibinfo {pages}
	{313} (\bibinfo {year} {1983})}\BibitemShut {NoStop}
	\bibitem [{\citenamefont {Shekhtman}\ \emph {et~al.}(1992)\citenamefont {Shekhtman} }]{Shekhtman}
\BibitemOpen
\bibfield  {author} {\bibinfo {author} {\bibfnamefont {L.}\ \bibnamefont {Shekhtman}}, \bibinfo {author} {\bibfnamefont {O.}\ \bibnamefont {Entin-Wohlman}} and \bibinfo {author} {\bibfnamefont {A.}\ \bibnamefont {Aharony}},} \bibfield{title}{{\bibinfo {title} {Moriya's anisotropic superexchange interaction, frustration, and Dzyaloshinsky's weak ferromagnetism,}}\ }\href {\doibase
	https://doi.org/10.1103/PhysRevLett.69.836} {\bibfield  {journal} {\bibinfo  {journal}
		{Phys. Rev. Lett.}\ }\textbf {\bibinfo {volume} {69}},\ \bibinfo {pages}
	{836} (\bibinfo {year} {1992})}\BibitemShut {NoStop}
	\bibitem [{\citenamefont {Aharony}\ \emph {et~al.}(1993)\citenamefont {Aharony} }]{Aharony}
\BibitemOpen
\bibfield  {author} {\bibinfo {author} {\bibfnamefont {L. }\ \bibnamefont {Shekhtman}}, \bibinfo {author} {\bibfnamefont {A.}~\bibnamefont
		{Aharony}}, and \bibinfo {author} {\bibfnamefont {O.}~\bibnamefont
		{Entin-Wohlman},} \ }\bibfield  {title} { {\bibinfo {title} {Bond-dependent symmetric and antisymmetric superexchange interactions in La2CuO4,}}\ }\href {\doibase
	10.1103/PhysRevB.47.174} {\bibfield  {journal} {\bibinfo  {journal}
		{Phys. Rev. B}\ }\textbf {\bibinfo {volume} {47}},\ \bibinfo {pages}
	{174} (\bibinfo {year} {1993})}\BibitemShut {NoStop}	
	\bibitem [{\citenamefont {Yildirim}\ \emph {et~al.}(1995)\citenamefont {Yildirim} }]{Yildirim}
\BibitemOpen
\bibfield  {author} {\bibinfo {author} {\bibfnamefont {T.}\ \bibnamefont {Yildirim}}, \bibinfo {author} {\bibfnamefont {A. B.}\ \bibnamefont {Harris}}, \bibinfo {author} {\bibfnamefont {A.}\ \bibnamefont {Aharony}}, and \bibinfo {author} {\bibfnamefont {O.}\ \bibnamefont {Entin-Wohlman}}, }\bibfield {title}{{\bibinfo {title} {Anisotropic spin Hamiltonians due to spin-orbit and Coulomb exchange interactions,}}\ }\href {\doibase
https://doi.org/10.1103/PhysRevB.52.10239} {\bibfield  {journal} {\bibinfo  {journal}
		{Phys. Rev. B}\ }\textbf {\bibinfo {volume} {52}},\ \bibinfo {pages}
	{10239} (\bibinfo {year} {1995})}\BibitemShut {NoStop}
	\bibitem [{\citenamefont {Yurischev}\ \emph {et~al.}(2020)\citenamefont {Yurischev} }]{Yurischev}
\BibitemOpen
\bibfield  {author} {\bibinfo {author} {\bibfnamefont {M. A.}\ \bibnamefont {Yurischev},}\ }\bibfield {title}{{\bibinfo {title} {On the quantum correlations in two-qubit XYZ spin chains with Dzyaloshinsky–Moriya and Kaplan–Shekhtman–Entin-Wohlman–Aharony interactions,}}\ }\href {\doibase
https://doi.org/10.1007/s11128-020-02835-x
} {\bibfield  {journal} {\bibinfo  {journal}
		{Quant. Inf. Process.}\ }\textbf {\bibinfo {volume} {19}},\ \bibinfo {pages}
	{336} (\bibinfo {year} {2020})}\BibitemShut {NoStop}
	\bibitem [{\citenamefont {Deniz}\ \emph {et~al.}(2020)\citenamefont {Deniz} }]{Deniz}
\BibitemOpen
\bibfield  {author} {\bibinfo {author} {\bibfnamefont {D.}\ \bibnamefont {Türkpençe},} and \bibinfo {author} {\bibfnamefont {F.}\ \bibnamefont {Altintas},}}\bibfield  {title} { {\bibinfo {title} {Coupled quantum Otto heat engine and refrigerator with inner friction,}}\ }\href {\doibase
https://doi.org/10.1007/s11128-019-2366-7
} {\bibfield  {journal} {\bibinfo  {journal} {Quant. Inf. Process.}\ }\textbf {\bibinfo {volume} {19}},\ \bibinfo {pages}
	{255} (\bibinfo {year} {2019})}\BibitemShut {NoStop}
	\bibitem [{\citenamefont {Moriya}\ \emph {et~al.}(1960)\citenamefont {Moriya} }]{Moriya}
\BibitemOpen
\bibfield  {author} {\bibinfo {author} {\bibfnamefont {T.}\ \bibnamefont {Moriya}},\ }\bibfield  {title}{{\bibinfo {title} {New mechanism of anisotropic superexchange interaction,}}\ }\href {\doibase
10.1103/PhysRevLett.4.228} {\bibfield  {journal} {\bibinfo  {journal}
		{Phys. Rev. Lett.}\ }\textbf {\bibinfo {volume} {4}},\ \bibinfo {pages}
	{228} (\bibinfo {year} {1960})}\BibitemShut {NoStop}
	\bibitem [{\citenamefont {Ergotropy}\ \emph {}(2004) \citenamefont {Ergotropy}}]{Ergotropy}
\BibitemOpen
\bibfield  {author} {\bibinfo {author} {\bibfnamefont {A. E.}\ \bibnamefont
		{Allahverdyan}}, \bibinfo {author} {\bibfnamefont {R.}~\bibnamefont
		{Balian}}, and \bibinfo {author} {\bibfnamefont {Th. M.}~\bibnamefont
		{Nieuwenhuizen},}\
}\bibfield  {title} { {\bibinfo {title} {Maximal work extraction from finite quantum systems,}}\ }\href {\doibase
https://doi.org/10.1209/epl/i2004-10101-2} {\bibfield {journal} {\bibinfo{journal}
		{Europhysics Letters (EPL),}\ }\textbf {\bibinfo {volume} {67}},\ \bibinfo {pages}
	{565} (\bibinfo {year} {2004})}\BibitemShut {NoStop}	
	\bibitem [{\citenamefont {Mendes}\ \emph {et~al.}(2021) \citenamefont {Mendes}}]{Mendes}
\BibitemOpen
\bibfield  {author} {\bibinfo {author} {\bibfnamefont {U. C.}\ \bibnamefont
		{Mendes}}, \bibinfo {author} {\bibfnamefont {J. S.}~\bibnamefont
		{Sales},} and \bibinfo {author} {\bibfnamefont { N. G.}\ \bibnamefont {de Almeida},}\
}\bibfield  {title}{{\bibinfo {title} {Quantum Otto thermal machines powered by Kerr nonlinearity,}}\ }\href {\doibase
https://doi.org/10.1088/1361-6455/ac291a} {\bibfield  {journal} {\bibinfo  {journal}{J. Phys. B: At. Mol. Opt. Phys.}\ }\textbf {\bibinfo {volume} {54}},\ \bibinfo {pages}
	{175504} (\bibinfo {year} {2021})}\BibitemShut {NoStop}
	\bibitem [{\citenamefont {Sourav}\ \emph {et~al.}() \citenamefont {Sourav}}]{Sourav}
\BibitemOpen
\bibfield  {author} {\bibinfo {author} {\bibfnamefont {S.}\ \bibnamefont
		{Karar}}, \bibinfo {author} {\bibfnamefont {S.}~\bibnamefont
		{Datta}}, \bibinfo {author} {\bibfnamefont {S.}~\bibnamefont
		{Ghosh}}, and \bibinfo {author} {\bibfnamefont { A. S.}\ \bibnamefont {Majumdar}},\
}\bibfield  {title}{{\bibinfo {title} {Anharmonicity can enhance the performance of quantum refrigerators,}}\ }\href {\doibase
https://doi.org/10.48550/arXiv.1902.10616} {\bibfield  {journal} {\bibinfo  {journal}{arXiv:1902.10616 [quant-ph]}\ }\textbf {\bibinfo {volume} {}},\ \bibinfo {pages}
	{}}\BibitemShut {NoStop}		
	\bibitem [{\citenamefont {Jarzynski1995}\ \emph {}(1995) \citenamefont {Jarzynski1997}}]{Jarzynski1997}
\BibitemOpen
\bibfield  {author} {\bibinfo {author} {\bibfnamefont {C.}\ \bibnamefont
		{Jarzynski}},\
}\bibfield  {title}{{\bibinfo {title} {Nonequilibrium equality for free energy differences,}}\ }\href {\doibase
	https://doi.org/10.1103/PhysRevLett.78.2690} {\bibfield  {journal} {\bibinfo  {journal}
		{Phys. Rev. Lett.}\ }\textbf {\bibinfo {volume} {78}},\ \bibinfo {pages}
	{2690} (\bibinfo {year} {1997})}\BibitemShut {NoStop}
\bibitem [{\citenamefont {Crooks}\ \emph {}(1999)\citenamefont {Crooks}}]{Crooks}
\BibitemOpen
\bibfield  {author} {\bibinfo {author} {\bibfnamefont {C. E.}\ \bibnamefont
		{Crooks}},\
}\bibfield  {title}{{\bibinfo {title} {Entropy production fluctuation theorem and the nonequilibrium work relation for free energy differences,}}\ }\href {\doibase
https://doi.org/10.1103/PhysRevE.60.2721} {\bibfield  {journal} {\bibinfo  {journal}{Phys. Rev. E,}\ }\textbf {\bibinfo {volume} {60}},\ \bibinfo {pages}
	{2721} (\bibinfo {year} {1999})}\BibitemShut {NoStop}	
	\bibitem [{\citenamefont {Campisi}\ \emph {}(2011) \citenamefont {Campisi}}]{Campisi}
\BibitemOpen
\bibfield  {author} {\bibinfo {author} {\bibfnamefont {M.}\ \bibnamefont
		{Campisi}}, \bibinfo {author} {\bibfnamefont {P.}~\bibnamefont
		{Hanggi}}, and \bibinfo {author} {\bibfnamefont {P.}~\bibnamefont
		{Talkner},}\
}\bibfield  {title}{{\bibinfo{title}{Quantum Fluctuation Relations: Foundations and Applications,}}\ }\href {\doibase
https://doi.org/10.1103/RevModPhys.83.771
} {\bibfield  {journal} {\bibinfo  {journal}
		{Rev. Mod. Phys.}\ }\textbf {\bibinfo {volume} {83}},\ \bibinfo {pages}
	{771} (\bibinfo {year} {2011})}\BibitemShut {NoStop}	
	\bibitem [{\citenamefont {Spekkens}\ \emph {}(2008) \citenamefont {Spekkens}}]{Spekkens}
\BibitemOpen
\bibfield  {author} {\bibinfo {author} {\bibfnamefont {G.}\ \bibnamefont
		{Gour}}, and \bibinfo {author} {\bibfnamefont {R. W.}~\bibnamefont
		{Spekkens},}\
}\bibfield  {title}{{\bibinfo{title}{The resource theory of quantum reference frames:
manipulations and monotones,}}\ }\href {\doibase
https://doi.org/10.1103/RevModPhys.91.025001
} {\bibfield  {journal} {\bibinfo  {journal}
		{New J. Phys.}\ }\textbf {\bibinfo {volume} {10}},\ \bibinfo {pages}
	{033023} (\bibinfo {year} {2008})}\BibitemShut {NoStop}	
	\bibitem [{\citenamefont {Brandao}\ \emph {}(2013) \citenamefont {Brandao}}]{Brandao}
\BibitemOpen
\bibfield  {author} {\bibinfo {author} {\bibfnamefont {F.}\ \bibnamefont
		{Brandão}}, \bibinfo {author} {\bibfnamefont {M.}\ \bibnamefont
		{Horodecki}}, \bibinfo {author} {\bibfnamefont {J.}\ \bibnamefont
		{Oppenheim}}, \bibinfo {author} {\bibfnamefont {J. M.}\ \bibnamefont
		{Renes},} and \bibinfo {author} {\bibfnamefont {R. W.}~\bibnamefont
		{Spekkens},}\
}\bibfield  {title}{{\bibinfo{title}{Resource theory of quantum states out of thermal equilibrium,}}\ }\href {\doibase
https://doi.org/10.1103/PhysRevLett.111.250404
} {\bibfield  {journal} {\bibinfo  {journal}
		{Phys. Rev. Lett.}\ }\textbf {\bibinfo {volume} {111}},\ \bibinfo {pages}
	{250404} (\bibinfo {year} {2013})}\BibitemShut {NoStop}		
	\bibitem [{\citenamefont {Baumgratz}\ \emph {}(2014) \citenamefont {Baumgratz}}]{Baumgratz}
\BibitemOpen
\bibfield  {author} {\bibinfo {author} {\bibfnamefont {T.}\ \bibnamefont
		{Baumgratz}}, and \bibinfo {author} {\bibfnamefont {M. B.}~\bibnamefont
		{Plenio},}\
}\bibfield  {title} { {\bibinfo {title} {Quantifying coherence,}}\ }\href {\doibase
https://doi.org/10.1103/PhysRevLett.113.140401
} {\bibfield  {journal} {\bibinfo  {journal}
		{Phys. Rev. Lett.}\ }\textbf {\bibinfo {volume} {113}},\ \bibinfo {pages}
	{140401} (\bibinfo {year} {2014})}\BibitemShut {NoStop} 	
	\bibitem [{\citenamefont {Winter}\ \emph {}(2014) \citenamefont {Winter}}]{Winter}
\BibitemOpen
\bibfield  {author} {\bibinfo {author} {\bibfnamefont {E.}\ \bibnamefont
		{Chitambar}}, \bibinfo {author} {\bibfnamefont {D.}\ \bibnamefont
		{Leung}}, \bibinfo {author} {\bibfnamefont {L.}\ \bibnamefont
		{Mancinska}}, \bibinfo {author} {\bibfnamefont {M.}\ \bibnamefont
		{Ozols}}, and \bibinfo {author} {\bibfnamefont {A.}~\bibnamefont
		{Winter},}\
}\bibfield  {title}{{\bibinfo{title}{Everything You Always Wanted to Know About LOCC (But Were Afraid to Ask)
,}}\ }\href {\doibase
https://doi.org/10.1007/s00220-014-1953-9
} {\bibfield  {journal} {\bibinfo  {journal}
		{Commun. Math. Phys.}\ }\textbf {\bibinfo {volume} {328}},\ \bibinfo {pages}
	{303} (\bibinfo {year} {2014})}\BibitemShut {NoStop}
	\bibitem [{\citenamefont {Brandao2}\ \emph {}(2015) \citenamefont {Brandao2}}]{Brandao2}
\BibitemOpen
\bibfield  {author} {\bibinfo {author} {\bibfnamefont {F.}\ \bibnamefont
		{Brandão}}, \bibinfo {author} {\bibfnamefont {M.}\ \bibnamefont
		{Horodecki}}, \bibinfo {author} {\bibfnamefont {N.}\ \bibnamefont
		{Ng}}, \bibinfo {author} {\bibfnamefont {J. }\ \bibnamefont
		{Oppenheim}}, and \bibinfo {author} {\bibfnamefont {S.}~\bibnamefont
		{Wehner},}\
}\bibfield  {title}{{\bibinfo {title}{The second laws of quantum thermodynamics,}}\ }\href {\doibase
https://doi.org/10.1073/pnas.1411728112
} {\bibfield  {journal} {\bibinfo  {journal}
		{Proc. Natl.
Acad. Sci.}\ }\textbf {\bibinfo {volume} {112}},\ \bibinfo {pages}
	{3275} (\bibinfo {year} {2015})}\BibitemShut {NoStop}	
	\bibitem [{\citenamefont {Gour}\ \emph {}(2019)\citenamefont {Gour}}]{Gour}
\BibitemOpen
\bibfield  {author} {\bibinfo {author} {\bibfnamefont {E.}\ \bibnamefont
		{Chitambar}}, and \bibinfo {author} {\bibfnamefont {G.}~\bibnamefont
		{Gour},}\
}\bibfield  {title}{{\bibinfo{title}{Quantum resource theories,}}\ }\href {\doibase
https://doi.org/10.1103/RevModPhys.91.025001
} {\bibfield  {journal} {\bibinfo  {journal}
		{Rev. Mod. Phys.}\ }\textbf {\bibinfo {volume} {91}},\ \bibinfo {pages}
	{025001} (\bibinfo {year} {2019})}\BibitemShut {NoStop}	
	\bibitem [{\citenamefont {Mauro}\ \emph {}(2016) \citenamefont {Mauro}}]{Mauro}
\BibitemOpen
\bibfield  {author} {\bibinfo {author} {\bibfnamefont {F.}\ \bibnamefont
		{Albarelli}}, \bibinfo {author} {\bibfnamefont {A.}~\bibnamefont
		{Ferraro}}, \bibinfo {author} {\bibfnamefont {M.}~\bibnamefont
		{Paternostro}}, and \bibinfo {author} {\bibfnamefont {M. G. A.}~\bibnamefont
		{Paris},}\
}\bibfield  {title} { {\bibinfo {title} {Nonlinearity as a resource for nonclassicality in anharmonic systems,}}\ }\href {\doibase
https://doi.org/10.1103/PhysRevA.93.032112
} {\bibfield {journal} {\bibinfo{journal}
		{Phys. Rev. A,}\ }\textbf {\bibinfo {volume} {93}},\ \bibinfo {pages}
	{032112} (\bibinfo {year} {2016})}\BibitemShut {NoStop}	
	\bibitem [{\citenamefont {Kurizki23}\ \emph {}(2016) \citenamefont {Kurizki23}}]{Kurizki23}
\BibitemOpen
\bibfield  {author} {\bibinfo {author} {\bibfnamefont {W.}\ \bibnamefont
		{Niedenzu}}, and \bibinfo {author} {\bibfnamefont {G.}~\bibnamefont
		{Kurizki},}\
}\bibfield  {title} {{\bibinfo {title} {Cooperative many-body enhancement of quantum thermal machine power,}}\ }\href {\doibase
https://doi.org/10.48550/arXiv.1806.10810
} {\bibfield {journal} {\bibinfo{journal}
		{New J. Phys.}\ }\textbf {\bibinfo {volume} {20}},\ \bibinfo {pages}
	{113038} (\bibinfo {year} {2018})}\BibitemShut {NoStop}		
	\bibitem [{\citenamefont {GWatanabe}\ \emph {}(2016) \citenamefont {GWatanabe}}]{GWatanabe}
\BibitemOpen
\bibfield  {author} {\bibinfo {author} {\bibfnamefont {G.}\ \bibnamefont
		{Watanabe}}, \bibinfo {author} {\bibfnamefont {B. P.}~\bibnamefont
		{Venkatesh}}, \bibinfo {author} {\bibfnamefont {P.}~\bibnamefont
		{Talkner}}, \bibinfo {author} {\bibfnamefont {M.-J.}~\bibnamefont
		{Hwang}}, and \bibinfo {author} {\bibfnamefont {A.}~\bibnamefont
		{del Campo},}\
}\bibfield  {title} { {\bibinfo {title} {Quantum Statistical Enhancement of the Collective Performance of Multiple Bosonic Engines,}}\ }\href {\doibase
https://doi.org/10.48550/arXiv.1904.07811
} {\bibfield {journal} {\bibinfo{journal}
		{Phys. Rev. Lett.}\ }\textbf {\bibinfo {volume} {124}},\ \bibinfo {pages}
	{210603} (\bibinfo {year} {2020})}\BibitemShut {NoStop}	
	\bibitem [{\citenamefont {Latune}\ \emph {}(2016) \citenamefont {Latune}}]{Latune}
\BibitemOpen
\bibfield  {author} {\bibinfo {author} {\bibfnamefont {C. L.}\ \bibnamefont
		{Latune}}, \bibinfo {author} {\bibfnamefont {I.}~\bibnamefont
		{Sinayskiy}}, and \bibinfo {author} {\bibfnamefont {F.}~\bibnamefont
		{Petruccione},}\
}\bibfield  {title} { {\bibinfo {title} {Collective heat capacity for quantum thermometry and quantum engine enhancements,}}\ }\href {\doibase
https://doi.org/10.48550/arXiv.2004.00032
} {\bibfield {journal} {\bibinfo{journal}
		{New J. Phys.}\ }\textbf {\bibinfo {volume} {22}},\ \bibinfo {pages}
	{083049} (\bibinfo {year} {2020})}\BibitemShut {NoStop}
	\bibitem [{\citenamefont {Kamimura}\ \emph {}(2016) \citenamefont {Kamimura}}]{Kamimura}
\BibitemOpen
\bibfield  {author} {\bibinfo {author} {\bibfnamefont {S.}\ \bibnamefont
		{Kamimura}}, \bibinfo {author} {\bibfnamefont {H.}~\bibnamefont
		{Hakoshima}}, \bibinfo {author} {\bibfnamefont {Y.}~\bibnamefont
		{Matsuzaki}}, \bibinfo {author} {\bibfnamefont {K.}~\bibnamefont
		{Yoshida}}, and \bibinfo {author} {\bibfnamefont {Y. }~\bibnamefont
		{Tokura},}\
}\bibfield  {title} { {\bibinfo {title} {Quantum-Enhanced Heat Engine Based on Superabsorption,}}\ }\href {\doibase
https://doi.org/10.48550/arXiv.2106.10813
} {\bibfield {journal} {\bibinfo{journal}
		{Phys. Rev. Lett.}\ }\textbf {\bibinfo {volume} {128}},\ \bibinfo {pages}
	{180602} (\bibinfo {year} {2022})}\BibitemShut {NoStop}	
	\bibitem [{\citenamefont {Souza}\ \emph {}(2016) \citenamefont {Souza}}]{Souza}
\BibitemOpen
\bibfield  {author} {\bibinfo {author} {\bibfnamefont {L. D. S. }\ \bibnamefont
		{Souza}}, \bibinfo {author} {\bibfnamefont {G.}~\bibnamefont
		{Manzano}}, \bibinfo {author} {\bibfnamefont {R.}~\bibnamefont
		{Fazio}}, and \bibinfo {author} {\bibfnamefont {F.}~\bibnamefont
		{Iemini},}\
}\bibfield  {title} { {\bibinfo {title} {Collective effects on the performance and stability of quantum heat engines,}}\ }\href {\doibase
https://doi.org/10.48550/arXiv.2106.13817
} {\bibfield {journal} {\bibinfo{journal}
		{Phys. Rev. E}\ }\textbf {\bibinfo {volume} {106}},\ \bibinfo {pages}
	{014143} (\bibinfo {year} {2022})}\BibitemShut {NoStop}	
\end{thebibliography}


\end{document}